\documentclass[aip,pre]{revtex4-1}
\usepackage{graphicx,calc,empheq,amssymb}
\usepackage{amsmath,amsthm}
\usepackage{amsfonts}
\usepackage{hyperref}
\usepackage{bm}
\usepackage{mathrsfs}
\usepackage{amsmath}
\usepackage{amsthm}
\usepackage{eucal}  
\usepackage{enumitem}  
\usepackage[T1]{fontenc} 
\usepackage[subnum]{cases}
\if@mathematic
   \def\vect#1{\ensuremath{\mathchoice
                     {\mbox{\boldmath$\displaystyle\mathbf{#1}$}}
                     {\mbox{\boldmath$\textstyle\mathbf{#1}$}}
                     {\mbox{\boldmath$\scriptstyle\mathbf{#1}$}}
                     {\mbox{\boldmath$\scriptscriptstyle\mathbf{#1}$}}}}
\else
   \def\vect#1{\ensuremath{\mathchoice
                     {\mbox{\boldmath$\displaystyle#1$}}
                     {\mbox{\boldmath$\textstyle#1$}}
                     {\mbox{\boldmath$\scriptstyle#1$}}
                     {\mbox{\boldmath$\scriptscriptstyle#1$}}}}
\fi

\if@vprodwedge
  
\else
  
\fi

\newcommand{\PP}{\mathbb{P}} 
\newcommand{\R}{\mathbb{R}} 
 
\newcommand{\N}{\mathbb{N}}

\newtheorem{theorem}{Theorem}{\bf}{\it}
\newtheorem{proposition}{Proposition}{\bf}{\it}
{\bf}{\it}
{\bf}{\it}
\newtheorem{remark}{Remark}{\bf}{\it}
\newtheorem{definition}{Definition}{\bf}{\it}
\theoremstyle{lemma}
\newtheorem{lemma}{Lemma}{\bf}{\bf}
\newenvironment{bew}[2]{\removelastskip\vspace{6pt}\noindent
 {\it Proof  #1.}~\rm#2}{\par\vspace{6pt}}             

\draft 
\begin{document}
\title{{About the foundation of the Kubo Generalized Cumulants theory. A revisited and corrected approach}}
\author{Marco Bianucci}
\email[]{marco.bianucci@cnr.it}
\affiliation{Istituto di Scienze Marine, Consiglio Nazionale delle Ricerche (ISMAR - CNR),\\
19032 Lerici (SP), Italy}
\author{Mauro Bologna}
\email[]{marco.bianucci@cnr.it}
\affiliation{Instituto de Alta Investigaci\'on, Universidad de Tarapac\'a, Casilla 6-D Arica, Chile}
\date{\today}
%

\begin{abstract}
More than fifty years ago, in a couple of seminal works~\cite{kuboGenCumJPSJ17,kuboGenCumJMP4} Kubo introduced the 
important idea of generalized cumulants, 
extending to stochastic operators this concept, implicitly introduced by Laplace in 1810.
Kubo's idea has been applied in several 
branches of physics, where the result of the average process is a Lioville operator or
an effective time evolution operator for the 
density matrix of spin systems or the reduced density matrix for boson-fermions etc. 
Despite this success, the theoretical developments in these Kubo works pose problems that were highlighted many 
years ago by Fox and van Kampen and never solved. These weaknesses and errors, in 
particular concerning the factorization property of exponentials of cumulants
and the explicit expressions that give generalized cumulants in terms of generalized moments and vice-versa,
caused some perplexity (and confusion) about the possible application of this procedure, limiting its use, in practice. 
In the present paper, we give a sound ground
to the approach to cumulant operators, working in a general framework that shows the potentiality 
of the old Kubo's idea, today not yet fully exploited. It results that for the same moment operators, 
different definitions of generalized cumulants can be adopted. A general Kubo-Meeron closed-form formula giving cumulant operators in terms of moment operators cannot be obtained, but the reverse one, cumulants in terms of operators, is given and, noticeably, formally it {\em does not} depend on the
specific nature of the moments, but just on the definition of the generalized cumulants.
\end{abstract}
\pacs{}

\maketitle 

\section{Introduction}
Cumulants where implicitly introduced by Laplace in 1810~\cite{Laplace1810,Laplace1812,Laplace1820}
in his proof of the central limit theorem, as coefficients of the power expansion of the logarithm of the characteristic function. 
It is noticeable that
nobody thought of defining the coefficients of this expansion as separate entities and to study their very important properties until Thiele  in 1889. Cumulants have been recognized as a very powerful tool in any field of random processes: cumulants change in a very simple way when the underlying random
variable is subject to an affine transformation, cumulants may be used in a simple way to describe the
difference between a distribution and its simplest Gaussian approximation and,
due to the factorization property of the exponential of a sum of commuting quantities, 
the cumulants have the useful fundamental property that they
are not vanishing if and only if they refer to  variables that are statistically  ``connected'' to each other.

In the early 60s of the past century, the great scientist R. Kubo had the idea  to generalize 
the definition of cumulants~\cite{kuboGenCumJPSJ17,kuboGenCumJMP4} (K62-63 hereafter), including stochastic operators, 
as we shall see hereafter.
At first sight, an extension of the cumulant concept to non-commuting stochastic processes
comes up against the fact that for non-commuting quantities the factorization property of exponentials does not hold.  
However, Kubo proposed a way to overcome this problem
by extending the definition of exponentials.
After this old work by Kubo, which, as we shall see, lacked some formal gaps and contained some errors (giving rise to a series of 
critical remarks~\cite{fJMP17,fJMP20}), the theory of cumulants made much progress. In particular, we consider very interesting 
the recent formal framework  where concepts related to combinatorial calculation, connected or crossed partitions etc. are used
and the old results on the M\"obius inversion on the lattice of partitions play a crucial role (e.g.,~\cite{nsDMJ92,lEJC23}). 
However, in this approach, references to the notion of probability, average, or characteristic function do
not enter, thus physics, or, more in general, the nature (and the definition) of the generalized moment generating function 
is a starting point, rather than a result. On the other hand, once the moment generating function is given and 
our formula that gives moments in terms of cumulants is obtained (see Eq.~(\ref{mukappaCombiMulti})),
we believe that the theory of lattice of partitions could be a powerful tool to obtain the inverse formula (cumulants in terms of moments) for some specific case of interest, a result that, as we shall show, cannot be achieved by our general approach.

In this work, we shall focus on the classical Kubo's method where (generalized)  moments arise by  problems 
in classical o quantum physics (and beyond physics, too) and the (generalized)  standard cumulants technique are introduced as a tool for some systematic expansion 
of these moment-operators of interest.
In these problems, the Kubo's original idea remains the pillar on which to base any
extension of the concept of cumulants and it is almost always referred (and cited)  as the basis of 
any  papers (also the more recent ones) concerning, e.g.,  the reduced or stochastic Liouville equation (see below), 
the evolution of density matrix of spin systems~\cite{fJCP49,ydfJCP62} or the reduced density matrix 
for many body boson or fermion
systems (e.g.,~\cite{smPRA88,ziesche2000,mCPL289,mIJQC70,jmJCP125,pbPRA90,rcspmJCP141,rmPRA92}). However, although mentioned, the general results of K62-63 
are not used, and a more analytical approach, case dependent, is adopted. A fairly recent example is the case of the above-mentioned works treating problems in the field of many-electron densities and reduced
density matrices and related Green functions. 
The analytical developments in these works mainly aim at obtaining 
a direct expression that gives reduced density matrices (or Green functions) in terms of generalized 
cumulants (see~\cite[TABLE II]{mIJQC70} or~\cite{mCPL289}), but, as it we shall show in this work,
this formula can be given in a much more general form (see 
Lemma~\ref{lemmaMeeron2}, Eq.~(\ref{DensMom_vs_DensCum})),
formally {\em independent}
of the specific nature of the generalized moments (in this case density matrices, Green functions etc..). 
This example will be extensively  discussed in Section~\ref{sec:fermions}.

The reason why, although so much cited, Kubo's work is not being fully exploited is precisely that, as some authoritative remarks have pointed out, some steps and conclusions are not well justified, and some results are actually wrong. To remedy this situation we shall review shortly 
 the Kubo's idea of generalized cumulants by using a specific example borrowed from classical statistical physics.
\section{The Kubo's idea for generalized cumulants and the related criticism\label{sec:kuboscumulant}}

 Let us assume to have a (classical or quantum) density function (DF) $P_{\xi}(\vect{x};t)$ where $\vect{x}$ are variables of interest or observables that we can
 measure in some way, while $\xi$ represents a collection of variables that we cannot measure and of which we just know a 
 DF $\rho(\xi)$ (or a probability measure $\text{d}\mu:=\rho(\xi) \text{d}\xi$) over which we can average. We assume that  $P_{\xi}(\vect{x};t)$  satisfies
 the following generic equation of motion:
 \begin{equation}
 \label{p(t)}
 \partial_t P_{\xi}(\vect{x};t)={\cal L}_0(\vect{x})\,P_{\xi}(\vect{x};t)+ {\cal L}_I(\vect{x},\xi)\, P_{\xi}(\vect{x};t)
 \end{equation}
 in which the   evolution operator ${\cal L}_0(\vect{x})$ depends only on the variables of interest, while the term ${\cal L}_I(\vect{x},\xi)$ 
 expresses the interaction with the rest of the system. 
 In interaction representation Eq.~(\ref{p(t)}) becomes
 \begin{equation}  
 \partial_t \tilde{P}_\xi (\vect{ x};t) 
 =\Omega(\vect{x},\xi;t) \tilde{P}_\xi (\vect{ x};t),  
 \label{app:liouvilleEq.1}  
 \end{equation}  
 where 
 \begin{equation} 
 \tilde{P}_\xi (\vect{ x};t):=  
 e^{-{\cal L}_0(\vect{x}) t}P_\xi ( \vect{ x};t),
 \end{equation}
 and
 \begin{equation}
 \label{LIt} 
 \Omega(\vect{x},\xi;t) := e^{-{\cal L}_0(\vect{x}) t}{\cal L}_I(\vect{x},\xi) e^{{\cal L}_0(\vect{x}) t}.
 \end{equation}
Integrating Eq.~(\ref{app:liouvilleEq.1}) and then averaging over the space of the non observable variables $\xi$, we get an equation for the reduced DF:
\begin{equation}  
\label{generalevolution3}
\tilde{P}(\vect{x};t) = 
\langle \stackrel{\longleftarrow}{\exp}  \left[\int_0^t  \text{d}u\,\Omega(u)\right]\rangle P(\vect{x};0),
\end{equation}  
 in which $ \Omega(u):= \Omega(\vect{x},\xi;u)$, $\stackrel{\longleftarrow}{\exp}[...]$ is the standard $t$-ordered exponential, ${P}(\vect{x};t):=\langle {P}_\xi (\vect{ x};t) \rangle$ 
 and we have assumed that ${P}_\xi (\vect{ x};t=0)={P} (\vect{ x};t=0)$: 
at the initial time the DF of the variables of interest does not depend on the hidden variables $\xi$. 
Usually, we aim at obtaining an equation of motion for the reduced DF
\begin{equation}
\label{eqmot}
\partial_t \tilde{P}(\vect{x};t)  = \mathfrak{G}\left(t,\tilde{P}(\vect{x};t)\right)
\end{equation}
or, more realistically, a systematic expansion of the local or non local operator $\mathfrak{G}\left(t,\tilde{P}(\vect{x};t)\right)$, in terms of the fluctuations of the variables over which we have averaged. This can be done formally by writing the solution of Eq.~(\ref{eqmot}) as 
\begin{equation}  
\label{generalevolution3_}
\tilde{P}(\vect{x};t) = 
\sum_{i=0}^\infty {\cal K}^{(i)}(\vect{x};t) P(\vect{x};0).
\end{equation}  
The problem is to find a relation between the operators ${\cal K}^{(i)}$ and the series that we obtain by expanding the $t$-ordered exponential of Eq.~(\ref{generalevolution3}).
For that,  Kubo~\cite{kuboGenCumJPSJ17} started by 
considering $ \Omega(\vect{x},\xi;t)$   as a ``random'' operator,  the statistics of which is determined by the probability measure $\text{d}\mu=\rho(\xi)\text{d}\xi$. Given that,  the time-evolution operator of Eq.~(\ref{generalevolution3}),  can be considered 
as a sort of
moment generating function ${\cal M}(t)$ for the q-number \footnote{For historical reason we shall use the definition
	of q-numbers as objects of a {\em non commutative} algebra, as opposed to c-numbers that are
	objects of a {\em  commutative} algebra. The operators considered in the present work are generally q-numbers.}
 ``stochastic process'' $\Omega(t)$:
 \begin{equation}
 \label{MG}
 {\cal M}(t):= \langle \stackrel{\longleftarrow}{\exp}  \left[\int_0^t \text{d}u\,\Omega(u)\right]\rangle.
 \end{equation}
The operator ${\cal M}(t)$ of Eq.~(\ref{MG}) can be written as 
 \begin{equation}
 \label{MnM}
 {\cal M}(t)=\sum_{n=0}^\infty {\cal M}_n(t),
 \end{equation}
 \begin{align}
 \label{defMn}
& {\cal M}_n(t)=\frac{1}{n!}\langle \left\{\left[\int_0^t \text{d}u\, \Omega(u)\right]^n
 \right\}_O\rangle \nonumber \\
&=\frac{1}{n!} \int_0^t \mbox{d}u_1 \int_0^t\mbox{d}u_2...\int_0^t\mbox{d}u_n
 \langle \left\{  \Omega(u_1) \Omega(u_2)... \Omega(u_n)\right\}_O \rangle\nonumber \\
 &=\int_0^t \mbox{d}u_1 \int_0^{u_1}\mbox{d}u_2...\int_0^{u_{n-1}}\mbox{d}u_n
 \langle  \Omega(u_1) \Omega(u_2)... \Omega(u_n) \rangle.
 \end{align}
 In Eq.~(\ref{defMn}) we have introduced the notation $\left\{...\right\}_O$ that hereafter means ``time ordering''; a definition that shall be better
 specified in the following, but that for the purpose of the present Section can be simply associated to the definition of the $t$-ordered exponential.  
 The average
 $\langle \left\{\Omega(u_1)\Omega(u_2)...\Omega(u_n) \right\}_O \rangle$ ($0\le u_i\le t$, $1\le i\le n$) is called $t$-ordered $n$-moment 
 density function.

Then, in analogy with c-number (objects of a {\em commutative} algebra) stochastic processes, to the  moment generating function ${\cal M}(t)$, Kubo associates the corresponding 
cumulant generating function ${\cal K}(t)$ (in general, still a q-number):
\begin{equation}
\label{CumulantDef}
\exp_M \left[{\cal K}\right] := \langle \stackrel{\longleftarrow}{\exp}  \left[\int_0^t \text{d}u\,\Omega(u)\right]\rangle.
\end{equation}
Expanding ${\cal K}$ in components, ${\cal K}(t)=\sum_{n=1}^\infty {\cal K}_n(t)$, and exploiting Eq.~(\ref{MnM}), we get
\begin{equation}
\label{KG}
\exp_M \left[\sum_{n=1}^\infty {\cal K}_n(t)\right]:=\sum_{n=0}^\infty {\cal M}_n(t).
\end{equation}
In the last two equations, $\exp_M$ is a generalized exponential, the proper definition and the discussion of which is the central task of the present work.
The ${\cal K}_n$  components are found by expanding the generalized exponential in the l.h.s. of Eq.~(\ref{KG}) and 
comparing terms of the same order on both sides of the same equation, starting from $n=1$. 
In this way the analogy with the standard definition for commuting processes seems complete. 
Remains to show {\em how and if} this procedure works, and this goes back to the  definition  
of the generalized exponential $\exp_M$ introduced in Eqs.~(\ref{CumulantDef}). As it is well known,
the fundamental characteristic of the cumulants is to be vanishing when including independent processes and,
for c-number ``stochastic variables'', this is directly related to the factorization property of the exponential 
function: $\exp(a+b)=\exp(a)\exp(b)$. 
To the best of our knowledge, no exponential function extension  (actually, no function) 
shares this property when  the arguments are q-numbers, instead of c-numbers.
  Thus, if ${\cal K}$ and ${\cal K}'$ are  non commuting 
  cumulant generating functions, then, in general, the following equation
\begin{align}  
\label{generalevolution4_3}
\exp_M\left[{\cal K}\right]\exp_M\left[{\cal K}'\right]=\exp_M\left[{\cal K}+{\cal K}'\right],
\end{align}  
{\em does not hold true, whatever the definition of the generalized exponential}. This was the main point in Fox~\cite{fJMP17}  criticism of the  general theoretical foundation of    
Kubo's results. In fact, concerning the factorization property of generalized exponentials of cumulants, he wrote
``it is not difficult to show that for
$T$ ordering, counter examples [...] can be constructed, and that even though the cluster property is
nevertheless true, Kubo's argument does not justify it.'' 

Fox criticized also the Kubo's closed-form formula that gives directly the generalized cumulants in terms of generalized moments. In fact, he 
wrote: ``Eq. (6.9) of Kubo's paper~\cite{kuboGenCumJPSJ17} purports to be a 
closed-form formula for ordered cumulants [...]'', but, for him, it is not: 
``The nature of the error is somewhat subtle, [...] but very significant''. 
Fox stated that in the case where the generalized exponential is the $t$-ordered one, the right
procedure to obtain cumulants from moments is that of van Kampen~\cite{vkP74a,vkP74b}, that he explicitly proved to be different from Kubo's.
However, van Kampen~\cite{vkP74a,vkP74b} and Fox himself, working directly (and heavily) with the 
analytical expressions of the 
cumulants defined by the $t$-ordered exponential, demonstrated the ``clustering'' property of cumulants operators, 
namely, a special case of the general Kubo's result. Something similar was demonstrated by Terwiel~\cite{tP74} 
working with the Zwanzig projection approach. 
Moreover,  van Kampen~\cite{vkP74a,vkP74b} obtained a  closed-form formula that gives cumulants in terms of moments, still
valid in the special (although important) case of $t$-ordered generalized exponential. In summary,  
Fox and van Kampen did not consider their results and the one of Terwiel, as particular cases of the Kubo's generalized cumulant theory.

As we have already stressed, the concept of generalized cumulants  
has been developed and applied in many different contexts of physics:  dynamical  mean-field  
theory~\cite{crbPRB85}, coupled bosons and valence electrons  coupled  to  
plasmons~\cite{hPS21,ahkPRL77},  multiple plasmons in 
photoemission~\cite{glsrgkrssrPRL107}, dynamic perturbations in metals~\cite{m-hrtPRB3},
one-electron Green function~\cite{krrPRB90}, entanglement between electronic 
domains and reduced density matrix for coupled electrons in general~\cite{smPRA88,ziesche2000,mCPL289,mIJQC70,jmJCP125,pbPRA90,rcspmJCP141,rmPRA92}, extending Wick theorem in particle 
physics~\cite{Mahan2008,Mukherjee1995,hkJCP401} and in many field of  statistical 
mechanics~\cite{hytsPA433,suzukiJPSJ22,sPTPS69,pppJPSJ82,Schneider89,ydfJCP62,fJCP49,vkP74a,vkP74b,tP74,fJMP17,fJMP20,TokuyamaPA102,TokuyamaPA109},
just to quote some among many 
others. 

In practice, when the generalized exponential 
 introduced by Kubo is just the time ordered one, the explicit calculus by van Kampen and Fox guarantee the cluster property of the
cumulants, namely that cumulants vanish if the variables are ``statistically un-connected''. 
In other cases, a specific formalism has been developed to address the generalization of the concept of cumulants to individual problems
 (e.g., in 
the already cited work on fermions). On the contrary, Kubo's theory aims to be very  general, but
 the weakness (and, unfortunately, some errors) in 
 the theoretical approach  originated some confusion about the proper way  to use it
 (e.g.,~\cite{fJCP49}, where Eqs.~(2.14a)-(2.14d) are not correct, or~\cite{sPTPS69} where 
 Eq.(4.9) is wrong because it is erroneously assumed that if the  fluctuating function $\eta(t)$ is Gaussian, then the 
 stochastic operator
  $\Omega(t)=\eta(t)Q(t)$ is also Gaussian in the Kubo's meaning).

Here we shall re-found  the K62-63 approach to cumulants, within a sounded formal scheme. Thus we shall show that, although 
in general the fundamental Eq.~(6.9) of Kubo's paper~\cite{kuboGenCumJPSJ17} is at least questionable,  the
 general approach (and idea) is correct. Moreover, we shall give the right  relation between moments and cumulants.
 
In the final section, we will shortly show to apply the present theoretical framework to some noticeable examples.

\section{Some formal definitions and results\label{formalpart}}
        Let $\mathfrak{B}$ be a vector space over a field with finite or infinite (typically) dimension. We indicate with 
        $\mathscr{B}=\{\bm{ \mathfrak{b}}_i\}$, ($i\in\N$, or $\R$) a selected basis of $\mathfrak{B}$, that we shall call
        the ``original special basis''.
For $n\in \N$, let
$$\mathfrak{B}^{\otimes n}:=\underbrace{\mathfrak{B}\otimes ...\otimes \mathfrak{B}}_{n\,times}$$
be the $n$-th tensor power of the $\mathfrak{B}$, of which, for any $n$, the special basis $\mathscr{B}^{\otimes n}$ is the $n$-th tensor powers of  $\mathscr{B}$.  We indicate with $\mathfrak{M}$ the vector space given by the direct sum of
the vector spaces $\mathfrak{B}^{\otimes n}$ for $1 \le n < \infty$:
\begin{equation}
\mathfrak{M}:=\mathfrak{B}\oplus \mathfrak{B}^{\otimes 2}\oplus \mathfrak{B}^{\otimes 3}...\oplus \mathfrak{B}^{\otimes k}
\oplus \mathfrak{B}^{\otimes (k+1)}...
\end{equation}
The special basis $\mathscr{M} $of $\mathfrak{M}$ is given by the direct sum of the special basis $\mathscr{B}^{\otimes n}$:
\begin{align*}\mathscr{M}&=\{\bm{\mathfrak{m}}_{\bm{i}}\}  :=\{\bm{\mathfrak {m}}_{i_1,i_2,...,i_n,...}\}\\
&:=\mathscr{B}\oplus \mathscr{B}^{\otimes 2}\oplus \mathscr{B}^{\otimes 3}...\oplus\mathscr{B}^{\otimes k}
\oplus\mathscr{B}^{\otimes (k+1)}...
\end{align*}
with $i_1,i_2,...,i_n \in \N$, or $\R$ and $n\in\N$.
We call  operators the elements of $\mathfrak{M}$ and, for  subsequent use, we have explicitly indicate with $\bm{\mathfrak{m}}_{\bm{i}}:=\bm{\mathfrak{m}}_{i_1,i_2,...,i_n,...}$ 
 the elements of the
special basis of $\mathfrak{M}$.
For convenience, let's associate also a subscript index (possibly a continuous parameter) to the generic
operators: ${\cal A}_i\in\mathfrak{M} $, such that ${\cal A}_i\ne {\cal A}_j$ if $i\ne j$. 
\begin{remark}
The tensor product is not, in general, commutative, but, by the universal property, it is associative: 
$${\cal A}_1\otimes ({\cal A}_2\otimes{\cal A}_3)=({\cal A}_1\otimes {\cal A}_2)\otimes{\cal A}_3.$$
\end{remark}
            Now we define the mapping that we shall use to introduce the generalized moment generating function.
\begin{definition}
        \label{definition1}
A  $M$-projection  $\{{\cal A}_i\}_M$ is a linear map $\phi_M\,:\,\mathfrak{M} \to  \mathfrak{M}_M$ where  $\mathfrak{M}_M$ is a subset of $\mathfrak{M}$, 
        such that 
        \begin{enumerate}[label=(\roman*)]
                \item\label{idempotent}  the $M$-projection is idempotent, namely, it is a projection map: 
                $\phi_M \circ \phi_M=\phi_M$, or  $\{\{{\cal A}_i\}_M\}_M=\{{\cal A}_i\}_M$;
                \item\label{defR2} the $M$-projection is defined by mapping the elements of the special 
                basis: if ${\cal A}_i=\sum_{i_1,i_2,...,i_n} a_{i_1,i_2,...,i_n}\bm{\mathfrak{m}}_{i_1,i_2,...,i_n}$, then  
                $\{{\cal A}_i\}_M:= \sum_{i_1,i_2,...,i_n} a_{i_1,i_2,...,i_n}\{\bm{\mathfrak{m}}_{i_1,i_2,...,i_n}\}_M$;
                \item \label{defR1} the order of the operators does not matter:
                                \begin{equation}
                \label{defR}
                \left\{{\cal A}_1\otimes{\cal A}_2\otimes...\otimes{\cal A}_n \right\}_M = 
                        \left\{{\cal A}_{1'}\otimes{\cal A}_{2'}\otimes...\otimes{\cal A}_{n'} \right\}_M 
                \end{equation}
                where $1',2',...,n'$ is any permutation of the indices $1,2,...,n$ (that meas any permutation of the operators inside the brackets);
                \end{enumerate}
\end{definition} 
 Of course, the time ordering defined by the $t$-ordered exponential of Eq.~(\ref{MG}) is an $M$-projection map. In physics literature, 
 it is the most common among the $M$-projection 
 maps that correspond to effective ordering respect to some parameters to which depend the operators ${\cal A}_i\in \mathfrak{M}$. 
 Considering the $M$-projection associated to  the $t$-ordered exponential, the point~\ref{defR2} is strictly related to the following well known  surprising ``paradox''
 (see~\cite{tlhPRE75}, where this fact
caused confusion and see~\cite{bJSP138}, instead, for an example where this apparent paradox is
properly considered): let ${\cal K}(t)$ a linear operator and consider the following trivial identity:
\begin{equation}
{\cal K}(t)-{\cal K}(0) = \int_0^t \text{d}u\, \partial_u {\cal K}(u).
\end{equation}
Now, if we apply the  $t$-ordered exponential to both sides of the above equations,
 the consequent equation
\begin{equation}
\label{TexpAntinomia}
\stackrel{\longleftarrow}{\exp} [{\cal K}(t)-{\cal K}(0)] = \stackrel{\longleftarrow}{\exp} \left[\int_0^t \text{d}u\,\partial_u {\cal K}(u)\right]
\end{equation}
{\em does not hold!}. The pitfall stays in a wrong association, to the  $t$-ordered exponential, of
the time ordering operator. The point~\ref{defR2} of Definition~\ref{definition1} says that to apply
the $M$-projection ordering map to the l.h.s. of Eq.~(\ref{TexpAntinomia}), the operator
${\cal K}(t)-{\cal K}(0)$ must be expanded in the special basis, that is defined in the r.h.s. of the same equation
by the time integral $\int_0^t \text{d}u\,\partial_u {\cal K}(u)$. 
See Appendix~\ref{app:timeordering} for an explicit example and a clarification about this point.

Besides the time ordering, others examples of $M$-projection maps are,  total symmetrization:
 \begin{equation}
 \label{TotalSymm}
\left\{ {\cal A}_1\otimes{\cal A}_2\otimes...\otimes{\cal A}_n\right\}_M=\frac{1}{n!}\sum_{permutations} {\cal A}_1\otimes{\cal A}_2\otimes...\otimes{\cal A}_n
 \end{equation}
 or anti-symmetrization, operations that are involved in quantum mechanics of identical particles (see the example in Section~\ref{sec:fermions}).
 \begin{remark}\label{remark2}
By Definition~\ref{definition1}, point~\ref{defR1}, combining  the $M$-projection with the tensor product $\otimes$ we have a commutative operation:
$\left\{{\cal A}_1 \otimes {\cal A}_2\right\}_M:={\cal A}_1 \otimes_M {\cal A}_2={\cal A}_2 \otimes_M {\cal A}_1$.
However, as for time ordering, caution must be
taken using this definition of $\otimes_M$, because, as it is clear by the points \ref{defR2}, the $M$-projection severely  affects the algebraic structure 
of the set  $\mathfrak{M}$, 
because the associativity and the distributivity  properties does
 not hold anymore. In particular, in general we have:
 \begin{align}        
 \label{noassociative}
 &\left\{{\cal A}_1\otimes{\cal A}_2\otimes...\otimes{\cal A}_m\otimes{\cal A}_{m+1}\otimes...\otimes{\cal A}_{m+n} \right\}_M 
\nonumber \\
&\ne 
 \left\{{\cal A}_1\otimes{\cal A}_2\otimes...\otimes{\cal A}_m\right\}_M\otimes \left\{{\cal A}_{m+1}\otimes...\otimes{\cal A}_{m+n} \right\}_M
 \end{align}
\end{remark}  
For example, if the $M$-projection is the total symmetrization operator of Eq.~(\ref{TotalSymm}), it is clear that while in the
l.h.s. of Eq.~(\ref{noassociative}) we have a totally symmetric result, in the r.h.s. the 
symmetrization is not complete. 
 \begin{remark}
 	Eq.~(\ref{noassociative}) stays at the basis of the Fox criticism~\cite{fJMP17,fJMP20} to the Kubo's theory of cumulants.
 \end{remark}
 \begin{definition}\label{ABunconnected}
 	Given an $M$-projection defined on $\mathfrak{M}$, two operators ${\cal A}$ and ${\cal B}\in \mathfrak{M}$ are 
 	$M$-unconnected to each other when 
 \begin{equation}
 \label{ABunconnectedEq}
\left\{{\cal A}\otimes{\cal B}\right\}_M=\left\{{\cal A}\right\}_M\otimes \left\{{\cal B}\right\}_M.
 \end{equation}	
 \end{definition}
Notice, Eq.~(\ref{ABunconnectedEq}) does not implies $\left\{{\cal A}\otimes{\cal B}\right\}_M= \left\{{\cal B}\right\}_M
\otimes \left\{{\cal A}\right\}_M$.

       For any element ${\cal A}\in \mathfrak{M}$ let 
$${\cal A}^{n} 
:=\underbrace{{\cal A}\otimes ...\otimes{\cal A}}_{n\,times}$$ be
the $n$-th tensor power of ${\cal A}$. Of course ${\cal A}^{n} \in \mathfrak{M}$.

\begin{definition}\label{unconnected}
	If there are two subsets $ \mathfrak{M}^{(A)}$, $ \mathfrak{M}^{(B)}\subset\mathfrak{M}$ such that for any operator  ${\cal A}_i\in  \mathfrak{M}^{(A)}$ and ${\cal B}_j\in  \mathfrak{M}^{(B)}$, and for any $n_i,m_j,p,q\in\N$ we have
	\begin{align}
	\label{Rassociative}
	&      \left\{{\cal A}_1^{n_1}\otimes{\cal A}_2^{n_2}\otimes...\otimes{\cal A}_p^{n_p}
	\otimes{\cal B}_{1}^{m_{1}}\otimes{\cal B}_{2}^{m_{2}}\otimes...\otimes{\cal B}_q^{m_q} \right\}_M\nonumber \\
	=& 
	\left\{{\cal A}_1^{n_1}\otimes{\cal A}_2^{n_2}\otimes...\otimes{\cal A}_p^{n_p}\right\}_M 
	\otimes
	\left\{{\cal B}_{1}^{m_{1}}\otimes{\cal B}_{2}^{m_{2}}\otimes...\otimes{\cal B}_q^{m_q}\right\}_M
	\end{align}
	 then the subsets  $ \mathfrak{M}^{(A)}$ and $\mathfrak{M}^{(B)}$ are called $M$-unconnected to each other.
	\end{definition}
\begin{remark}
	\label{baseNOspazio}
	The two subsets $ \mathfrak{M}^{(A)}$ and $ \mathfrak{M}^{(B)}$ as in Definition~\ref{unconnected}
	are vector spaces. In fact 
\begin{align*}
	&\left\{\sum_i{\cal A}_i\otimes \sum_j{\cal B}_j \right\}_M=
	\sum_{i,j}\left\{{\cal A}_i\otimes {\cal B}_j \right\}_M\\
&=
	\sum_{i,j}\left\{{\cal A}_i\right\}_M\otimes\left\{ {\cal B}_j \right\}_M=\left\{\sum_i{\cal A}_i\right\}_M \otimes \left\{\sum_j{\cal B}_j \right\}_M
\end{align*}
	thus we shall use the therm $M$-unconnected ``subspaces'' instead of $M$-unconnected subsets. Moreover, 
 by Eq.~(\ref{Rassociative}) we see that the two unconnected sets can be infinitely extended by including all the possible powers of their original elements. 
 \end{remark}
\begin{remark}
	In the r.h.s. of Eq.~(\ref{Rassociative}), the ordering between the ${\cal A}_i$ and the ${\cal B}_j$ elements is usually important (here, the ${\cal A}_i$ stay at the left).
\end{remark}

\begin{proposition}\label{unconnected_bis}
	In Definition~(\ref{unconnected}), we can substitute Eq.~(\ref{Rassociative}) with the following equivalent one:
	\begin{equation}
	\label{Rassociative_bis}
	\left\{{\cal A}_i^{n}\otimes {\cal B}_j^{m}\right\}_M=\left\{{\cal A}_i^{n}\right\}_M \otimes \left\{ {\cal B}_j^{m}\right\}_M
	\;\;\forall n,m\in \N.
	\end{equation}
\end{proposition}
At first sight the condition given in Eq.~(\ref{Rassociative_bis}) looks less restrictive than that of Eq.~(\ref{Rassociative}),
however, the equivalence can be demonstrated by using the fact that the sets $\mathfrak{M}^{(A)}$ and $\mathfrak{M}^{(B)}$ are 
infinite dimensional vector spaces:
\begin{bew}{of Proposition~\ref{unconnected_bis}}
	 To demonstrate the proposition, let us write the operators ${\cal A}_i$ and ${\cal B}_j$ in terms of the elements of the special basis of $\mathfrak{M}^{(A)}$ and $\mathfrak{M}^{(B)}$, respectively:
	\begin{align}
	{\cal A}_i&=\sum_{k=1}^p a_{\bm{i}_k} \bm{\mathfrak{m}}_{\bm{i}_k} \nonumber \\
	{\cal B}_j&=\sum_{h=1}^q b_{\bm{j}_h} \bm{\mathfrak{m}}_{\bm{j}_h}
	\end{align}
inserting these expansions in Eq.~(\ref{Rassociative_bis}) we get:
\begin{align}
	&\left\{\left(\sum_{k=1}^p a_{\bm{i}_k}  \bm{\mathfrak{m}}_{\bm{i}_k}\right)^{n} \otimes  
	\left(\sum_{h=1}^q b_{\bm{j}_h}  \bm{\mathfrak{m}}_{\bm{j}_h}\right)^{m} \right\}_M \nonumber \\&
=
	 \left\{\left(\sum_{k=1}^p a_{\bm{i}_k}  \bm{\mathfrak{m}}_{\bm{i}_k}\right)^{n}\right\}_M \otimes
	 \left\{\left(\sum_{h=1}^q b_{\bm{j}_h}  \bm{\mathfrak{m}}_{\bm{j}_h}\right)^{m}\right\}_M,
\end{align}
by using the multinomial theorem the above equation becomes ($r_k$ and $s_h$ are such that $\sum_{k=1}^p r_k=n$ and $\sum_{h=1}^q s_h=m$, 
respectively)
\begin{align} 
\label{unconnected_bis_demo_NO}
&\sum_{r_k}\;\sum_{s_h} \frac{n!}{r_1!r_2!...r_p}\frac{m!}{s_1!s_2!...s_q}
a_{\bm{i}_1}^{r_1}a_{\bm{i}_2}^{r_2}...a_{\bm{i}_{r_p}}^{r_p}
b_{\bm{j}_1}^{s_1}b_{\bm{j}_2}^{s_2}...b_{\bm{j}_{s_q}}^{s_q} \nonumber \\
&\times\left\{\bm{\mathfrak{m}}_{\bm{i}_{1}}^{r_1} \otimes \bm{\mathfrak{m}}_{\bm{i}_{2}}^{r_2}\otimes ... \bm{\mathfrak{m}}_{\bm{i}_{p}}^{r_p}
 \otimes \bm{\mathfrak{m}}_{\bm{j}_{1}}^{s_1} \otimes \bm{\mathfrak{m}}_{\bm{j}_{2}}^{s_2}\otimes ... \bm{\mathfrak{m}}_{\bm{j}_{q}}^{r_q}
 \right\}_M\nonumber \\
 &=\sum_{r_k}\;\sum_{ s_h}
 \frac{n!}{r_1!r_2!...r_p}\frac{m!}{s_1!s_2!...s_q}a_{\bm{i}_1}^{r_1} a_{\bm{i}_2}^{r_2}...a_{\bm{i}_{r_p}}^{r_p}
 b_{\bm{j}_1}^{s_1}b_{\bm{j}_2}^{s_2}...b_{\bm{j}_{s_q}}^{s_q}\nonumber \\
&\times \left\{\bm{\mathfrak{m}}_{\bm{i}_{1}}^{r_1} \otimes \bm{\mathfrak{m}}_{\bm{i}_{2}}^{r_2}\otimes ... \bm{\mathfrak{m}}_{\bm{i}_{p}}^{r_p}
 \right\}_M
 \otimes \left\{\bm{\mathfrak{m}}_{\bm{j}_{1}}^{s_1} \otimes \bm{\mathfrak{m}}_{\bm{j}_{2}}^{s_2}\otimes ... \bm{\mathfrak{m}}_{\bm{j}_{q}}^{r_q}
 \right\}_M
\end{align}
from which,
\begin{align} 
\label{unconnected_bis_demo}
&\sum_{r_k}\;\sum_{s_h}  \frac{n!}{r_1!r_2!...r_p}\frac{m!}{s_1!s_2!...s_q}a_{\bm{i}_1}^{r_1}a_{\bm{i}_2}^{r_2}...a_{\bm{i}_{r_p}}^{r_p}
b_{\bm{j}_1}^{s_1}b_{\bm{j}_2}^{s_2}...b_{\bm{j}_{s_q}}^{s_q} \nonumber \\
&\times \left(
\left\{\bm{\mathfrak{m}}_{\bm{i}_{1}}^{r_1} \otimes \bm{\mathfrak{m}}_{\bm{i}_{2}}^{r_2}\otimes ... \bm{\mathfrak{m}}_{\bm{i}_{p}}^{r_p}
\otimes \bm{\mathfrak{m}}_{\bm{j}_{1}}^{s_1} \otimes \bm{\mathfrak{m}}_{\bm{j}_{2}}^{s_2}\otimes ... \bm{\mathfrak{m}}_{\bm{j}_{q}}^{r_q}
\right\}_M  \right.\nonumber \\
&- 
\left. \left\{\bm{\mathfrak{m}}_{\bm{i}_{1}}^{r_1} \otimes \bm{\mathfrak{m}}_{\bm{i}_{2}}^{r_2}\otimes ... \bm{\mathfrak{m}}_{\bm{i}_{p}}^{r_p}
\right\}_M
\otimes \left\{\bm{\mathfrak{m}}_{\bm{j}_{1}}^{s_1} \otimes \bm{\mathfrak{m}}_{\bm{j}_{2}}^{s_2}\otimes ... \bm{\mathfrak{m}}_{\bm{j}_{q}}^{r_q}
\right\}_M \right) \nonumber \\
&=0.
\end{align}
Because the vector spaces $\mathfrak{M}^{(A)}$, $\mathfrak{M}^{(B)}$ are infinite-dimensional, $n$ and $m$ can be any integer,
thus the same it holds for all the powers $r_k$, $s_h$. Considering that  the coefficients  $a_{\bm{i}_k}$ and $ b_{\bm{j}_h} $ are arbitrary,
it follows that Eq.~(\ref{unconnected_bis_demo}) implies
\begin{align} 
\label{unconnected_bis_demo_2}
&\left\{\bm{\mathfrak{m}}_{\bm{i}_{1}}^{r_1} \otimes \bm{\mathfrak{m}}_{\bm{i}_{2}}^{r_2}\otimes ... \bm{\mathfrak{m}}_{\bm{i}_{p}}^{r_p}
\otimes \bm{\mathfrak{m}}_{\bm{j}_{1}}^{s_1} \otimes \bm{\mathfrak{m}}_{\bm{j}_{2}}^{s_2}\otimes ... \bm{\mathfrak{m}}_{\bm{j}_{q}}^{r_q}
\right\}_M\nonumber \\
&=\left\{\bm{\mathfrak{m}}_{\bm{i}_{1}}^{r_1} \otimes \bm{\mathfrak{m}}_{\bm{i}_{2}}^{r_2}\otimes ... \bm{\mathfrak{m}}_{\bm{i}_{p}}^{r_p}
\right\}_M
\otimes \left\{\bm{\mathfrak{m}}_{\bm{j}_{1}}^{s_1} \otimes \bm{\mathfrak{m}}_{\bm{j}_{2}}^{s_2}\otimes ... \bm{\mathfrak{m}}_{\bm{j}_{q}}^{r_q}
\right\}_M
\end{align}
$\forall r_k,s_h\in \N,\,p,q \in \N$, which, in turn, implies Eq~(\ref{Rassociative}). \qed
\end{bew}
        If $f(x)$ is an analytical function in the argument $x$ and $f^{(n)}(x):=\text{d}^n/\text{d}x^n f(x)$, then we can define
        $f({\cal A}):=\sum_{n=0}^\infty \frac{1}{n!}f^{(n)}(0)){\cal A}^{n}$.       
       Thus, we can easily define the generalized exponential for non commuting operators:
\begin{definition}\label{Rexponential}
        For any ${\cal A}\in \mathfrak{M}$ the $M$-exponential  $\exp_M\left[{\cal A}\right]$ 
        is defined as the $M$-projection applied to the formal tensor power expansion of $\exp\left[{\cal A}\right]$:
                \begin{align}
        \label{defExpR}
        \exp_M\left[{\cal A}\right]&:=\left\{\exp\left[{\cal A}\right]\right\}_M=
        \left\{\sum_{n=0}^\infty \frac{1}{n!}{\cal A}^{n}\right\}_M \nonumber \\
       & =
        \sum_{n=0}^\infty \frac{1}{n!} \left\{ {\cal A}^{n}\right\}_M 
        \end{align}     
\end{definition}
\begin{theorem}\label{theo1}    
        Let  ${\cal A}_i\in \mathfrak{M}$, $1\le i \le N$, then 
        the following factorization property holds:
        \begin{align}
        \label{defRexample}
        \exp_M\left[\sum_{i=1}^N{\cal A}_i\right]=\left\{\prod_{i=1}^N \exp\left[{\cal A}_i\right]\right\}_M.
        \end{align}
\end{theorem}
\begin{bew}{of Theorem~\ref{theo1}}
        The demonstration is trivial because by the definition of the $M$-projection, properties~\ref{defR1}, 
         the argument of the $M$-projection can be formally freely rearranged, without caring about the order of the factors, thus the multinomial theorem can be used 
        \begin{align}
\label{profTheoEq}
&\exp_M\left[\sum_{i=1}^N{\cal A}_i\right]:=
\sum_{n=0}^\infty \frac{1}{n!} \left\{   \left[\sum_{i=1}^N{\cal A}_i\right]^{n}\right\}_M \nonumber \\
&=
\sum_{m_1=0}^\infty\sum_{m_2=0}^\infty...\sum_{m_N=0}^\infty \frac{1}{m_1!} \frac{1}{m_2!}.. \frac{1}{m_N!}\nonumber \\
&\times
\left\{{\cal A}_1^{m_1}\otimes {\cal A}_2^{m_2}\otimes...\otimes {\cal A}_N^{m_N}\right\}_M
=\left\{\prod_{i=1}^N \exp\left[{\cal A}_i\right]\right\}_M.
\end{align}
\begin{remark}
       Although  Theorem~\ref{theo1} shows that the $M$-exponential shares a fundamental property with the usual exponential, this fact
        does {\em not} imply that  Eq.~(\ref{generalevolution4_3}) holds true. We think that this is the principal source of confusion in the work of Kubo~\cite{kuboGenCumJPSJ17}, and the basis of the Fox criticism.
        However,  for a proper definition of cumulants,  Eq.~(\ref{generalevolution4_3}) is not required to hold in general, but \textit{only when the processes    $\Omega(u),\,u\in T$ and $\Omega(u'),\,u'\in T'$, to which the generating cumulant functions ${\cal K}$ and ${\cal K}'$, respectively, are related, are ``independent'' to each other}. As we shall show in the next section, with a proper definition of independent stochastic processes,  this is the case, thanks to the following result:
\end{remark}

\end{bew}
\begin{theorem}\label{theoBASE}
        Two vector subspaces $ \mathfrak{M}^{(A)}$, $ \mathfrak{M}^{(B)}\subset \mathfrak{M}$, are  $M$-unconnected to each other, 
        if and only if, for any 
        ${\cal A} \in  \mathfrak{M}^{(A)}$ and ${\cal B} \in  \mathfrak{M}^{(B)}$ we have:
        \begin{align}
        \label{theoBASEEq}
        \exp_M\left[{\cal A}+{\cal B}\right]&=\exp_M\left[{\cal A}\right]\otimes\exp_M\left[{\cal B}\right]\;\;or\nonumber \\
        \exp_M\left[{\cal A}+{\cal B}\right]&=\exp_M\left[{\cal B}\right]\otimes\exp_M\left[{\cal A}\right]
        \end{align}  
\end{theorem}
\begin{bew}{of Theorem~\ref{theoBASE}}
        According to Definition~\ref{Rexponential},
        the r.h.s of Eq.~(\ref{theoBASEEq})  can be rewritten as (let's consider only  the ordering given in the first line, for simplicity):
        \begin{align}
        \label{prooftheo2Eq_1}
        &\exp_M\left[{\cal A}\right]\otimes\exp_M\left[{\cal B}\right]:=
        \sum_{n=0}^\infty \frac{1}{n!} \left\{    {\cal A}^n\right\}_M \otimes
        \sum_{m=0}^\infty \frac{1}{m!}\left\{    {\cal B}^m\right\}_M \nonumber \\
        &=
        \sum_{n,m=0}^\infty \frac{1}{n!} \frac{1}{m!} \left\{{\cal A}^n\right\}_M\otimes \left\{ {\cal B}^m\right\}_M,
        \end{align}
while from Theorem~\ref{theo1} the l.h.s. is given by
                \begin{align}
        \label{prooftheo2Eq_2}
        \exp_M\left[{\cal A}+{\cal B}\right]=
        \sum_{n,m=0}^\infty \frac{1}{n!} \frac{1}{m!} \left\{{\cal A}^n \otimes {\cal B}^m\right\}_M.
        \end{align}
Comparing Eq.~(\ref{prooftheo2Eq_1}) with Eq.~(\ref{prooftheo2Eq_2}), exploiting  Definition~\ref{unconnected} of $M$-unconnected sets
 and considering that ${\cal A}$ and ${\cal B}$ are generic 
elements of  $ \mathfrak{M}^{(A)}$ and $ \mathfrak{M}^{(B)}$, respectively, it follows that if $ \mathfrak{M}^{(A)}$ and $ \mathfrak{M}^{(B)}$ are
$M$-unconnected, then Eq.~(\ref{theoBASEEq}) holds. The reverse  follows straightforward from Proposition~\ref{unconnected_bis}. \qed
\end{bew}
Now we introduce the map that we shall use to define the generalized cumulants:
\begin{definition}
	\label{weakordering}
	an $M$-ordering map is an $M$-projection map subjected to the condition that 
	 \textit{if}  for some ${\cal A},{\cal B}\in  \mathfrak{M}$ we have  
	$\left\{ {\cal A}\right\}_M\otimes\left\{ {\cal B}\right\}_M=
	\left\{ \left\{ {\cal A}\right\}_M\otimes\left\{ {\cal B}\right\}_M\right\}_M$ 
	(thus,	$\left\{ {\cal A}\right\}_M\otimes\left\{ {\cal B}\right\}_M\in \mathfrak{M}_M$, 
	the co-domain of the $M$-projection map) \textit{then} $\left\{ {\cal A}\right\}_M\otimes\left\{ {\cal B}\right\}_M=
	\left\{ {\cal A}\otimes {\cal B}\right\}_M$ (namely, ${\cal A}$ and ${\cal B}$ are $M$-unconnected to each other).
\end{definition}
It is clear that an  $M$-projection map corresponding to some ordering/symmetrization respect to some index/parameter labeling the operators of $\mathfrak{M}$ 
is an $M$-ordering map. However, in general an $M$-projection map involving some average procedure is not an $M$-ordering map. 
For example, if some subspace $\mathfrak{M}^{(A)}$
of $\mathfrak{M}$ is spanned by the parameters $i\in I_i$, $y\in I_y$, $I_i\subset \N$, $I_y\subset \R$ (or $\N$), namely, if 
$\forall i\in I_i$, $y\in I_y$ we have ${\cal A}(i,y)\in \mathfrak{M}^{(A)}$,
then, the $M$-projection map defined as the combined action of some ordering respect to the $i$ parameter (e.g.
$i$ decreasing from left to right) and a sum (or average) respect to the $y$ one:
\begin{align*}
&\left\{{\cal A}(i_1,y_1)\otimes{\cal A}(i_2,y_2)\otimes...\otimes{\cal A}(i_n,y_n)\right\}_M  \\
&=
 \int_{I_y} \left\{{\cal A}(i_1,y)\otimes{\cal A}(i_2,y)\otimes...\otimes{\cal A}(i_n,y)\right\}_{O_i} \text{d}y
\end{align*}
is not an $M$-ordering map. In fact,
	$\left\{ {\cal A}\right\}_M\otimes\left\{ {\cal B}\right\}_M=
\left\{ \left\{ {\cal A}\right\}_M\otimes\left\{ {\cal B}\right\}_M\right\}_M$ 
 is satisfied for any ${\cal A}={\cal A}(i_1,y_1)\otimes{\cal A}(i_2,y_2)...$ and
 ${\cal B}={\cal A}(j_1,y_1)\otimes{\cal A}(j_2,y_2)...$ such that $i_k>j_h$ $\forall h,k$, but $\left\{ {\cal A}\right\}_M\otimes\left\{ {\cal B}\right\}_M\ne
 \left\{ {\cal A}\otimes {\cal B}\right\}_M$  (in general, the product of averages is not equal to the average of the product). 

In the following 
the $M$-ordering map  applied to a generic operator ${\cal A}\in\mathfrak{M}$ shall be indicated
by $\left\{{\cal A}\right\}_{M_O}$.

The following result follows now without need of demonstration:
\begin{lemma}\label{lemma1}
Let $ \mathfrak{M}^{(A)}$ and $\mathfrak{M}^{(B)}\subset \mathfrak{M}$. If $\forall\,{\cal A}_i\in  \mathfrak{M}^{(A)}$ and 
$\forall\,{\cal B}_j\in  \mathfrak{M}^{(B)}$ we have ($n_i,m_i,p,q\in\N$):
\begin{align}
&\left\{{\cal A}_1^{n_1}\otimes{\cal A}_2^{n_2}\otimes...\otimes{\cal A}_p^{n_p}\right\}_{M_O} 
\otimes
\left\{{\cal B}_{1}^{n_{1}}\otimes{\cal B}_{2}^{m_{2}}\otimes...\otimes{\cal B}_q^{m_q}\right\}_{M_O} \nonumber \\
=&\left\{
\left\{{\cal A}_1^{n_1}\otimes{\cal A}_2^{n_2}\otimes...\otimes{\cal A}_p^{n_p}\right\}_{M_O} 
\otimes
\left\{{\cal B}_{1}^{n_{1}}\otimes{\cal B}_{2}^{m_{2}}\otimes...\otimes{\cal B}_q^{m_q}\right\}_{M_O}\right\}_{M_O},
\end{align}
then $ \mathfrak{M}^{(A)}$ and $ \mathfrak{M}^{(B)}$ are $M$-unconnected to each other vector spaces.
\end{lemma}
\section{The case of cumulants}

Having defined in a proper way the $M$-exponential, we proceed now with a less formal pathway. 

\subsection{General settings\label{CumIntro}}
In the definition of the moment generating function of Eqs.~(\ref{MG})-(\ref{defMn}) we have used 
the usual $t$-ordered exponential. As already observed, the $M$-projection map associated with the $t$-ordered exponential,
fulfills both the 
requirements~\ref{idempotent}-\ref{defR1} of the 
definition of $M$-projection and also the more restrictive Definition~\ref{weakordering} of $M$-ordering.
To identify this specific mapping we use the notation $M=O$,
namely $\exp_O[...]:=\stackrel{\longleftarrow}{\exp}[...]$.
We  generalize  Eqs.~(\ref{MG}) and (\ref{defMn}) by substituting the  
$t$-ordered exponential with another $M$-exponential associated to a general $M$-projection map, 
that we shortly indicate with the $M$ map. 
\begin{definition}
	Assuming that  $\Omega$ is a random q-number means here that in
	the set of the possible ``instances'' $\Omega$ 
	we have defined
	a projection map, that we call ``average operation (or process)'' and that we indicate with $\langle ...\rangle$. 
	Of course, the above definition does not coincide with the usual one for random numbers, 
	in fact $\Omega$ could be a deterministic (instead of really random) operator that depends on some parameter/variables $\xi$ and
	$\langle...\rangle$ could correspond to the integration of $\xi$ by using some defined measure $\text{d}\mu(\xi)$ (see the example of Eqs.~(\ref{p(t)})-(\ref{LIt})). 
\end{definition}
 The generalized cumulants that concern this work are related to the definition of independent processes through the factorization of the generalized moment generating function.
  Namely, 
 \begin{definition}\label{independent}
 	if
 	$\Omega_1$ and $\Omega_2$ are two independent  random q-numbers  with 
 	generalized moment generating functions ${\cal M}_1:=\langle \exp_M[\Omega_1]\rangle$ and 
 	${\cal M}_2:=\langle \exp_M[\Omega_2]\rangle$, respectively, then 
 	$\Omega_1$ and $\Omega_2$ are defined statistically independent   if and only if the  
 	generalized moment generating function ${\cal M}_{1\oplus2}:=\langle \exp_M[\Omega_1\oplus\Omega_2]\rangle$ of the random q-number
 	$\Omega_1\oplus\Omega_2$  is given by ${\cal M}_1\otimes{\cal M}_2$.
 	
 \end{definition}
Now we introduce some notations and comments for further use:
\begin{enumerate}
	\item\label{uinU}
 $\CMcal{U}$ is some (generally continuous) set of values for the parameter $u$ and, for \textit{any} fixed value of $u$, $\Omega(u)$ is a random operator, namely  $\Omega(u)$ is
 a q-number (or operator) stochastic process. 
The parameter $u$  does not need to be a time, it could be a temperature value, as in 
classical and quantum statistical mechanics, or some other index or parameter for the set of stochastic operators $\Omega(u)$. 
\item\label{u_not_Omega} 
By  point~\ref{uinU}, a value of $u\in\CMcal{U}$ does not uniquely  individuate an operator $\Omega(u)$ (for example, in the case of Eq.~(\ref{p(t)})-(\ref{LIt}), the 
$\Omega(u)=\Omega(\vect{x},\xi;u)$ operator depends on the time $u$, 
but also on the set of variables $\xi$ that we integrate in the average process). 
We assume that the set of  all the possible instances $\Omega$ are the original special basis
$\mathscr{B}=\{\bm{ \mathfrak{b}}_i\}$ of 
a vector space $\mathfrak{B}$, and the tensor product $\otimes$ by which a space $\mathfrak{M}$ can be defined  (see the beginning of 
section~\ref{formalpart}) is the product among the operators $\Omega$. From now on, we shall denote with  ``$M$ map''an $M$-projection  defined on {\em this} $\mathfrak{M}$ space.
Exploiting again the example of Eq.~(\ref{p(t)})-(\ref{LIt}), 
for any $u\in \CMcal{U}$, and $\xi$ ranging in a proper space where it is defined, we have
$ \Omega(\vect{x},\xi;u)\in  \mathscr{B}$, and  
$ \Omega(\vect{x},\xi';u_1) \Omega(\vect{x},\xi'';u_2)\Omega(\vect{x},\xi''';u_3) ... \in \mathscr{M}$. 
\item 
Because the average process $\langle...\rangle$ is a projection,  the composition of an $M$-projection map $M$ with the average operation is, in turn, an $M$-projection map, that we  indicate with $\tilde M$:
$\left\{ \dots\right\}_{\tilde M}:= \langle \dots \rangle\circ \left\{\dots\right\}_{M}$
(or, shortly, $\tilde M=\langle.\rangle\circ M$). 
\item  
The generalized moment generating function ${\cal M}(\CMcal{T})$ is defined as
 \begin{equation}
\label{MGgen}
{\cal M}(\CMcal{T}):= \langle \exp_{M}  \left[\int_\CMcal{T} \text{d}u\,\Omega(u)\right]\rangle
:=\exp_{\tilde M} \left[\int_\CMcal{T} \text{d}u\,\Omega(u)\right] .
\end{equation}
where  $\CMcal{T}$ is a subset of  $\CMcal{U}$.
\item 
The generalized moment generating components ${\cal M}_n(\CMcal{T})$ are given by
\begin{align}
\label{defMnGen}
&{\cal M}_n(\CMcal{T})=\frac{1}{n!} \left\{\left[\int_\CMcal{T} \text{d}u\, \Omega(u)\right]^n
\right\}_{\tilde M} \nonumber \\
&=\frac{1}{n!}\int_\CMcal{T} \mbox{d}u_1 \int_\CMcal{T}\mbox{d}u_2...\int_\CMcal{T}\mbox{d}u_n
 \left\{ \Omega(u_1) \Omega(u_2)... \Omega(u_n)\right\}_{\tilde M}.
\end{align}
Of course we have ${\cal M}(\CMcal{T}):=\sum_{n=0}^\infty {\cal M}_n(\CMcal{T})$ as in Eq.~(\ref{MnM}). From the operative side, the last side of Eq.~(\ref{defMnGen}) 
is mandatory because of the point~\ref{defR2} of the definition of
$M$-projection map.
\item 
The generalized  $n$-moment density function is defined as
\begin{equation}
\label{defmu}
	\mu^{( \tilde M)}_n(u_1,u_2,...,u_n):= \left\{ \Omega(u_1) \Omega(u_2)... \Omega(u_n)\right\}_{\tilde M},
\end{equation}
thus, the generalized $n$-moments density functions are elements of the co-domain of the $\tilde M$ map, and, in particular, each one corresponds to the mapping of a different element of the special basis $\mathscr{M}$ of the space $\mathfrak{M}$ where the $\tilde M$ map is defined.
\end{enumerate}

	According to Definition~\ref{independent}, we specify what we mean for independence of stochastic operator processes:
\begin{definition}\label{independentprocesses}
 If $\CMcal{T}$ and $\CMcal{T}'$ are two different subsets of $\CMcal{U}$, given the Definition~\ref{independent} and the Theorem~\ref{theoBASE}, 
	we say that 
	two q-number stochastic processes $\Omega(u)$ and $\Omega(u')\in\mathfrak{B}$, with
	 $u\in \CMcal{T}$, $u'\in \CMcal{T}'$, are $\mathfrak{u}$-independent to each other if  the subsets
	 $\CMcal{T}$ and $\CMcal{T}'$ define two vector spaces that are
	 $M$-unconnected to each other respect to the $\tilde M$ map.
\end{definition}	
\begin{remark}\label{independentpartitions}	
	According to Theorem~\ref{theoBASE}, from Definition~\ref{independentprocesses} it follows that the subsets
	$\CMcal{T}$ and $\CMcal{T}'$ are  $\mathfrak{u}$-independent to each other if and only if	
	the following factorization of the moment generating function holds true:
			\begin{equation}
	\label{CumulantDefGenFact}
	\exp_{\tilde M}  \left[\int_{\CMcal{T}\cup\CMcal{T'}} \text{d}u\,\Omega(u)\right] =  \exp_{\tilde M}  \left[\int_{\CMcal{T}} \text{d}u\,\Omega(u)\right] \,\exp_{\tilde M} 
	\left[\int_{\CMcal{T}'} \text{d}u\,\Omega(u)\right].
	\end{equation}
	\end{remark}
Having generalized the  definition of the moment generating function by introducing the $\tilde M$ mapping, 
 we have now to generalize also the definition of the cumulant generating function. 
For reasons that will become clear later, this is done by using an $M$-ordering map, instead of a more general $M$-projection map\footnote{%
	As done by Kubo~\cite{kuboGenCumJPSJ17}, we could consider the case where the integral, in the r.h.s. of Eq-~(\ref{CumulantDefGen}) is substituted with 
	a sum over the components of a $N$ dimensional vector of stochastic operators; however, for the sake of simplicity, we shall consider only the case of stochastic processes 
	depending on a continuous parameter, e.g. the time. Adapting the results to the case of discrete, finite or infinite sets of 
	stochastic operators is straightforward.%
}:
\begin{equation}
\label{CumulantDefGen}
\exp_{M_O} \left[{\cal K}(\CMcal{T})\right] := {\cal M}(\CMcal{T}):= \exp_{\tilde M}  \left[\int_\CMcal{T} \text{d}u\,\Omega(u)\right],
\end{equation}
in which $M_O$ is the $M$-ordering map (the ``$M_O$'' map, shortly) that applies to the generalized cumulants. Notice that often both the 
the $M_O$ and the $M$ maps are the chronological time ordering that give rise to the $t$-ordered exponential, but {\em in general they can be different
to each other} (see below, around Eqs.~(\ref{KgenDef1_})-(\ref{defmuRR})).
According to Kubo~\cite{kuboGenCumJPSJ17} the generalized cumulant generating function ${\cal K}(\CMcal{T})$ can be
conveniently written as 
\begin{equation}
\label{kumulant_generator}
{\cal K}(\CMcal{T})=\langle \exp_{M'}\left[ \int_\CMcal{T} \text{d}u\,\Omega(u)\right]-1 \rangle_c,
\end{equation}
in which $M'$ is some other $M$-projection map we shall identify hereafter. Eq.~(\ref{kumulant_generator}) at the same time defines the ``cumulant average'' $\langle ...\rangle_c$, and  gives rise to the series  ${\cal K}=\sum_{n=1}^\infty {\cal K}_n$ of cumulant generating components, of which the 
$n$-th term is
\begin{align}
\label{defKnGenTemp}
{\cal K}_n &:= \frac{1}{n!}\,\int_\CMcal{T} \text{d}u_1 \int_\CMcal{T} \text{d}u_2... \int_\CMcal{T} \text{d}u_n\,
\langle \left\{ \Omega(u_1)\Omega(u_2)...\Omega(u_n) \right\}_{M'} \rangle_c \nonumber \\
&:= \frac{1}{n!}\,\int_\CMcal{T} \text{d}u_1 \int_\CMcal{T} \text{d}u_2... \int_\CMcal{T} \text{d}u_n\,
 \left\{ \Omega(u_1)\Omega(u_2)...\Omega(u_n) \right\}_{\tilde M_c'}.
\end{align}
where, extending a little the already introduced convention, $\tilde M_c'=\langle.\rangle_c\circ M'$.
In the Kubo's papers~K62-63 there is not an indication about the relations among the 
$M$, $M_O$ and $M'$ maps. We shall remedy hereafter to that.
It is apparent that if in the l.h.s. of Eq.~(\ref{CumulantDefGen}) we take the first order on the
power series that defines the $M_O$ exponential and if we exploit Eq.~(\ref{kumulant_generator}), for consistency we must have $M'=M$. Thus 
\begin{align}
\label{defKnGen}
{\cal K}_n&:= \frac{1}{n!}\,\int_\CMcal{T} \text{d}u_1 \int_\CMcal{T} \text{d}u_2... \int_\CMcal{T} \text{d}u_n\,
\langle \left\{ \Omega(u_1)\Omega(u_2)...\Omega(u_n) \right\}_{M} \rangle_c \nonumber \\
&:= \frac{1}{n!}\,\int_\CMcal{T} \text{d}u_1 \int_\CMcal{T} \text{d}u_2... \int_\CMcal{T} \text{d}u_n\,
\left\{ \Omega(u_1)\Omega(u_2)...\Omega(u_n) \right\}_{\tilde M_c} 
\end{align}
The generalized $n$-cumulant density function is then defined as 
\begin{equation}
\label{defkappa}
 \kappa_n^{(\tilde M,M_O)}(u_1,u_2,...,u_n):= \left\{ \Omega(u_1)\Omega(u_2)...\Omega(u_n) \right\}_{\tilde M_c} .
\end{equation}
\begin{remark}\label{basiscumulant}
	The generalized cumulant density functions of Eq.~(\ref{defkappa}) are the
	``original special basis''   
	$\mathscr{B}=\{\bm{ \mathfrak{b}}_i\}$ (possible redundant) of the vector space $\mathfrak{B}$,
	that, togheter with the tensor product $\otimes$, gives rise to the space $\mathfrak{M}$ 
	where the $M_O$ map is defined.
\end{remark}

By Eq.~(\ref{defkappa}), the generalized $n$-cumulant density functions are elements of the co-domain of the $\tilde M_c$ map.

Because $\left\{...\right\}_{M_O}$ is a projection (namely, $\left\{\left\{...\right\}_{M_O}\right\}_{M_O}=\left\{...\right\}_{M_O}$ ), we can apply
the $M_O$ map to the l.h.s. of Eq.~(\ref{CumulantDefGen})  without affecting the result, thus the same should hold for the r.h.s. of the same equation.
Then, using  Definition~\ref{Rexponential} we have
\begin{align}
\label{KgenDef1_}
\left\{\left( \int_\CMcal{T} \text{d}u\,\Omega(u)  \right)^n \right\}_{\tilde M} =
\left\{  \left\{\left( \int_\CMcal{T} \text{d}u\,\Omega(u)  \right)^n \right\}_{\tilde M}  \right\}_{M_O}.
\end{align}
The above equation is the constraint we have to take into account when in  choosing the  $M_O$ map  once the  $\tilde M$ one is
 given (as is common in real problems), namely, it is the relationship between these two mappings we were 
 looking for. 
 In practice, exploiting Eqs.~(\ref{defMnGen})-(\ref{defmu}), Eq.~(\ref{KgenDef1_})
says that the generalized moments must not be affected by the $M_O$ map (they belong to
the co-domain of $M_O$):
\begin{align}
\label{defmuRR}
\left\{\mu^{( \tilde M)}_n(u_1,u_2,...,u_n)\right\}_{M_O}&
= \left\{\left\{ \Omega(u_1) \Omega(u_2)... \Omega(u_n)\right\}_{\tilde M}\right\}_{M_O}\nonumber \\
&=\left\{ \Omega(u_1) \Omega(u_2)... \Omega(u_n)\right\}_{\tilde M}=\mu^{( \tilde M)}_n(u_1,u_2,...,u_n).
\end{align}
Of course, the most trivial way to satisfy Eq.~(\ref{defmuRR}) is to choose $M_O\equiv M$ (the equivalence symbol is used, instead of the equality one, because the $\mathfrak{M}$ 
spaces of the two mappings
are not the same), but we stress again that this is not the only possible choice (e.g., see Section~\ref{TTOC}).

By using both Eq.~(\ref{kumulant_generator}) and the definition of the $\tilde M_c$ map, Eq.~(\ref{CumulantDefGen}) can be also written as:
\begin{equation}
\label{CumulantDefGen2}
\exp_{M_O} \left[ \exp_{\tilde M_c}\left[ \int_\CMcal{T} \text{d}u\,\Omega(u)\right]-1 \right] =  \exp_{\tilde M}  \left[\int_\CMcal{T} \text{d}u\,\Omega(u)\right].
\end{equation}

If $M=O$, namely if it is the usual time ordering map (increasing time from right to left), we have 
\begin{equation}
\label{mukappa_O}
\left. \begin{array}{lcl} 
\mu_n^{(\tilde O)}(u_1,u_2,...,u_n)&=&\langle  \Omega(u_1)\Omega(u_2)...\Omega(u_n) \rangle\\
\kappa_n^{(\tilde O,M_O)}(u_1,u_2,...,u_n)&=&\langle  \Omega(u_1)\Omega(u_2)...\Omega(u_n) \rangle_c
\end{array} \right\} 
\text{with}\; u_1\ge u_2 \ge ...\ge u_n, 
\end{equation}
that, inserted in Eqs.~(\ref{defMnGen}) and (\ref{defKnGen}) gives,
\begin{equation}
\label{defMnGen_O}
{\cal M}_n:= \,\int_0^t \text{d}u_1 \int_{0}^{u_1} \text{d}u_2... \int_{0}^{u_{n-1}} \text{d}u_n\,
\langle \Omega(u_1)\Omega(u_2)...\Omega(u_n)  \rangle.
\end{equation}
\begin{equation}
\label{defKnGen_O}
{\cal K}_n:= \int_0^t \text{d}u_1 \int_{0}^{u_1} \text{d}u_2... \int_{0}^{u_{n-1}} \text{d}u_n\,
\langle \Omega(u_1)\Omega(u_2)...\Omega(u_n)  \rangle_c,
\end{equation}
respectively.
It is worth stressing that  respect to the ``time'' parameter $u$, the result
of the ordering $\left\{...\right\}_O$ associated to the $t$-ordered exponential, is different for the case of the $M$ 
respect to the case of the $M_O$ mappings. This is because these mappings act on different spaces. In fact, 
in the $\mathfrak{M}$ vector space of which the basis is given by the tensor products of $\Omega(u_i)$ 
(we recall that the time parameter $u$ does not uniquely identify the base element $\Omega$), $M=O$ leads to 
an ordering of the basis respect to the time $u$, e.g., 
\begin{equation}
\label{exR=O}
\left\{\Omega(u_i)\Omega(u_j)\right\}_{O}=
\begin{cases} 
\;\Omega(u_i)\Omega(u_j) \,\,\,\mbox{for $u_i>u_j$}  \\
\;\Omega(u_j)\Omega(u_i) \,\,\,\mbox{for $u_j>u_i$}.
\end{cases}
\end{equation}
Therefore, in this space, each one of the elements identified by the parameter $u$ is $M$-unconnected to any other element identified by a paramter $u'\ne u$.
On the other hand, the mapping $M_O=O$, defined by assuming that $\exp_{M_O}[...]$  of Eq.~(\ref{CumulantDefGen}) is a $t$-ordered exponential, acts on the vector space with basis given by 
the tensor products of cumulants $ \kappa_n^{(\tilde M,\tilde O)}(u_{i_1},u_{i_2},...,u_{i_n}):=\left\{ \Omega(u_1)\Omega(u_2)...\Omega(u_n) \right\}_{\tilde M_c}$ (see Remark~\ref{basiscumulant}) that depend on $n$ time parameters and where 
the ordering is respect to the {\em largest} time inside each cumulant, e.g. (assuming $u_{i_1}>u_{i_k}$ and 
 $u_{j_1}>u_{j_k}$, $k>1$):
\begin{align}
\label{cum_o}
&\left\{ \left\{\Omega(u_{i_1})\Omega(u_{i_2})...\Omega(u_{i_n}) \right\}_{\tilde M_c} \otimes\left\{ \Omega(u_{j_1})\Omega(u_{j_2})...\Omega(u_{j_m}) \right\}_{\tilde M_c} \right\}_{O}\nonumber \\
=&\begin{cases} 
\;\left\{ \Omega(u_{i_1})\Omega(u_{i_2})...\Omega(u_{i_n})\right\}_{\tilde M_c} \otimes \left\{\Omega(u_{j_1})\Omega(u_{j_2})...\Omega(u_{j_m}) \right\}_{\tilde M_c} \,\,\,\mbox{for $u_{i_1}>u_{j_1}$}  \\
\;\left\{  \Omega(u_{j_1})\Omega(u_{j_2})...\Omega(u_{j_m})\right\}_{\tilde M_c} \otimes \left\{\Omega(u_{i_1})\Omega(u_{i_2})...\Omega(u_{i_n}) \right\}_{\tilde M_c} \,\,\,\mbox{for $u_{j_1}>u_{i_1}$},
\end{cases}
\end{align}
no matter the relationship among the other times. Therefore $M_O=O$ gives a partial ordering respect to the time $u_i$ 
(see the examples in Section~\ref{examples}).

\subsection{The fundamental property for cumulants}\label{fundamentalTeorem}
Within the framework defined in the previous sections, the approach to generalized cumulants results 
well-founded because, as it is shown hereafter, it follows that a
cumulant is vanishing if it refers to $\mathfrak{u}$-independent stochastic processes (or, that is the same, to $M$-unconnected sets). 
In this respect, the following Lemma is central for the present work and, after  Definition~\ref{independentprocesses}, 
 Lemma~\ref{lemma1} and Theorem~\ref{theoBASE}, it goes without the need of demonstration:
\begin{lemma}
	\label{TeoFundamental}	
	For any $u\in \CMcal{T}\subset  \CMcal{U}$ and $u'\in \CMcal{T}'\subset  \CMcal{U}$ two stochastic processes $\Omega (u)$ and $\Omega(u')$ 
	are reciprocally $\mathfrak{u}$-independent if and only if
	we have
	\begin{equation}
	\label{CumulantDefGenFact3}
	\exp_{M_O} \left[{\cal K}(\CMcal{T})\right]\exp_{M_O} \left[{\cal K}(\CMcal{T}')\right]=
	\exp_{M_O} \left[{\cal K}(\CMcal{T})+
	{\cal K}(\CMcal{T}')\right].
	\end{equation}
\end{lemma} 
\begin{remark}\label{KuUnconnected}
By Theorem~\ref{theoBASE}, from Eq.~(\ref{CumulantDefGenFact3}) it follows that	for any 
$u\in \CMcal{T}\subset  \CMcal{U}$ and 
$u'\in \CMcal{T}'\subset  \CMcal{U}$ we have that ${\cal K} (u)$ and ${\cal K}(u')$ are $M$-unconnected to each other. 
Thus, we can restate Lemma~\ref{TeoFundamental} as follows: 
{\em
	using the $M_O$ map to define the generalized cumulant generating function leads to a bijective mapping between couples of reciprocally $M$-unconnected sets in the $\mathfrak{M}$ space of the stochastic process $\Omega(u)$ and 
	 couples of reciprocally $M$-unconnected sets  in the $\mathfrak{M}$ space of the generalized cumulant generating function ${\cal K}(u)$. 
}
\end{remark}
\begin{definition}
	The {\em fundamental property of cumulants} is the statement that in the same assumption of Lemma~\ref{TeoFundamental}, namely if
	the  factorization of Eq.~(\ref{CumulantDefGenFact}) occurs,
	then  any cumulant mixing the average of the $\Omega$ random operators of the two sets $\CMcal{T}$
	and $\CMcal{T'}$ is vanishing.
\end{definition} 
\begin{lemma}\label{LemmaTeoFound1}
	Under the same assumption of Lemma~\ref{TeoFundamental}, the fundamental property of cumulants holds.
\end{lemma}
\begin{remark}	
	Lemma~\ref{LemmaTeoFound1}, that is one of the principal results of the present work, states that, although 
	for non-commuting processes Eq.~(\ref{CumulantDefGenFact3}), in general, does not hold,  it {\em automatically} holds when the 
	stochastic processes are $\mathfrak{u}$-independent according to Definition~\ref{independentprocesses}. 
	\end{remark}
\begin{remark}
	 Usually,  the definition of the $\mathfrak{M}$ space of the stochastic process $\Omega(u)$, the introduction of the generalized moment 
	 generating function and of the related $M$-projection map, cam naturally from the ``starting facts'' of the specific real problem 
	 (e.g. from physics) we 
	 are interested in, while cumulants are introduced by hand for convenience in different ways (provided Eq.~(\ref{KgenDef1_})
	  is satisfied). One might wonder why  for cumulants we have restrict the choice of the $M$-projection map to
	  the less general $M$-ordering type. This is because thanks to this constraint 
	  	Eq.~(\ref{CumulantDefGenFact3}) is {\em automatically fulfilled} for independent 
	  stochastic processes $\Omega(u)$ and $\Omega(u')$.
	  This is not always the case for cumulants defined by using the a
	  $M$-projection map without restrictions (it should be checked case by case)
\end{remark}	
	After Lemma~\ref{TeoFundamental} the  demonstration of Lemma~\ref{LemmaTeoFound1} is trivial being the same holding for standard c-number stochastic processes:
	comparing Eq.~(\ref{CumulantDefGen}) with Eq.~(\ref{CumulantDefGenFact}) we see that 
	 Eq.~(\ref{CumulantDefGenFact3}) means that ${\cal K}(\CMcal{T}\cup\CMcal{T'})= {\cal K}(\CMcal{T})+
	{\cal K}(\CMcal{T}')$, thus ${\cal K}(\CMcal{T}\cup\CMcal{T'})$ does not have terms mixing the averages of the $\Omega$ stochastic operators of the two sets $\CMcal{T}$
	and $\CMcal{T'}$.  
\begin{remark}\label{remarkTeoFound2}  
	Lemma~\ref{TeoFundamental} and Lemma~\ref{LemmaTeoFound1} are based on the definition of $M$-unconnected sets that, by 
	Theorem~\ref{theoBASE}, is strictly related to the factorization of the exponential of sums (and, in turn, to the 
	Definition~\ref{independentprocesses} of $\mathfrak{u}$-unconnected stochastic processes). 
	Shortly (in the frame of the present formal approach) the fundamental property of cumulants holds if the ``statistical'' independence is defined by
	 the factorization of the averages given in Eq.~(\ref{CumulantDefGenFact}). 
	 The lack of clarity
	 about the definition of statistical independence for q-numbers has been the main source of confusion
	in the debate about the validity of the Kubo's 
	approach to generalized cumulants. In fact, in his seminal work~\cite{kuboGenCumJPSJ17} Kubo enunciated his  Theorem~I in the following way:
	``.. a cumulant is zero if the elements are divided into two or more groups which are statistically
	independent.''
	And, in the following Corollary:
	``A cumulant is zero if one of the variables in it is independent of the others.
	Conversely, a cumulant is not zero if and only if the variables in it are statistically connected.'' 
		Both the theorems and the corollary are true for commuting random processes. 
	For q-numbers they are also true  only if we identify the statistical independence with the 
	factorization of the moment generator function as in Definition~\ref{independent}. The extension to q-number stochastic processes
	leads also to the introduction of the $\mathfrak{u}$-independence as in Definition~\ref{independentprocesses}.
	In our opinion Fox misunderstood this point, because his ``counterexample''~\cite{fJMP20}, that (for him) should invalidate
	the Kubo's theorems, simply is an archetypal case where the statistical independence of the stochastic coefficients (c-numbers) of an operator is not enough to guarantee
	the factorization of the moment generating function. 	
\end{remark}
\begin{remark}	Another point of the Kubo's work that has originated some confusion is the not so clear definition of the $M$ maps. In particular,
	it has not be clarified which are the domain spaces  where these maps are defined. As we have shown in an example here above (see
	 Eqs.~(\ref{mukappa_O})-(\ref{cum_o})), the $M=0$ map ($t$-ordering), acts in different ways in different $\mathfrak{M}$ vector spaces 
	 (see e.g.,~\cite{fJCP49}, where Eqs.~(2.14a)-(2.14d) are not correct because of a not proper definition of the $M_O=O$ operator).
\end{remark}

Before to pass to the second formal part of the present work, concerning the explicit relationship between generalized moments and generalized cumulants, we discuss a little about the above results. Although formally it sounds well, one could (and should) wonder if,  in any ``realistic'' model of some interest, the assumption
	of Lemma~\ref{TeoFundamental} (factorization of the generalized moment generating function) is fulfilled and, if it is the case, if it has a clear
	physical (or whatever interesting) interpretation. There are cases  in physics 
	where this happens. A class of cases it that treated in Section~\ref{sec:fermions}, where the generalized moments are reduced density matrices and they naturally factorize for non-interacting particles (or, asymptotically, for parts of a very large system).  Another situation, probably the most common in Physics, it that for which the $M$ mapping is the time ordering: $M=O$ (also treated in detail in Sections~\ref{PTOC}-\ref{TTOC}). In this case, as already remarked after Eq.~(\ref{exR=O}), 
	each $\Omega(u)\in\mathfrak{B} $
	is $M$-unconnected to any other 
	element of $\mathfrak{B} $, with a different value of the parameter $u$. It follows that in $\mathfrak{B}$, the $M=O$ map defines an 
	ordered partition of $M$-unconnected subsets $\left\{\Omega(u)\right\}$, {\em each one associated with a different value of the ``time'' parameter} 
	$u$: in the space $\mathfrak{M}$ the $M=O$ map defines subsets with different ``time-scales'' that are $M$-unconnected to each other. More precisely, for any couple of
	 subsets $\CMcal{T}$, $\CMcal{T}'$ of $\CMcal{U}$ such that $\forall u\in \CMcal{T}$, $u'\in \CMcal{T}'$, we have $u>u'$,
	   then
	$\{\Omega(u_{1})\Omega(u_{2})...\Omega(u_{n})\Omega(u_{1}')\Omega(u_{2}')...\Omega(u_{m}')\}_O=
	\{\Omega(u_{1})\Omega(u_{2})...\Omega(u_{n})\}_O\, \{\Omega(u_{1}')\Omega(u_{2}')...\Omega(u_{m}')\}_O$. Now, if the
	average process $\langle...\rangle$ is done by using a measure that factorizes in two separate measures for the two subsets with different ``time-scales''
	 $\CMcal{T}$ and $\CMcal{T}'$, then, the same two subsets of $\mathfrak{M}$ are $M$-unconnected respect to the  composite map $\tilde M=\langle. \rangle \circ M$. Thus, by Theorem~\ref{theoBASE} and Definition~\ref{independentprocesses}, it follows that the factorization of Eq.~\ref{CumulantDefGenFact} is satisfied.

	Let us see that explicitly. We assume that 
	 $\Omega(t)={\cal L}(t) \xi(t)$, where ${\cal L}(t)$ is some time-dependent operator (e.g., a matrix, a differential operator, a Lie-adjoint operator etc.), that does not 
	 commute with itself for different times: $[{\cal L}(t_1),{\cal L}(t_2)]\ne 0$ and $\xi(t)$ is some c-number stochastic process. 
	 We indicate with
	 $p(\xi_1,t_1;\xi_2,t_2;...;\xi_n,t:n)$ the joint probability density function to have (with some abuse of notation) $\xi(t_1)=\xi_1$ and  
	 $\xi(t_2)=\xi_2$ and...and $\xi(t_n)=\xi_n$.
	 Let us assume now that there is a time $\bar{t}$ such that the instances
	  of $\xi(t)$ at times $t>\bar{t}$ are statistically uncorrelated with the instances of $\xi(t)$ at  times $t<\bar{t}$. 
	  Thus, if $t_1>\bar{t}$ and $t_2<\bar{t}$ we have
	    $p(\xi_1,t_1;\xi_2,t_2)=p(\xi_1,t_1)p(\xi_2,t_2)$. Then, the generalized moment generating function for the stochastic operator $\Omega(t_1),\Omega(t_2)$
	    is
	    \begin{align}
	    \label{factorO}
	    \langle \left\{e^{{\cal L}(t_1) \xi(t_1)\text{d}t+{\cal L}(t_2) \xi(t_2)\text{d}t}\right\}_O\rangle &
	    :=\int \text{d}\xi_1\text{d}\xi_2 p(\xi_1,t_1;\xi_2,t_2)\left\{e^{{\cal L}(t_1) \xi(t_1)\text{d}t+{\cal L}(t_2) \xi(t_2)\text{d}t}\right\}_O\nonumber \\
        &=\left\{\int \text{d}\xi_1\text{d}\xi_2 p(\xi_1,t_1)p(\xi_2,t_2) e^{{\cal L}(t_1) \xi(t_1)\text{d}t+{\cal L}(t_2) \xi(t_2)\text{d}t}\right\}_O\nonumber \\
        &= \langle e^{{\cal L}(t_1) \xi(t_1)\text{d}t}\rangle  \langle e^{{\cal L}(t_2) \xi(t_2)\text{d}t}\rangle 
	    \end{align}
	And, for $t>\bar{t}$, we also have
		    \begin{align}
	\label{factorO_2}
	\langle \exp_O\left[\int_0^{t}{\cal L}(u) \xi(u)\text{d}u\right] \rangle &
	:=\lim_{\epsilon\to 0}
	 \langle \exp_O\left[\int_{\bar{t}+\epsilon}^{t}{\cal L}(u) \xi(u)\text{d}u\right] \rangle
	 \langle \exp_O\left[\int_0^{\bar{t}-\epsilon}{\cal L}(u) \xi(u)\text{d}u\right]\rangle
	\end{align}
\subsection{A closed-form formula for cumulants}\label{closedformula}
\subsubsection{Background}
As it is well known, for commuting stochastic processes a closed-form formula was worked out by Meeron~\cite[Appendix]{mJCP27}.
In Kubo's basic paper~\cite{kuboGenCumJPSJ17}, a similar formula has been proposed also for non-commuting processes. 
That was strongly criticized by Fox~\cite{fJMP17} for the case of time-ordered cumulants. 
In a subsequent paper, Apresyan~\cite{Apresyan1978}, supporting  Kubo's work, proposed  a Meeron-Kubo's equivalent formula, that, in turn,
Fox~\cite{fJMP20} criticized again, presenting what he purports to be a counterexample. 
As we have already observed in Remark~\ref{remarkTeoFound2}, in the Fox criticism there was confusion (actually due to some mistake in 
the original Kubo's paper) generated by the identification of the partition defined by unconnected sets and the partition given by the sets of statistically independent stochastic operators. Moreover, more confusion, supporting the
Fox arguments, was introduced by Fox himself~\cite{fJMP16} and by other authors (e.g.,~\cite{fJCP49}), which made a too generalized use of the 
Meeron-Kubo formula for cumulants.

In this section, we shall show that using the definitions of cumulants given in Eq.~(\ref{CumulantDefGen}), a Meeron formula 
holds, but only for the expression the gives the operator moments in terms of operator cumulants. A general inverse formula
(namely, valid for any $M$-projection mapping), that gives generalized cumulants
  in terms of generalized moments, does not hold true for operators (or, if it exists,  is not so 
straightforward). Actually, in some  cases of physical interest,  it is possible to find such a closed-form 
formula (see the examples in Section~\ref{examples}).
Thus Fox, despite that he was wrong when stated~\cite{fJMP17} that
 Kubo's approach to generalized cumulants does not lead to the validity of the  fundamental property of cumulants, 
he was partially right affirming that the Meeron-Kubo's closed-form formula cannot be applied.
\subsubsection{Main results}
We start inserting in Eq.~(\ref{CumulantDefGen}) the series ${\cal M}=\sum_{n=0}^\infty {\cal M}_n$ and ${\cal K}=\sum_{n=1}^\infty {\cal K}_n$:
\begin{align}
\label{KgenDef2}
\sum_{m=0}^\infty \frac{1}{m!}\left\{\left(\sum_{r=1}^\infty {\cal K}_r\right)^m\right\}_{M_O}=
\sum_{n=0}^\infty {\cal M}_n,
\end{align}
The Meeron closed-form formula that gives the generalized moment generating components ${\cal M}_n$ in terms of sums of products of generalized cumulant generating components ${\cal K}_r$, $r\le n$, is obtained by rearranging
the sums in the l.h.s. of  Eq.~(\ref{KgenDef2}). By Definition~\ref{definition1}, point~\ref{defR2}, the $M$-projection map does not interfere with this kind of 
arrangement of the addends of the sums. Thus, the formula {\em is the same as that of 
	Meeron~\cite[Eq.~(A7)]{mJCP27}, apart the $M$-projection mapping applied to each 
	addend of the sums}: 
\begin{proposition}\label{Mn_vsKn}
	The generalized moment generating component of order $n$ is given in terms of the generalized cumulant generating components of order $r\le n$ by the following expression:
	\begin{equation}
	\label{Meeron-Kubo2}
	{\cal M}_n=\sum_{Part.\,of\,n} \left\{\prod_{r=1}^{n} \frac{1}{s_r!}\left({\cal K}_r\right)^{s_r}\right\}_{M_O},
	\end{equation}
	in which the sum is over all the possible ways to partition  $n$ ``elements'' such that: one partition is made of $s_1=n$ 
	groups with $r=1$ element, another partition is made of only one group ( $s_n=1$ ) with $r=n$ elements, a generic partition is made of $s_1$ groups with $r=1$ elements and $s_2$ groups with $r=2$ elements and etc..., such that $\sum_{r=1}^n r\, s_r=n$. 
\end{proposition}
We report shortly the demonstration in Appendix~\ref{app:meeron}.

For what concerns the inverse Meeron-like formula:
\begin{equation}
\label{Meeron-Kubo}
{\cal K}_n=\sum_{Part.\,of\,n}  (-1)^{p-1} (p-1)!\prod_{r=1}^{n} \frac{1}{s_r!}\left({\cal M}_r\right)^{s_r},\;\;\;
p=\sum_{r=1}^n s_r,\;\;\;{\cal M}_r\;\text{are c-numbers}
\end{equation}
{\bf it cannot be generalized to q-number stochastic processes} just by applying  one of the $M$-projection mappings ($M_O$ and/or $M$) to the addends.
Actually, for non-commuting stochastic processes we are not able to find such a general formula. 
The reason is the following:
for standard commuting stochastic processes to obtain such a 
result we start from   Eq.~(\ref{CumulantDefGen}) and, to get rid of the exponential in the l.h.s.,
we take the logarithm of both sides of the equation. Then, in the r.h.s. of the resulting equation, we 
expand this logarithm in power series of its argument. For non-commuting processes in the l.h.s. of  
Eq.~(\ref{CumulantDefGen}) we have a  $M$-exponential instead of 
a standard exponential, and, because it is a projection map, it is not usually invertible. In conclusion, we guess that
there is not a general definition for a generalized logarithm that corresponds to the inverse of the $M$-exponential
(however,  we do hope that we will be proved wrong). 

In our opinion this is  the real mistake in the Kubo's paper~\cite{kuboGenCumJPSJ17} that 
has originated the Fox criticism and
that, to the best of our knowledge, has not been cured so far. 
Case by case, the formula that gives directly generalized cumulants from generalized moments can be found, as it will be shown in Section~\ref{examples}. A way to face the problem (still case by case) is to consider
the relation between cumulants and moments in the frame of a combinatoric calculus on  partition lattices. 
By expressing in terms of set-partitions of $n$ distinguishable objects the 
formula that gives the $n$-moment in terms of cumulants (see Eq.~(\ref{Moments_vs_Cumulants})), 
and taking into account 
the specific $M$ and $M_O$ maps that define the rule for managing (e.g., by ordering) the elements in the partitions, then, at least in principle, 
 the M\"obius inversion result on the partition lattice should lead to
the inverse formula~\cite{ahlvAM282} for the specific case of interest.

Also when this formula is known, often the easier practical way to get
${\cal K}_n$ from ${\cal M}_r$ is to iteratively use Eq.~(\ref{Meeron-Kubo2}). We explicitly do that by running some iterations: we start with ${\cal K}_1={\cal M}_1$ and
\begin{align}
\label{K2}
{\cal K}_2&={\cal M}_2- \frac{1}{2} \left\{({\cal K}_1)^2\right\}_{M_O} 
\end{align}
in which ${\cal K}_1={\cal M}_1$,
\begin{align}
\label{K3}
{\cal K}_3&={\cal M}_3 -  \left\{{\cal K}_1 \otimes {\cal K}_2 \right\}_{M_O}- \frac{1}{3!} \left\{({\cal K}_1)^3\right\}_{M_O} 
\end{align}
in which ${\cal K}_1={\cal M}_1$ and ${\cal K}_2$ is obtained by Eq.~(\ref{K2}),
\begin{align}
\label{K4}
{\cal K}_4&={\cal M}_4- \frac{1}{2} \left\{({\cal K}_2)^2 \right\}_{M_O} - \frac{1}{2}\left\{({\cal K}_1)^2 \otimes {\cal K}_2\right\}_{M_O}- \frac{1}{4!} \left\{({\cal K}_1)^4 \right\}_{M_O}
\end{align}
in which ${\cal K}_1={\cal M}_1$ and ${\cal K}_2$ and ${\cal K}_3$ are derived from Eq.~(\ref{K2}) and Eq.~(\ref{K3}), respectively,
\begin{align}
	\label{K5}
{\cal K}_5&={\cal M}_5-  \left\{{\cal K}_2\otimes {\cal K}_3\right\}_{M_O}
-\frac{1}{2}  \left\{{\cal K}_2^2 \otimes {\cal K}_1 \right\}_{M_O} \nonumber \\
&	-  \left\{{\cal K}_1\otimes {\cal K}_4\right\}_{M_O}
-\frac{1}{2}  \left\{{\cal K}_1^2 \otimes {\cal K}_3 \right\}_{M_O}
-\frac{1}{3!}  \left\{{\cal K}_1^3 \otimes {\cal K}_2 \right\}_{M_O}
-\frac{1}{5!}  \left\{{\cal K}_1^5 \right\}_{M_O}
\end{align}
where  ${\cal K}_1={\cal M}_1$ and ${\cal K}_2$, ${\cal K}_3$ and  ${\cal K}_4$ are derived from Eq.~(\ref{K2}), Eq.~(\ref{K3}) and Eq.~(\ref{K4}), respectively; and so on. Notice that 	if  ${\cal M}_1=0$
  then   
	${\cal K}_2={\cal M}_2$ and ${\cal K}_3={\cal M}_3$, {\em whatever the  rule $M_O$}.
To go ahead with the above iterative procedure we have to  bear in mind how 
the  $M_O$ map acts on the space of tensor products of the cumulant density functions of Eq.~(\ref{defkappa}) (on which it is defined).  

The same arguments that lead to {\em Proposition}~\ref{Mn_vsKn}  
can also be used to write
the analog closed-form formula that gives directly density moments in terms of density cumulants:
\begin{lemma}\label{lemmaMeeron2}
	The generalized $n$-moment density function $\mu_n^{(\tilde M)}(u_1,u_2,...u_n)$  is given in terms of the generalized cumulant density functions $\kappa_n^{(\tilde M,M_O)}(u_1,u_2,...u_m)$ 
	by the standard Meeron formula, taking care to apply  the $M$-ordering map to the result:
	\begin{align}
	\label{Moments_vs_Cumulants}
	\mu_n^{(\tilde M)}(u_1,u_2,...u_n)&=
	\sum_{p=1}^{ n} \left( \sum_{\sum_{i=1}^p m_i={ n}}  
	\left\{
	\prod_{i=1}^{p}
					\kappa_{m_i}^{(\tilde M,M_O)}
	\right\}_{M_O} \right)
	\end{align}
		where $\kappa_{m_i}^{(\tilde M,M_O)}=\left\{\Omega(u_{j_1},\Omega(u_{j_2}),...,\Omega(u_{j_{m_i}})\right\}_{\tilde M_c}$ is the
	generalized $m_i$-cumulant density function, and $\left\{m_i\right\}$ is a subset of  ${n}:=[1,2,...,n]$, with $m_i$ elements.
\end{lemma}
An equivalent but more compact way to write Eq.~(\ref{Moments_vs_Cumulants}) is the following:
\begin{equation}
\label{mukappaCombiMulti}
\left\{\Omega(u_{1}),\Omega(u_{2}),...,\Omega(u_{n})\right\}_{\tilde M}=\sum_{\pi(n)}  \left\{\prod_{B\in\pi(n)} \left\{ \prod_{i\in B}  \Omega(u_{i}) \right\}_{\tilde M_c}\right\}_{M_O}
\end{equation}
where $\pi(n)$ runs through the list of al set-partitions (or grouping) of $n$ distinguishable objects and  $B$ runs through the list of all blocks of the partition $\pi(n)$.
\begin{remark}
	The formal algebraic expression of Eq.~(\ref{Moments_vs_Cumulants}) that gives generalized moments in terms of generalized cumulants
	{\em does not} depend on the nature of the generalized moments (that is related to the $\tilde M$ map), but it depends on the choice of the $M_O$ map. Of course, the choice of the $M_O$ map is not completely independent of $\tilde M$ but it is constrained by Eq.~(\ref{KgenDef1_}).
\end{remark}
	From a practical point of view, it is more convenient to rephrase the above lemma by adapting to the present general case the recipe of Roednik~\cite[pag.27, ``reverse transformation'']{rPA109} (the 
	Roednik-van Kampen  analytical work on the operator-cumulants defined by the $t$-ordered exponential  can be considered as an anticipation of the modern combinatorial approach to this subject):
	\begin{enumerate}[label=(\roman*)]
		\item\label{n-dots} Write a sequence of $n$-dots.
		\item\label{partition} Partition them into non crossing subsequences $ \left\{.\,.\,.\,.\,.\,.\right\}_{\tilde M_c}$ (cumulant averages inside of which the $M$ mapping is applied) by inserting 
		these brackets in all possible ways (excluding empty subsequences).
		\item For each partition write 1 on the first dot, and any permutation of
		the numerals $2, 3, . . . . .,n$ on the remaining dots, subject to the condition
		that subsequences  with  different ordering of the {\em same} numerals are considered equivalent (or, that is the same,  take just one kind of ordering, for example, not decreasing).
		\item\label{replace} Replace each numeral ``$i$''  
		by ``$\Omega(u_i)$''.
		\item\label{applyR} Apply the $M$-ordering map (the $M_O$ map, shortly) to each addend so obtained.
	\end{enumerate} 
Here we explicitly write down the first four generalized density moments in terms of generalized density cumulants (notice, $\langle \Omega(u)\rangle=\langle \Omega(u)\rangle_c$ whatever the $M_O$ and $M$ mappings,  we also shall use the shorthand notation $\Omega(u_i)\to i$ and $\Omega(u_i)\Omega(u_j)\to i\cdot j$, thus  $ \left\{ i_1 \cdot i_2\cdot...\cdot i_k \right\}_{\tilde M_c}:=
\langle \left\{ i_1 \cdot i_2\cdot...\cdot i_k\right\}_{M}\rangle_c $):
\begin{eqnarray}
\label{DensMom_vs_DensCum}
 \left\{ i \cdot j \right\}_{\tilde M}  &=&
 \left\{ i \cdot j \right\}_{\tilde M_c}  +\big\{ \langle i \rangle \otimes \langle j \rangle \big\}_{M_O}\nonumber\\
 \left\{ i \cdot j \cdot r \right\}_{\tilde M}&=&
 \left\{ i \cdot j \cdot r \right\}_{\tilde M_c}
+\big\{ \langle i \rangle \otimes \langle j \rangle \otimes \langle r \rangle \big\}_{M_O}\nonumber\\
&   +&\underbrace{\big\{  \left\{ i \cdot j \right\}_{\tilde M_c} \otimes  \langle r \rangle \big\}_{M_O}+
	\big\{  \left\{ i \cdot r \right\}_{\tilde M_c} \otimes  \langle j \rangle \big\}_{M_O}+
	\big\{  \left\{ j \cdot r \right\}_{\tilde M_c} \otimes  \langle i \rangle \big\}_{M_O}
}_{:=\left\{  \left\{ i \cdot j \right\}_{\tilde M_c} \otimes  \langle r \rangle \right\}_{M_O}[3]}\nonumber\\
 \left\{ i \cdot j \cdot r \cdot s \right\}_{\tilde M}&=&
 \left\{ i \cdot j \cdot r \cdot s \right\}_{\tilde M_c}
+\big\{ \langle i \rangle \otimes \langle j \rangle \otimes \langle r \rangle \otimes \langle s \rangle\big\}_{M_O}\nonumber\\
&   +&\big\{  \left\{ i \cdot j \cdot r \right\}_{\tilde M_c} \otimes  \langle s \rangle \big\}_{M_O}[4]+
\big\{  \left\{ i \cdot j \right\}_{\tilde M_c} \otimes  \left\{ r \cdot s \right\}_{\tilde M_c} \big\}_{M_O}[3]\nonumber\\
&  +&\big\{  \left\{ i \cdot j \right\}_{\tilde M_c} \otimes  \langle r \rangle \otimes  \langle s \rangle \big\}_{M_O}[6]
\end{eqnarray}
Each number between  square parentheses indicates a sum over distinct partitions
having the same block sizes, so the fourth-order moment is, in general, a sum of 15 distinct density cumulant products.
In some cases however the $M_O$ map could impose to discard some of terms we obtain in the r.h.s. of the  Eq.~(\ref{DensMom_vs_DensCum}).
For example, in the case where $M_O=G$, illustrated in section~\ref{examples}, we take only the partitions 
where the operators are totally ordered respect to the time parameter $u$, see Eq.~(\ref{DensMom_vs_DensCum_O_2}). Another example is where
the $M_O$-projection includes a fully anti-symmetrizer operator (see Section~\ref{sec:fermions}). 
In this case the  relations of Eq.~(\ref{DensMom_vs_DensCum}) have been
extensively used to  express reduced densities matrices of electrons in terms of cumulants that cancel  quantities that are not size-extensive (see Eqs.~(2.25)-(2.28) of the Ziesche seminal paper of~\cite{ziesche2000} and~\cite{smPRA88,mCPL289,mIJQC70,jmJCP125,smPRA88,rmPRA92}).

\subsection{A few important classes of cases\label{examples}}
In the most common cases  $M=O$ (time ordering of the operators $\Omega(u)$) and we shall treat some of them in detail. However, 
before that, we want to discuss, albeit  briefly, the application of our results to the context of quantum mechanics, and in particular
to many body boson or fermion
systems. 
\subsubsection{Many identical particles\label{sec:fermions}}
For the sake of simplicity, we shall deal only with fermions, but the extension to bosons is straightforward. 
This is a field subjected to a very dynamic research activity, in particular concerning the reduced density matrix (RDM) approach and 
the related reduced density matrix cumulants (RDMC) technique, adopted by many research 
groups 
(e.g,~\cite{kmJCP6,smPRA88,ziesche2000,mCPL289,mIJQC70,jmJCP125,pbPRA90,rcspmJCP141,rmPRA92,MazziottiWiley2007}).
We think that the results of the present work can contribute to a simple systematic development of this research field. 
For the reader skilled in this 
matter, it should be enough to observe that the antisymmetric Grassmann (or exterior) product is commutative 
 (when applied between tensors with the same number  of  upper  and  lower  indices) and satisfy the conditions of 
 Definition~\ref{weakordering} for an $M$-ordering map. Thus, if we use it as the $M_O$ map introduced in Section~\ref{CumIntro}, we see that 
whatever the definition of the generating function for the RDM, Lemma~\ref{lemmaMeeron2}, and the explicit expansion of Eq.~(\ref{DensMom_vs_DensCum})
gives  the same relation between RDM and RDMC we find in the specific literature (e.g., see TABLE II of~\cite{mIJQC70}). Notice that in our treatment we don't need to introduce any additional Grassmann function/Schwinger probes.

Let us now give some more (few) details for not expert in this matter. Let us work in the frame of the second quantization, and let us define de $p$ particles RDM as
\begin{equation}
\label{RDM}
\bm{D}_{\bm p}:=\bm{D}_{j_1,...,j_p}^{i_1,...,i_p}:=\frac{1}{p!} \langle \Psi_N| \hat{\bm a}^\dagger_{i_1}\cdots  \hat{\bm a}^\dagger_{i_p} \hat{\bm a}_{j_p}
\cdots  \hat{\bm a}_{j_1}  |\Psi_N\rangle
\end{equation}
where  the number of indices implicitly
specifies the tensor rank, a convention that shall be followed hereafter, $\Psi_N$ represent some state of the whole system of $N$ particles and, finally, as usual
$\hat{\bm a}^\dagger_{i_p}$ ($\hat{\bm a}_{j_p}$) is the creation (annihilation) operator, for the quantum number $i_p$ ($j_p$). For example, if
$\Psi_{N}$ is the pure state of a $N$-electrons system,
the $N$-electron density matrix for the state $\Psi$ is the following projector:
\begin{equation}
\label{RDMPure}
\bm{D}_{\bm{N}}:= |\Psi_N \rangle  \langle \Psi_N|=
\frac{1}{N!} \langle \Psi_N| \hat{\bm a}^\dagger_{i_1}\cdots  \hat{\bm a}^\dagger_{i_N} \hat{\bm a}_{j_N}
\cdots  \hat{\bm a}_{j_1}  |\Psi_N\rangle
\end{equation}
of which a partial trace operation, applied to the indices from $p+1$ ($p<N$) to $N$ (by a factor $N!/p!$), gives, by definition,  the $p$-RDM.
An old result by L\"owdin~\cite{lPR97} makes the RDM interesting: due to the antisymmetry of $\Psi$, $\bm{D}_{p}$ can be used to compute the exact
expectation value of any $p$-electron operator that treats all $p$ electrons equivalently.
Since electrons are indistinguishable according to the postulates of quantum mechanics,
any valid observable must correspond to such an operator. Moreover, it is a consequence of  empirical facts that all the $N$ particle operators used 
in quantum mechanics are the sum of one or two pairs operators. 
We shall not go deeper in describing the advantages of the RDM approach to many-particle systems, the interested reader can consult some 
of the above-cited literature. Here we are focused on the possibility of decomposing the RDM in generalized cumulants. There are many reasons for 
which this decomposition is 
advantageous. We just cite the fact that in the thermodynamic limit ($N\to \infty$) RDMs are not extensive quantities (RDMCs are), so they do not 
necessarily become additively separable in the limit of noninteracting subsystems (RDMs are multiplicatively
separable rather than additively separable). This is easily seen: consider a composite system of identical particles made of two noninteracting
subsystems, one with $p$ electrons (subsystem $A$) and the other with
$q= N-p$ electrons (subsystem $B$).  Thus the $\bm{D}_{ 2}$ RDM is given by 
$\bm{D}_{2}=\frac{1}{2} \langle \Psi_A\Psi_B| \hat{\bm a}^\dagger_{i_1}\hat{\bm a}^\dagger_{i_2} \hat{\bm a}_{j_2}
 \hat{\bm a}_{j_1}  |\Psi_A\Psi_B\rangle$, and it is clear that its matrix elements scales as $N^2$. On the other hand, the second
reduced density matrix cumulant (RDMC) must scale as $N$ because, from the fundamental property of cumulants, the RDMC of the whole system 
is given by the {\em sum} of the RDMC of the two noninteracting parts.

To introduce the RDMC in the treatment of the present work, we start observing that  Eq.~(\ref{RDM}) is written taking care of the normal ordering for 
the products of creation (left position) and annihilation (right position) operators, and we associate to this rule the $M$ map defined in 
Section~\ref{CumIntro}. Moreover, the average process, which combined with $M$ defines the $\tilde M$ map, makes fully antisymmetric the 
 RDM, respect to all the indices.  
Comparing Eq.~(\ref{RDM}) with Eqs.~(\ref{MGgen})-(\ref{defmu}) it is clear that 
the  RDM-generating function is given by 
 \begin{equation}
\label{MGgenRDM}
{\cal M}:= \langle \Psi|\exp_{M}  \left[ \sum_{i_p} \sum_{j_p} \hat{\bm a}^\dagger_{i_p} \hat{\bm a}_{j_p}\right] |\Psi\rangle
:=\exp_{\tilde M} \left[ \sum_{i_p} \sum_{j_p} \hat{\bm a}^\dagger_{i_p} \hat{\bm a}_{j_p}\right].
\end{equation}
Following the procedure of the present work, we write the RDMC as:
\begin{equation}
\label{RDMC}
\bm{\Delta}_{\bm p}:=\bm{\Delta}_{j_1,...,j_p}^{i_1,...,i_p}:=\frac{1}{p!} \langle \hat{\bm a}^\dagger_{i_1}\cdots  \hat{\bm a}^\dagger_{i_p} \hat{\bm a}_{j_p}
\cdots  \hat{\bm a}_{j_1}\rangle_c:=\frac{1}{p!} \left\{\hat{\bm a}^\dagger_{i_1}\cdots  \hat{\bm a}^\dagger_{i_p} \hat{\bm a}_{j_p}
\cdots  \hat{\bm a}_{j_1}\right\}_{\tilde M_c}
\end{equation}
Following the 
definition of Eq.~(\ref{CumulantDefGen}), the RDMC-generating function in the present case can be written as:
\begin{equation}
\label{RDMCDefGen}
\exp_{M_O} \left[{\cal K}\right] := {\cal M}:=\exp_{\tilde M} \left[ \sum_{i_p} \sum_{j_p} \hat{\bm a}^\dagger_{i_p} \hat{\bm a}_{j_p}\right],
\end{equation}
For the $M_O$ map, from which depends the specific definition of the RDMC, we know  that we have different possibilities, but Eq.~(\ref{KgenDef1_}) 
must be satisfied. Because the RDM are fully antisymmetric respect to all the indices, the natural choice for the $M_O$ map is the operator that fully antisymmetrizes the indices of tensors, namely the Grassmann product, that, for two tensors 
${\cal A}_{\bm p}$ and ${\cal B}_{\bm q} $ is   defined by:
\begin{align}
\label{Grassmann}
{\cal A}_{\bm p} \wedge {\cal B}_{\bm q} :={\cal A}_{j_1,...,j_p}^{i_1,...,i_p}\wedge {\cal B}_{j_1,...,j_q}^{i_1,...,i_q}
:= \frac{1}{[(p+q)!]^2}\sum_{\pi,\sigma}\,\pi\sigma 
\epsilon(\pi)\epsilon(\sigma){\cal A}_{j_1,...,j_p}^{i_1,...,i_p} {\cal B}_{j_1,...,j_q}^{i_1,...,i_q}
\end{align}
where $\pi$ represents all permutations of the upper indices and $\sigma$ represents all permutations of the lower indices; the function 
$\epsilon(\pi)$ gives $+1$ for even permutations and $-1$ for odd permutations.  
The Grassmann product is linear and commutative (when applied between tensors with the same number  of  upper  and  lower  indices). Thus we make the following choice for $M_O$:
$$\left\{\bm{\Delta}_{\bm p}\bm{\Delta}_{\bm q}\bm{\Delta}_{\bm s}...\right\}_{M_O}:=\bm{\Delta}_{\bm p}\wedge \bm{\Delta}_{\bm q}\wedge\bm{\Delta}_{\bm s}\wedge...$$
With this choice, by exploiting Lemma~\ref{lemmaMeeron2} or Eq.~(\ref{DensMom_vs_DensCum}) we get exactly the same relation between RDM and RDMC we can find in the specific literature (e.g.~\cite{MazziottiWiley2007,mIJQC70}):
\begin{eqnarray}
\label{RDMvsRDMC}
\bm{D}_{1}&=&\bm{\Delta}_{1}\nonumber \\
\bm{D}_{2}&=&\bm{\Delta}_{2}+\bm{\Delta}_{1}\wedge\bm{\Delta}_{1}\nonumber \\
\bm{D}_{ 3}&=&\bm{\Delta}_{3}+(\bm{\Delta}_{1})^{\wedge 3}+3\bm{\Delta}_{2}\wedge\bm{\Delta}_{1}\nonumber \\
\bm{D}_{ 4}&=&\bm{\Delta}_{4}+(\bm{\Delta}_{1})^{\wedge 4}+6\bm{\Delta}_{2}\wedge(\bm{\Delta}_{1})^{\wedge 2}
+4\bm{\Delta}_{3}\wedge\bm{\Delta}_{1}+3 (\bm{\Delta}_{2})^{\wedge 2}\nonumber \\
\vdots
\end{eqnarray}
where ${\cal A}^{\wedge n}:=\underbrace{{\cal A}\wedge {\cal A}\wedge ..\wedge {\cal A}}_{n\, times}$. Notice that in the expression for 
$\bm{D}_{2}$ in the r.h.s. of Eq.~(\ref{RDMvsRDMC}), the term $\bm{\Delta}_{1}\wedge\bm{\Delta}_{1}$ is the one that
for $N\to \infty$ scales with $N^2$, while the second RDMC represents the correlations, that scales with $N$, because for uncorrelated particles $\bm{\Delta}_{2}$
is additive.

In the same way, we can treat the $p$-density operators (or the $p$-density functions) or the Green's functions of electrons, which
 are the time-dependent version (in Heisenberg picture) of the RDMs~\cite{ziesche2000}.

\subsubsection{The classical time ordered cumulants\label{PTOC}}
In statistical mechanics, both concerning the stochastic Langevin equation (see Eqs.~(\ref{p(t)})-(\ref{MG})) or the treatment of spin systems~\cite{fJCP49},  the case,
 $M=O$ (chronological ordering of the original basis constituted by the operators $\Omega(u)$) is dominant.  In this case the
formal expression of the generalized moment and cumulant density functions are given in Eq.~(\ref{mukappa_O}), and, from the cumulant side,
the original basis is given by the generalized cumulant density functions:
\begin{align}
	\label{Original_Basis_Cum_O} 
\kappa_n^{(\tilde O,M_O)}(u_1,u_2,...,u_n)=\langle \left\{ \Omega(u_1)\Omega(u_2)...\Omega(u_n)\right\}_O \rangle_c.
\end{align}
In terms of this basis, the formal expression for the cumulant generating components are (see Eq.~(\ref{defKnGen_O})):
\begin{align}
\label{_OK}
{\cal K}(t)=& 1+\int_0^t \text{d}u \, \langle \Omega(u) \rangle
+\int_0^t \text{d}u_1 \int_{0}^{u_1} \text{d}u_2 \,\langle \Omega(u_1)\Omega(u_2)\rangle_c\nonumber \\
+...+&\int_0^t \text{d}u_1 \int_{0}^{u_1} \text{d}u_2... \int_{0}^{u_{n-1}} \text{d}u_n\,
\langle \Omega(u_1)\Omega(u_2)...\Omega(u_n)  \rangle_c+...
\end{align}
 Eq.~(\ref{mukappaCombiMulti}) in this case becomes:
\begin{equation}
\label{mukappaCombiMulti_O}
\langle \left\{\Omega(u_{1}),\Omega(u_{2}),...,\Omega(u_{n})\right\}_{O}	\rangle =\sum_{\pi(n)}  \left\{\prod_{B\in\pi(n)} \langle \left\{ \prod_{i\in B}  \Omega(u_{i}) \right\}_{O}\rangle_c\right\}_{M_O}.
\end{equation}
Once we have set $M=O$, the relationship between cumulants and moments depends on the specific $M_O$ map we choose. 
In fact for $M=O$, the constraint in Eqs.~(\ref{KgenDef1_})-(\ref{defmuRR}) leaves some freedom.  
This is made clear by using the following argument: $M=O$ means ``chronological ordering'' of the elements of the original basis given by the stochastic operators
$\Omega(u)$. Because these operators depend only on one time parameter, this ordering is well defined. However, from the cumulant side,  
the elements of the original basis  depend on many time parameters (see Eq.~(\ref{Original_Basis_Cum_O})), thus we have to specify (or choose) 
which of them is involved in the ordering process associated to the $M_O$ map. Two different choices are here presented, the first one in the present Section, the second one in the next Section.
 
If we know (or if we want) that the generalized moment generating function satisfies
a local time differential equation, like
\begin{align}
\label{MEO}
\partial_t {\cal M}= \mathfrak{F}(t)  {\cal M},
\end{align}
we have
\begin{align}
\label{MO_temp}
{\cal M}(t)&= \stackrel{\longleftarrow}{\exp} \left[\int_0^t  \text{d}u \, \mathfrak{F}(u)\right]
\end{align}
thus, 
\begin{align}
\label{MO_2}
{\cal M}(t)&=\stackrel{\longleftarrow}{\exp} \left[\int_0^t  \text{d}u \, \mathfrak{F}(u)\right]
:=\exp_{M_O}[{\cal K}].
\end{align}
From Eq.~(\ref{MO_2}) it follows that $\mathfrak{F}(t)=\left(\partial_t {\cal K}^{\scriptscriptstyle}(t)\right)$, namely
\begin{align}
\label{MO}
{\cal M}(t)&=\exp_{M_O}[{\cal K}]=  \stackrel{\longleftarrow}{\exp} \left[\int_0^t  \text{d}u \,\left( \partial_u {\cal K}(u)\right)\right]\nonumber \\
&=1+\int_0^t \text{d}u_1 \,
\left(\partial_{u_1} {\cal K}(u_1) \right)+
\int_0^t\, \text{d}u_1\int_0^{u_1} \text{d}u_2 \,\left( \partial_{u_1} {\cal K}(u_1)\right)\otimes\left(\partial_{u_2} {\cal K}(u_2)\right) \nonumber \\
&+\int_0^t \text{d}u_1\int_0^{u_1} \text{d}u_2 \int_0^{u_2} \text{d}u_3 \,\left( \partial_{u_1} {\cal K}(u_1)\right) \otimes \left(\partial_{u_2} {\cal K}(u_2)\right)
\otimes \left(\partial_{u_3} {\cal K}(u_3)\right)+...
\end{align}
Exploiting Eq.~(\ref{_OK}) it is easy to see that the series of Eq.~(\ref{MO}) leads to choose $M_O=O$ 
defined in this way: in any tensor product ($u_i\ge u_j$ if $j\ge i$) such as
$$..\otimes \langle \Omega(u_i)\Omega(u_{i+1})...\Omega(u_{i+n-1})\rangle_c
\otimes \langle \Omega(u_j)\Omega(u_{j+1})...\Omega(u_{i+m-1})\rangle_c\otimes ...$$
the $M_O$ map  imposes
	a decreasing time ordering from left to right between the 
	 {\em first} parameter $u_{i}$ of any cumulant density function and  
	the  first parameter $u_j$ of the next cumulant density function, namely, $u_i\ge u_j$.
This kind of ordering for the product of 
cumulant density functions is called Partial Time Ordering (PTO). Using this rule and reminding that the $M=O$ mapping imposes a time ordering inside any average, 
by exploiting Eq.~(\ref{DensMom_vs_DensCum}) ($u_i\ge u_j$ if $j\ge i$) it is easy to write the first four density moments in terms of cumulants:
\begin{eqnarray}
\label{DensMom_vs_DensCum_O}
\langle 1 \cdot 2  \rangle &=&
\langle  1 \cdot 2 \rangle_c +\langle 1 \rangle \otimes \langle 2 \rangle\nonumber\\
\langle 1 \cdot 2 \cdot 3 \rangle&=&
\langle 1 \cdot 2 \cdot 3 \rangle_c+
 \langle 1 \rangle \otimes \langle 2 \rangle \otimes \langle 3 \rangle \nonumber\\
	\text{$[3$ terms:$]$} &   +& \langle  1 \cdot 2 \rangle_c \otimes  \langle 3 \rangle +
	 \langle 1 \cdot 3 \rangle_c \otimes  \langle 2 \rangle +
	  \langle 1 \rangle  \otimes \langle  2 \cdot 3 \rangle_c \nonumber\\
	  \langle  1 \cdot 2 \cdot 3 \cdot 4 \rangle&=&
	  \langle 1 \cdot 2 \cdot 3 \cdot 4 \rangle_c
	  +\langle 1 \rangle \otimes \langle 2 \rangle \otimes \langle 3 \rangle \otimes \langle 4 \rangle \nonumber\\
	\text{$[4$ terms:$]$} &   +& \langle 1 \cdot 2 \cdot 3 \rangle_c \otimes  \langle 4 \rangle
	  +   \langle 1 \rangle \otimes \langle 2 \cdot 3 \cdot 4 \rangle_c
	  + \langle 1 \cdot 3 \cdot 4 \rangle_c \otimes  \langle 2 \rangle
	  + \langle 1 \cdot 2 \cdot 4 \rangle_c \otimes  \langle 3 \rangle \nonumber\\
	\text{$[3$ terms:$]$}  &   +& \langle 1 \cdot 2 \rangle_c \otimes \langle 3 \cdot 4 \rangle_c+ \langle 1 \cdot 3 \rangle_c \otimes \langle 2 \cdot 4 \rangle_c
	  + \langle 1 \cdot 4 \rangle_c \otimes \langle 2 \cdot 3 \rangle_c
	  \nonumber\\
	\text{$[6$ terms:$]$}  &  +& \langle 1 \cdot 2 \rangle_c \otimes  \langle 3 \rangle \otimes  \langle 4 \rangle 
	  +\langle 1 \cdot 3 \rangle_c \otimes  \langle 2 \rangle \otimes  \langle 4 \rangle
	  +\langle 1 \cdot 4 \rangle_c \otimes  \langle 2 \rangle \otimes  \langle 3 \rangle \nonumber \\
	  &+&\langle 1 \rangle   \otimes  \langle 2 \cdot 3 \rangle_c \otimes  \langle 4 \rangle
	  +\langle 1 \rangle   \otimes  \langle 2 \rangle \otimes  \langle 3\cdot 4 \rangle_c
	  +\langle 1 \rangle   \otimes  \langle 2 \cdot 4 \rangle_c \otimes  \langle 3 \rangle
\end{eqnarray}
in which the partial ordering of the numerals is apparent. 
Eq.~(\ref{DensMom_vs_DensCum_O}) is the same we have in the case of commuting stochastic processes,
but here the terms are re-arranged following two rules (that are ineffective in the case of c-numbers): 
\begin{itemize}
	\item inside each group (or average) both in the l.h.s. and in the r.h.s. of Eq.~(\ref{DensMom_vs_DensCum_O})
	it is important to keep the ordering of numerals, increasing from left to right (this is due to the $M=O$ map);
	\item  in each addend, the ordering in the sequence of groups multiplication  is such that the first numeral of
	any group is lesser than the first numeral of the next group (this is because of the $M_O$ mapping
	chosen for this case).
\end{itemize} 
By using the combinatorial approach of set partitions, we can exploit the  
M\"{o}bius inversion formula (see for example~\cite{lEJC23}) to invert Eq.~(\ref{mukappaCombiMulti}) and obtain: 
	\begin{equation}
	\label{kappamuCombiMulti}
\langle \left\{\Omega(u_{1}),\Omega(u_{2}),...,\Omega(u_{n})\right\}_O \rangle_c=\sum_{\pi(n)}   (-1)^{|\pi|-1} 
\sum_{P(B)} P
 \prod_{B\in\pi(n)} \langle \left\{ \prod_{i\in B}  \Omega(u_{i})\right\}_O \rangle
	\end{equation}
	where, again, $\pi$ runs through the list of al set-partitions (or grouping) of $n$ distinguishable objects,  $B$ runs through the list of all blocks of the partition $\pi$,
	$|\pi|$ is the number of blocks in the partition $\pi$ (the number of blocks, corresponding to $p$ in Eq.~(\ref{Moments_vs_Cumulants})),
		$P$ is the permutation operator acting on the  sequence of $|\pi|$ blocks  $B$, but keeping fixed to the left side
		the one that contains (and then starts) with the numeral $1$. Of course, the number of these
		permutations are $(|\pi|-1)!$
	 We stress that, unlike Eq.~(\ref{mukappaCombiMulti_O}), Eq.~(\ref{kappamuCombiMulti})
	  is not the Meeron formula, 
	  modified by simply inserting some $M$-projection/ordering map. In fact here we have a sum over the permutations of blocks that is not present in the
	  standard commuting processes case (where
	  we have a factor  $(|\pi|-1)!$ in place of the same sum).
An example may serve for illustration: 	from Eq.~(\ref{kappamuCombiMulti}) we obtain, for the third cumulant density function, the following expression
\begin{align}
\langle 1 \cdot 2 \cdot 3 \rangle_c=&
\langle 1 \cdot 2 \cdot 3 \rangle
	- \langle  1 \cdot 2 \rangle_c \otimes  \langle 3 \rangle -
\langle 1 \cdot 3 \rangle \otimes  \langle 2 \rangle -
\langle 1 \rangle  \otimes \langle  2 \cdot 3 \rangle \nonumber \\
&+\langle 1 \rangle \otimes \langle 2 \rangle \otimes \langle 3 \rangle
+\langle 1 \rangle \otimes \langle 3 \rangle \otimes \langle 2 \rangle;
\end{align}	
the last two addends came out from the sum of the permutations of the two last groups (made of just one element, 2 or 3). In the case of c-number stochastic processes (and, unfortunately, also in the 
wrong closed-form formula, Eq.~(6.9) of Kubo~\cite{kuboGenCumJPSJ17}) we can collect them in only one addend with a factor $(3-1)!=2$.

Eq.~(\ref{kappamuCombiMulti}) is a compact way to express the procedure  found by 
van Kampen~\cite{vkP74a,vkP74b} and then Roednik~\cite[pag.27]{rPA109} by using a direct calculus approach. 
At the end of the next section  we shall reformulate this result in a slightly different, but equivalent, way.

\subsubsection{Green operator and totally ordered cumulants\label{TTOC}}
Again with $M=O$, let assume now that the generalized moment generating function, instead of satisfying a local equation of motion as~(\ref{MEO}), satisfies an integral (or master) equation:
\begin{equation}
\label{MEG}
\partial_t {\cal M}=\int_0^t \text{d}u\,G(t,u)\otimes {\cal M}(u).
\end{equation} 
This equation is, of course, very common in physics (and not only in physics). For example, it is
obtained by using a  Zwanzig projection
operator technique (e.g.,~\cite{bENTROPY19,bJSTAT2015,grigo_memory,tP74}).  Concerning this case, note that while Terwiel~\cite{tP74} 
proved a cluster property for the ``partial kernels'' obtained using the Zwanzig projection approach, for Fox~\cite{fJMP17} ``his proof does not
establish directly the factorization property for the
ordered cumulants''. The following procedure and Lemma~\ref{TeoFundamental} remedies to this situation.

By a recursive integration of the non-local equation~(\ref{MEG}) we get
\begin{align}
\label{MGG}
{\cal M}&=1+\int_0^t \text{d}u_1\int_0^{u_1} \text{d}u_1'\,G(u_1,u_1') \nonumber \\
&+\int_0^t \text{d}u_1\int_0^{u_1} \text{d}u_1'\,G(u_1,u_1')\otimes\int_0^{u_1'} \text{d}u_2\int_0^{u_2} \text{d}u_2'\,G(u_2,u_2')\nonumber \\
&+\int_0^t \text{d}u_1\int_0^{u_1} \text{d}u_1'\,G(u_1,u_1')\otimes\int_0^{u_1'} \text{d}u_2\int_0^{u_2} \text{d}u_2'\,G(u_2,u_2')\otimes
\int_0^{u_2'} \text{d}u_3\int_0^{u_3} \text{d}u_3'\,G(u_3,u_3')+...\nonumber \\
&:=\exp_G\left[\int_0^t \text{d}u\int_0^{u} \text{d}u'\,G(u,u') \right]:=\exp_G\left[{\cal K}(t) \right].
\end{align}
Thus, $M_O=G$. Assuming the expansion $G(u,u')=\sum_1^\infty G_n(u,u')$ and making the association 
\begin{equation}
\label{associationG}
\int_0^t\text{d}u\int_0^{u} \text{d}u'\,G_n(u,u')= {\cal K}_n(t)
\end{equation}
 we have:
\begin{equation}
\int_0^t \text{d}u \int_0^{u} \text{d}u'G_1(u,u')={\cal K}_1(t)={\cal M}_1(t),
\end{equation}
from which, with some arbitrariness, we can set,
\begin{equation}
\label{G1} G_1(u,u')=\delta(u-u') \langle \Omega(u)\rangle
\end{equation}
and it follows that
\begin{align}
\label{K2G}
&\int_0^t \text{d}u \int_0^{u} \text{d}u'G_2(u,u')={\cal K}_2(t)\nonumber \\
&=\int_0^t \text{d}u_1  \int_0^{u_1} \text{d}u_2
\langle \Omega(u_1) \Omega(u_2)\rangle_c.
\end{align}
Eq.~(\ref{K2G}) leads to the following identifications:
$u=u_1$ and $u'=u_2$, and
\begin{equation}
\label{G2} 
G_2(u,u')=\langle \Omega(u) \Omega(u')\rangle_c.
\end{equation}
Going ahead in this way, we see that for $n\ge 3 $ there is ambiguity in the 
identification of the time $u'$ of $G_n(u,u')$ with one of the times $u_2,u_3,...,u_n$ of the $n$-cumulant density function. 
 The ambiguity is made clear by deriving respect to the time $t$ both sides of Eq.~(\ref{associationG}), from which we get  
\begin{equation}
\label{GCumulants}
\int_0^t du' G_n(t,u') =\int_0^{t} \text{d}u_2\int_0^{u_2} \text{d}u_3... \int_0^{u_{n-1}} \text{d}u_n
\langle \Omega(t)\Omega(u_2)\Omega(u_3)...\Omega(u_n)\rangle_c.
\end{equation}
From the above equation it is clear that the first time parameter $u$ of $G_n(u,u')$ is uniquely associated to the first one, $u_1$, of the 
generalized cumulant density function $ \Omega(t)\Omega(u_2)\Omega(u_3)...\Omega(u_n)\rangle_c$. 
On the other hand, 
for $n>2$ we have $n-1$  different possibilities for the second time parameter $u'$: $u'=u_2$, $u'=u_3$,...,$u'=u_n$ (see Appendix~\ref{switchintegrals}).
Let us make the choice $u'=u_n$, thus (see Eq.~(\ref{Gchoicei}) of Appendix~\ref{switchintegrals}):
\begin{align}
\label{Gchoicen}
& G_n(u,u') =\int_{u'}^{u} \text{d}u_2 \int_{u'}^{u_2} \text{d}u_3...\int_{u'}^{u_{n-2}} \text{d}u_{n-1}
\,\langle \Omega(u)\Omega(u_2)\Omega(u_3)...\Omega(u_{n-1})\Omega(u')\rangle_c,
\end{align}
or 
\begin{align}
\label{Kn_Gn}
{\cal K}_n(t)&=\int_0^t\text{d}u\int_0^{u} \text{d}u'\,G_n(u,u') \nonumber \\
&=\int_0^t\text{d}u
\int_0^{u} \text{d}u_2\int_0^{u_2} \text{d}u_3... \int_0^{u_{n-1}} \text{d}u'
\langle \Omega(u)\Omega(u_2)\Omega(u_3)...\Omega(u')\rangle_c
\end{align}
Exploiting Eq.~(\ref{Kn_Gn}) in the series of Eq.~(\ref{MGG}),
we see that in this case the $M_O$ projection map must be defined in the following way: in any tensor product such as ($n$, $m\in \N$)
$$..\otimes \langle \Omega(u_i)\Omega(u_{i+1})...\Omega(u_{i+n-1})\rangle_c\otimes \langle \Omega(u_j)\Omega(u_{j+1})...\Omega(u_{i+m-1})\rangle_c\otimes ...$$
 the $M_O$ map imposes
a decreasing time ordering from left to right between the 
 {\em last} parameter $u_{i+n-1}$ of any generalized cumulant density function and  
the first parameter $u_j$ of the next cumulant density function, namely, $u_{i+n-1}\ge u_j$.
Because inside the cumulant density function the operators are ordered from left to right (we are considering the case where $M=O$), 
this $M_O$ map leads to a total time ordering (TTO) for the operators $\Omega(u)$ in any expression connecting generalized moments to generalized cumulants and vice-versa. Thus, Eq.~(\ref{Moments_vs_Cumulants}) becomes:
(again, we assume  $u_i> u_j$ for $j>i$):
   \begin{numcases}{ \label{m_nVSk_n}
 	\langle \Omega(u_1)\Omega(u_2)...\Omega(u_n) \rangle=  }
 \label{m_nVSk_n_a}
 \sum_{compositions}  
 \underset{\langle...\rangle_c}{grouping}\left[ \Omega( u_1)\Omega( u_2)...\Omega( u_{ n})\right] 
 \\
 \label{m_nVSk_n_c} 
 \langle \Omega( u_1)(1+\PP_c)\Omega( u_2)(1+\PP_c)\Omega( u_3)....(1+\PP_c)\Omega( u_{ n})\rangle
 \end{numcases}
    In Eq.~(\ref{m_nVSk_n_a}) the sum is over all the $2^{{ n}-1}$ possible ordered compositions of 
   $  \Omega( u_1)\Omega( u_2)....\Omega( u_{ n})$, with grouping made by using the  average
  $\langle...\rangle_c$.  In Eq.~(\ref{m_nVSk_n_c})  the projection operator $\PP_c$ breaks a cumulant average in the product of two  averages: ``$...\PP_c...=...\rangle_c \otimes \langle...$''.  
   A generic term of the sum is given by the product of $p$ cumulant density functions:
  \begin{align}
 &\langle \Omega(u_1)...\Omega(u_{m_1})\rangle_c \otimes \langle \Omega(u_{m_1+1})...\Omega(u_{m_1+m_2})\rangle_c\otimes ... \nonumber \\
  &...\otimes \langle \Omega(u_{m_1+m_2+...+m_{p-1}+1})...\Omega(u_{m_1+m_2+...+m_{p-1}+m_p})\rangle_c
  \end{align}
  such that $m_1+m_2+...+m_{p-1}+m_p=n$ (i.e., the last time index must be equal to $n$). 
  For fixed $p$ groups we have $\frac{({ n}-1)!}{(p-1)!({ n}-p)!}$ possible ways of grouping, of course, $\sum_{p=1}^{ n}\frac{({ n}-1)!}{(p-1)!({ n}-p)!}=2^{{ n}-1}$. 
The first four density moments in terms of cumulants  can be  written by using Eq.~(\ref{m_nVSk_n}), or, more easily,
we can take advantage of the partial ordered result of Eq.~(\ref{DensMom_vs_DensCum_O}), and discard, in the same equation, all the terms that are non totally ordered:
\begin{eqnarray}
\label{DensMom_vs_DensCum_O_2}
\langle 1 \cdot 2  \rangle &=&
\langle  1 \cdot 2 \rangle_c +\langle 1 \rangle \otimes \langle 2 \rangle\nonumber\\
\langle 1 \cdot 2 \cdot 3 \rangle&=&
\langle 1 \cdot 2 \cdot 3 \rangle_c+
\langle 1 \rangle \otimes \langle 2 \rangle \otimes \langle 3 \rangle \nonumber\\
\text{$[3$ terms $\Rightarrow 1$ term:$]$} &   +& \langle  1 \cdot 2 \rangle_c \otimes  \langle 3 \rangle +
\langle 1 \rangle  \otimes \langle  2 \cdot 3 \rangle_c \nonumber\\
\langle  1 \cdot 2 \cdot 3 \cdot 4 \rangle&=&
\langle 1 \cdot 2 \cdot 3 \cdot 4 \rangle_c
+\langle 1 \rangle \otimes \langle 2 \rangle \otimes \langle 3 \rangle \otimes \langle 4 \rangle \nonumber\\
\text{$[4$ terms $\Rightarrow 2$ terms:$]$} &   +& \langle 1 \cdot 2 \cdot 3 \rangle_c \otimes  \langle 4 \rangle
+   \langle 1 \rangle \otimes \langle 2 \cdot 3 \cdot 4 \rangle_c \nonumber\\
\text{$[3$ terms $\Rightarrow 1$ term:$]$}  &   +& \langle 1 \cdot 2 \rangle_c \otimes \langle 3 \cdot 4 \rangle_c
\nonumber\\
\text{$[6$ terms $\Rightarrow 3$ terms:$]$}  &  +& \langle 1 \cdot 2 \rangle_c \otimes  \langle 3 \rangle \otimes  \langle 4 \rangle  \nonumber\\
&+&\langle 1 \rangle   \otimes  \langle 2 \cdot 3 \rangle_c \otimes  \langle 4 \rangle
+\langle 1 \rangle   \otimes  \langle 2 \rangle \otimes  \langle 3\cdot 4 \rangle_c
\end{eqnarray}
The relation of Eq.~(\ref{m_nVSk_n}) (or of Eq.~(\ref{DensMom_vs_DensCum_O_2})) is easily invertible and gives:
 \begin{align}
 \label{k_nVSm_n}
 	\langle \Omega(u_1)\Omega(u_2)...\Omega(u_n) \rangle_c=  
\langle \Omega( u_1)(1-\PP)\Omega( u_2)(1-\PP)\Omega( u_3)....(1-\PP)\Omega( u_{ n})\rangle
 \end{align}
 in which the projection operator $\PP$ is similar to $\PP_c$ but acts to standard moment averages, instead of cumulant ones
  ``$...\PP...=...\rangle \langle...$''.
As for Eq.~(\ref{m_nVSk_n}), the r.h.s. of Eq.~(\ref{k_nVSm_n}) is the sum of 
$2^{n-1}$ terms, each one corresponding to one of the possible different ways to make a partition of the ordered sequence of $n$ operators 
$\Omega(u_i)$, $0\le i\le n$, in $p\le n$ groups (the moments), such that the time ordering
is fully preserved (namely $u_i>u_j$ for $j>i$) and suppling the partition with a factor  $(-1)^{p+1} $. A generic term of the sum is given by
\begin{align}
(-1)^{p+1}&\langle \Omega(u_1)...\Omega(u_{m_1})\rangle \langle \Omega(u_{m_1+1})...\Omega(u_{m_1+m_2})\rangle... \nonumber \\
&...\langle \Omega(u_{m_1+m_2+...+m_{p-1}+1})...\Omega(u_{m_1+m_2+...+m_{p-1}+m_p})\rangle
\end{align}
where $m_1+m_2+...+m_{p-1}+m_p=n$. 
As we have reported at the beginning of this Section, we are now in the position to make a direct 
link between the cumulant approach and the Zwanzig~\cite{Zwanzig2001,grigo_memory} perturbation projection procedure:
from Eq.~(\ref{k_nVSm_n}), the $n$th cumulant generating component 
${\cal K}_n=\int_0^t\text{d}u_1
\int_0^{u_1} \text{d}u_2... \int_0^{u_{n-1}} \text{d}u_n
\langle \Omega(u_1)...\Omega(u_n)\rangle_c$ corresponds to the $n$th order of the series expansion we get from the Zwanzig projection method.
This fact, and the fundamental property for cumulants (Lemma~\ref{TeoFundamental}), prove the connection
between the Terwiel~\cite{tP74} 
cluster property for the ``partial kernels'' arising from the Zwanzig approach and 
 the factorization property of the
ordered cumulants.

In the case where the stochastic process is stationary,  the integral form of Eq.~(\ref{MEG}) becomes a convolution, thus it can
easily Fourier-transformed. From the relation of Eq.~(\ref{associationG}), we have $\hat G(\nu)=\nu^2 \hat {\cal K}(\nu)$.
In general, the TTO cumulant approach can be useful for analysis performed in the frequency domain, sometimes the only possible way for anomalous diffusion processes. 

Eq.~(\ref{DensMom_vs_DensCum_O}) differs from Eq.~(\ref{DensMom_vs_DensCum_O_2}) because the former considers also cross partitions of the
 numerals ${1,2...,n}$, while the latter only ordered (non-crossing) ones. Thus, starting from the 
 reverse relation of Eq.~(\ref{k_nVSm_n}),  by using combinatorial (or graph) arguments, 
 it is easy to obtain  the procedure of Roerdnik~\cite{rPA109}, which gives cumulants in terms of moments also for
  the previous example, 
 where the  moment generating function satisfies the local time equation of Eq.~(\ref{MEO}):
\begin{enumerate}[label=(\roman*)]
\item Write a sequence of $n$-dots.
\item Write $1$ on the first dot, and any permutation of
the numerals $2, 3, . . . . .,n$ on the remaining dots.
\item Partition each of the $(n-1)!$ permutations of numerals into subsequences 
by inserting TTO cumulant averages $\langle.\,.\,.\,.\,.\,.\rangle_c$ with the constraint that two successive numerals belong to the same
subsequence {\em if and only if} the first one is smaller than the second.
\item For each partition consisting of $p$ subsequences supply a factor $(-1)^{p+1}$.
\item Replace each numeral ``$i$'' 
by ``$\Omega(u_i)$''.
\end{enumerate} 

Eqs.~(\ref{k_nVSm_n}) and the related above procedure of Roerdnik are here obtained in a very general context, and can always be
applied when the $M$ map is the time ordering corresponding to a $t$-ordered exponential and when the moment generating function 
satisfies an integral (or time-convolution) equation or a time local (or time-convolutionless) equation, respectively. 
For example, the expansion formulas of Eqs.~(5.9)-(5.10) and Eqs.~(5.17)-(5.18) of~\cite{ucsfPRE60} can be directly obtained by using
Eqs.~(\ref{k_nVSm_n}) and the above Roerdnik's procedure, respectively.

\section{Conclusions}
In this work, we revisit the classical Kubo's approach~\cite{kuboGenCumJPSJ17,kuboGenCumJMP4} (referred as K62-63 in this paper) that introduces  
moment-operators and cumulant-operators.  The Kubo's idea of extending to non-commuting quantities the concept of 
cumulants has been   used by many researchers in the last fifty years as
a tool for some systematic expansion of, for example, the Liouville equation of both classical systems~\cite{vkP74a,vkP74b,fJMP15,tP74}
and spin systems (e.g.~\cite{fJCP49,ydfJCP62}), or to
separate the non extensive parts of the reduced density matrix 
for many body boson or fermion
systems (e.g.,~\cite{smPRA88,ziesche2000,mCPL289,mIJQC70,jmJCP125,pbPRA90,rcspmJCP141,rmPRA92,kmJCP6}). 
However, apart from the general idea, usually the results of these old works of Kubo have not been so exploited. A more case-dependent and analytical 
approach has been instead adopted. This is because, as it was pointed out, for example, by 
Fox~\cite{fJMP17,fJMP20}, in K62-63 there are some theoretical gaps/flaws and, unfortunately, at least questionable results, that
have never been fully cured so far. 

In extending to non-commuting quantities the concept of cumulants
a way to generalize the definition  of exponential of operators must be introduced, so as to recover
the factorization property of the exponential of a sum. Here we give indications on how to do that. 

The generalized
moment generating function is usually identified by the ``facts'' of the specific problem (e.g., from physics) we are interested in. 
For example, 
in the case of fermions a natural $\tilde M$ map  is the physical constraints that the creation operators stay at the left of the 
annihilation ones and that the average
is made with a totally antisymmetric function. Because in this case  we are interested in the reduced density matrices (RDM see~\ref{sec:fermions}),
this $\tilde M$ map naturally leads to Eq.~(\ref{MGgenRDM})  as the RDM-generating function.

Once the generalized moments have been defined, the generalized exponential function that define the cumulant generating function can 
be chosen in different ways, depending on the specific expansion we are interested in. 
The fundamental requirement is that any cumulant that mixes independent ``processes'' should be zero. A property which, 
for non-commuting cumulants, is closely linked to the factorization property of the exponentials of the sums.  
In this work we faced the problem of how to introduce the generalized exponential for cumulant operators, in such a way to 
automatically meet this strong requirement. With Lemma~\ref{TeoFundamental} and Lemma~\ref{LemmaTeoFound1} we provide a solution to this 
problem, suitable for a wide class of practical cases.

Moreover, we also clarify  that, in spite of what stated in K62-63, a general Meeron formula 
that gives generalized cumulants in terms of generalized moments cannot be given, but the reverse is easily obtained (see Lemma~\ref{lemmaMeeron2} and 
Eqs.~(\ref{Moments_vs_Cumulants})-(\ref{mukappaCombiMulti})). It is noticeable that its formal expression does not depend on the
specific definition of generalized moments, but only on the chosen definition of generalized cumulants (the $M_O$ map).

Applications of the  results of the present paper to different problems in physics are briefly illustrated, emphasizing  how it is possible to recover, 
in an unique theoretical framework, many different results that have been obtained by using case-dependent analytical approaches.

Other applications to new problems are under work.

\appendix
\section{A precise definition of ``time ordering''\label{app:timeordering}}
Let us consider the $M$-projection map corresponding to the chronological ordering of operators, from right (smaller times) to
left (larger times). Thus, using the notation
$\left\{...\right\}_O$ to indicate time ordering of the argument ``$...$'', for any two time dependent operators ${\cal A}(t)$ and ${\cal B}(t)$ we have
\begin{equation}
\label{defO}
\left\{ {\cal A}(t_1)\otimes{\cal B}(t_2)
\right\}_O :=
\begin{cases} 
{\cal A}(t_1)\otimes{\cal B}(t_2) \,\,\,\mbox{for $t_1>t_2$}  \\
\frac{{\cal B}(t_1)\otimes{\cal A}(t_1)+{\cal A}(t_1)\otimes{\cal B}(t_1)}{2}  \,\,\,\mbox{for $t_2=t_1$} \\
{\cal B}(t_2)\otimes{\cal A}(t_1)  \,\,\,\mbox{for $t_2>t_1$}.
\end{cases}
\end{equation}
Now, we assume that $t_1>t_2$ and let us consider the sums $S_{\cal A}(t_1,t_2):={\cal A}(t_1)+{\cal A}(t_2)$
and  $S_{\cal B}(t_1,t_2):={\cal B}(t_1)+{\cal B}(t_2)$. One would be naturally tempted to write: 
\begin{align}
\label{SASB}
\left\{S_{\cal A}(t_1,t_2)\otimes S_{\cal B}(t_1,t_2)\right\}_O
=& {\cal A}(t_1)\otimes{\cal B}(t_1)+ {\cal A}(t_1)\otimes{\cal B}(t_2)+{\cal B}(t_1) \otimes{\cal A}(t_2)+ {\cal A}(t_2)\otimes{\cal B}(t_2).
\end{align}
The above definition of time order looks clear, but it present some pitfalls, in fact it invalidates the  distributive 
property of the usual algebra associated to  operators of interest (like differential operators). This is clearly seen with the following two examples.
In the first one we have ${\cal A}(t)={\cal B}(t)= t{\cal C}+{\cal D}$, in which the operators ${\cal C}$ and ${\cal D}$ {\em does not} 
depend on time. From the l.h.s. of Eq.~(\ref{SASB}) we have
\begin{align}
\label{exS1}
&\left\{\big( (t_1{\cal C}+{\cal D})+(t_2{\cal C}+{\cal D})\big)\otimes \big( (t_1{\cal C}+{\cal D})+(t_2{\cal C}+{\cal D})\big)\right\}_O\nonumber \\
&=\left\{\big( (t_1+t_2){\cal C}+2{\cal D})\big)\big( (t_1+t_2){\cal C}+2{\cal D})\big)\right\}_O\nonumber \\
&=(t_1+t_2)^2 {\cal C}^2 +4 {\cal D}^2 +2 (t_1+t_2) ({\cal C}\otimes{\cal D} + {\cal D}\otimes{\cal C})
\end{align}
while from the r.h.s. of Eq.~(\ref{SASB}) we have
\begin{align}
\label{exS2}
&(t_1+t_2)^2 {\cal C}^2 +4 {\cal D}^2 + (t_1+t_2) ({\cal C}\otimes{\cal D} + {\cal D}\otimes{\cal C})+2 t_1({\cal C}\otimes{\cal D}  +2 t_2 {\cal D}\otimes{\cal C}).
\end{align}
Eq.~(\ref{exS1}) differs from (\ref{exS2})  by the term $(t_1+t_2)({\cal D}\otimes{\cal C}-{\cal C}\otimes{\cal D})$. 
In words, the difference is due to the fact that if we {\em first} make the sum ${\cal S}$ of the operators evaluated at different times 
(${\cal A}(t_1) +{\cal A}(t_2)+..)$, as in Eq.~(\ref{exS1}), we loose the detailed information of the value of the operators at the partial times $t_1,t_2,...$, thus the $M$-projection map is not involved. 

The second example is an explicit case of the first one, where ${\cal C}=t\, y\partial_x$ and ${\cal D}=\partial_y$,
namely ${\cal A}(t)=t\, y\partial_x +\partial_y$. Moreover, we assume also that the sum 
$S_{\cal A}$ is extended to
an infinite series of times: $t_0=0, t_1=t/N,t_2=2t/N ,..,t_N=Nt/N=t$, $N\to \infty$. Therefore we have 
$$S_{\cal A}(0,t):=\int_0^t {\cal A}(u)\text{d}u=\int_0^t (u\, y\partial_x +\partial_y)\,\text{d}u=\frac{t^2}{2} y \partial_x + t\,\partial_y.$$
In the time ordered product $\left\{S_{\cal A}(t)S_{\cal A}(t)\right\}_O$, if we first solve the integral, then we have just the square of an operator 
evaluated at the final time $t$ and the $M$-projection map is not effective: 
\begin{align}
\label{TOdef_no}
\left\{S_{\cal A}(0,t)\otimes S_{\cal A}(0,t)\right\}_O&=\left\{\int_0^t {\cal A}(u)\text{d}u\otimes \int_0^t {\cal A}(u)\text{d}u\right\}_O\nonumber \\
&=\left(\frac{t^2}{2} y \partial_x + t\,\partial_y\right)^2=\frac{t^4}{4} y^2 \partial_x^2
+\frac{t^3}{2}(y\partial_x \partial_y+\partial_y y \partial_x) + t^2\,\partial_y^2.
\end{align}
But if the integrals are  made   after the time ordering, we get:
\begin{align}
\label{TOdef}
\left\{S_{\cal A}(0,t)\otimes S_{\cal A}(0,t)\right\}_O&=\left\{\int_0^t {\cal A}(u)\text{d}u\otimes\int_0^t {\cal A}(u)\text{d}u\right\}_O\nonumber \\
&=\int_0^t\text{d}u_1\int_0^t\text{d}u_2 \left\{(u_1\, y\partial_x +\partial_y)
(u_2\, y\partial_x +\partial_y)\right\}_O\nonumber \\
&=2\int_0^t\text{d}u_1\int_0^{u_1}\text{d}u_2 \,(u_1\, y\partial_x +\partial_y)
(u_2\, y\partial_x +\partial_y)\nonumber \\
&=\frac{t^4}{4} y^2 \partial_x^2
+\frac{t^3}{3}(2y\partial_x \partial_y+\partial_y y \partial_x) + t^2\,\partial_y^2,
\end{align}
that is different from Eq.~(\ref{TOdef_no}).
The $t$-ordered exponential is affected by this problem because is a series of terms like $\left\{S_{\cal A}(0,t)^n\right\}_O$. On the other hand, if in the above example we state that  the action of the 
time ordering map  $\{...\}_O$ is defined  only on the special basis generated by the tensor products of ${\cal A}(u)$, $u\in [0,t]$, 
then, when it is applied to the vector $S_{\cal A}(0,t)$, this last must be first decomposed in the special basis:
	$S_{\cal A}(0,t):=\int_0^t {\cal A}(u)\text{d}u$. In this way we avoid any apparent paradox.
This is the reason for which we have
introduced the point~\ref{defR2} in the Definition~\ref{definition1} of the $M$-projection map.
\begin{remark}
	It is trivial but important to notice that in the case where the times of integration (or the sums) are not overlapping, for example 
	$\left\{S_{\cal A}(t_1,t)S_{\cal A}(0,t_1)\right\}_O$ with $0\le t_1 \le t$, we recover the  distributive property of the tensor multiplication combined with the
	$M$-projection map. This is because in this case the non overlapping time intervals defines a partition of $M$-unconnetced sets on $\mathfrak{M}$ (see Definition~\ref{unconnected}).
\end{remark}
\section{A demonstration of the Meeron formula\label{app:meeron}}
	\begin{bew}{of Proposition~\ref{Mn_vsKn}}
		The demonstration  is straightforward and well known for commuting stochastic processes, for which the $M$-projection is just the identity mapping. Here the sketch:  exploiting the multinomial theorem, Eq.~(\ref{KgenDef2}) becomes
		\begin{align}
		\label{KgenDef3}
		\sum_{n=0}^\infty {\cal M}_n=\sum_{m=0}^\infty 
		\sum_{Comp.\,of\,m}\left\{
		\prod_{r=1}^{\infty} \frac{1}{s_r!}\left({\cal K}_r\right)^{s_r} \right\}_{M_O}
		\end{align}
		where the sum of compositions of $m$ means sum over any infinite set of non negative integers $s_1,s_2,..$ with the constraint 
		$\sum_{r=1}^\infty s_r=m$. Because $m$ can assume any value, we can rearrange the r.h.s. of Eq.~(\ref{KgenDef3}) as a sum over partitions of $n$ defined by $\sum_{r=1}^n r\, s_r=n$, as in the statement of the theorem:
		\begin{align}
		\label{KgenDef4}
		\sum_{n=0}^\infty {\cal M}_n=\sum_{n=0}^\infty 
		\sum_{Part.\,of\,n}\left\{
		\prod_{r=1}^{n} \frac{1}{s_r!}\left({\cal K}_r\right)^{s_r} \right\}_{M_O}
		\end{align}
		In the r.h.s. of the above equation, for any fixed $n$ we have sums of terms of order $\sum_{r=1}^n r\, s_r=n$, thus, by equating
		terms of the same order of both sides of Eq.~(\ref{KgenDef4}) the demonstration end. \qed
	\end{bew}
	
\section{How to switch the time integrations\label{switchintegrals}}
Here we demonstrate that the parameter $u'$ of the l.h.s. of Eq.~(\ref{GCumulants}) 
can be associated with anyone of the $n-1$ parameters $u_2,...,u_n$ of the r.h.s. of the same equation. 

Getting rid of the first integral in both the sides of Eq.~(\ref{GCumulants}), one could be tempted to define $u'=u_2$, from which
\begin{align}
\label{Gchoice1}
&G_n(u,u') =\int_0^{u'} \text{d}u'... \int_0^{u_{n-1}} \text{d}u_n\,
\langle \Omega(u)\Omega(u')\Omega(u_3)...\Omega(u_n)\rangle_c.
\end{align}
With this choice the time ordering associated with the generalized exponential defined by the series~(\ref{MGG}), with the identification
written in Eq.~(\ref{associationG}),
would involve the first time $u=u_1$ and to the second time $u'=u_2$ of the density cumulants $\langle \Omega(u_1)\Omega(u_2)..\Omega(u_n)\rangle_c$. 
However, before to get rid of the first integral, in the r.h.s. of 
Eq.~(\ref{GCumulants}) we can change the order of integration between any couple integrals, let us say that involving $u_i$ and $u_{i-1}$:
\begin{equation}
\label{switchintegration}
\int_0^{u_{i-2}} \text{d}u_{i-1}\int_0^{u_{i-1}} \text{d}u_{i}...=
\int_0^{u_{i-2}} \text{d}u_{i}\int_{u_{i}}^{u_{i-2}} \text{d}u_{i-1}...
\end{equation}
and, subsequently repeating this integral switch, we get (we have already set $u_1=u$)
\begin{align}
\label{GCumulants2}
&\int_o^u du' G_n(u,u') =\int_0^{u} \text{d}u_i\int_{u_{i}}^{u} \text{d}u_2 \int_{u_{i}}^{u_2} \text{d}u_3
...\int_{u_{i}}^{u_{i-2}} \text{d}u_{i-1}
\int_{0}^{u_{i}} \text{d}u_{i+1}...\nonumber \\
&...\int_0^{u_{n-2}} \text{d}u_{n-1}\int_0^{u_{n-1}} \text{d}u_n\,
\langle \Omega(u)\Omega(u_2)\Omega(u_3)...\Omega(u_i)...\Omega(u_n)\rangle_c.
\end{align}
Now, getting rid the first integral in both sides of this equation, we are lead  to set $u'=u_i$, from which 
\begin{align}
\label{Gchoicei}
& G_n(u,u') =\int_{u'}^{u} \text{d}u_2 \int_{u'}^{u_2} \text{d}u_3...\int_{u'}^{u_{i-2}} \text{d}u_{i-1}
\int_{0}^{u'} \text{d}u_{i+1}...\nonumber \\
&...\int_0^{u_{n-2}} \text{d}u_{n-1}\int_0^{u_{n-1}} \text{d}u_n\,
\langle \Omega(u)\Omega(u_2)\Omega(u_3)...\Omega(u')...\Omega(u_n)\rangle_c,
\end{align}
\qed
\bibliography{BiblioCentrale}

 \newcommand{\noop}[1]{}
\begin{thebibliography}{56}%
\makeatletter
\providecommand \@ifxundefined [1]{%
 \@ifx{#1\undefined}
}%
\providecommand \@ifnum [1]{%
 \ifnum #1\expandafter \@firstoftwo
 \else \expandafter \@secondoftwo
 \fi
}%
\providecommand \@ifx [1]{%
 \ifx #1\expandafter \@firstoftwo
 \else \expandafter \@secondoftwo
 \fi
}%
\providecommand \natexlab [1]{#1}%
\providecommand \enquote  [1]{``#1''}%
\providecommand \bibnamefont  [1]{#1}%
\providecommand \bibfnamefont [1]{#1}%
\providecommand \citenamefont [1]{#1}%
\providecommand \href@noop [0]{\@secondoftwo}%
\providecommand \href [0]{\begingroup \@sanitize@url \@href}%
\providecommand \@href[1]{\@@startlink{#1}\@@href}%
\providecommand \@@href[1]{\endgroup#1\@@endlink}%
\providecommand \@sanitize@url [0]{\catcode `\\12\catcode `\$12\catcode
  `\&12\catcode `\#12\catcode `\^12\catcode `\_12\catcode `\%12\relax}%
\providecommand \@@startlink[1]{}%
\providecommand \@@endlink[0]{}%
\providecommand \url  [0]{\begingroup\@sanitize@url \@url }%
\providecommand \@url [1]{\endgroup\@href {#1}{\urlprefix }}%
\providecommand \urlprefix  [0]{URL }%
\providecommand \Eprint [0]{\href }%
\providecommand \doibase [0]{http://dx.doi.org/}%
\providecommand \selectlanguage [0]{\@gobble}%
\providecommand \bibinfo  [0]{\@secondoftwo}%
\providecommand \bibfield  [0]{\@secondoftwo}%
\providecommand \translation [1]{[#1]}%
\providecommand \BibitemOpen [0]{}%
\providecommand \bibitemStop [0]{}%
\providecommand \bibitemNoStop [0]{.\EOS\space}%
\providecommand \EOS [0]{\spacefactor3000\relax}%
\providecommand \BibitemShut  [1]{\csname bibitem#1\endcsname}%
\let\auto@bib@innerbib\@empty
\bibitem [{\citenamefont {Kubo}(1962)}]{kuboGenCumJPSJ17}%
  \BibitemOpen
  \bibfield  {author} {\bibinfo {author} {\bibfnamefont {R.}~\bibnamefont
  {Kubo}},\ }\bibfield  {title} {\enquote {\bibinfo {title} {Generalized
  cumulant expansion method},}\ }\href {\doibase 10.1143/JPSJ.17.1100}
  {\bibfield  {journal} {\bibinfo  {journal} {Journal of the Physical Society
  of Japan}\ }\textbf {\bibinfo {volume} {17}},\ \bibinfo {pages} {1100--1120}
  (\bibinfo {year} {1962})},\ \Eprint
  {http://arxiv.org/abs/https://doi.org/10.1143/JPSJ.17.1100}
  {https://doi.org/10.1143/JPSJ.17.1100} \BibitemShut {NoStop}%
\bibitem [{\citenamefont {Kubo}(1963)}]{kuboGenCumJMP4}%
  \BibitemOpen
  \bibfield  {author} {\bibinfo {author} {\bibfnamefont {R.}~\bibnamefont
  {Kubo}},\ }\bibfield  {title} {\enquote {\bibinfo {title} {Stochastic
  liouville equations},}\ }\href {\doibase 10.1063/1.1703941} {\bibfield
  {journal} {\bibinfo  {journal} {Journal of Mathematical Physics}\ }\textbf
  {\bibinfo {volume} {4}},\ \bibinfo {pages} {174--183} (\bibinfo {year}
  {1963})},\ \Eprint {http://arxiv.org/abs/https://doi.org/10.1063/1.1703941}
  {https://doi.org/10.1063/1.1703941} \BibitemShut {NoStop}%
\bibitem [{\citenamefont {Laplace}({\natexlab{a}})}]{Laplace1810}%
  \BibitemOpen
  \bibfield  {author} {\bibinfo {author} {\bibfnamefont {P.}~\bibnamefont
  {Laplace}},\ }\href@noop {} {\emph {\bibinfo {title} {M\'emoire sur les
  approximations des formules qui sont fonctions de tr\`es grands nombres et
  sur leur application aux probabilit\'e}}},\ edited by\ \bibinfo {editor}
  {\bibfnamefont {M.~A.~S.}\ \bibnamefont {Paris}}\BibitemShut {NoStop}%
\bibitem [{\citenamefont {Laplace}({\natexlab{b}})}]{Laplace1812}%
  \BibitemOpen
  \bibfield  {author} {\bibinfo {author} {\bibfnamefont {P.}~\bibnamefont
  {Laplace}},\ }\href@noop {} {\emph {\bibinfo {title} {Th\'eorie analytique
  des probabilit\'es}}},\ \bibinfo {edition} {2nd}\ ed.,\ edited by\ \bibinfo
  {editor} {\bibnamefont {Coucir}}\BibitemShut {NoStop}%
\bibitem [{\citenamefont {Laplace}({\natexlab{c}})}]{Laplace1820}%
  \BibitemOpen
  \bibfield  {author} {\bibinfo {author} {\bibfnamefont {P.}~\bibnamefont
  {Laplace}},\ }\href@noop {} {\emph {\bibinfo {title} {Th\'eorie analytique
  des probabilit\'es}}},\ \bibinfo {edition} {3rd}\ ed.,\ edited by\ \bibinfo
  {editor} {\bibnamefont {Coucir}}\BibitemShut {NoStop}%
\bibitem [{\citenamefont {Fox}(1976)}]{fJMP17}%
  \BibitemOpen
  \bibfield  {author} {\bibinfo {author} {\bibfnamefont {R.~F.}\ \bibnamefont
  {Fox}},\ }\bibfield  {title} {\enquote {\bibinfo {title} {Critique of the
  generalized cumulant expansion method},}\ }\href {\doibase 10.1063/1.523041}
  {\bibfield  {journal} {\bibinfo  {journal} {Journal of Mathematical Physics}\
  }\textbf {\bibinfo {volume} {17}},\ \bibinfo {pages} {1148--1153} (\bibinfo
  {year} {1976})},\ \Eprint
  {http://arxiv.org/abs/https://doi.org/10.1063/1.523041}
  {https://doi.org/10.1063/1.523041} \BibitemShut {NoStop}%
\bibitem [{\citenamefont {Fox}(1979)}]{fJMP20}%
  \BibitemOpen
  \bibfield  {author} {\bibinfo {author} {\bibfnamefont {R.~F.}\ \bibnamefont
  {Fox}},\ }\bibfield  {title} {\enquote {\bibinfo {title} {Time ordered
  operator cumulants: Statistical independence and noncommutativity},}\ }\href
  {\doibase 10.1063/1.524055} {\bibfield  {journal} {\bibinfo  {journal}
  {Journal of Mathematical Physics}\ }\textbf {\bibinfo {volume} {20}},\
  \bibinfo {pages} {2467--2470} (\bibinfo {year} {1979})},\ \Eprint
  {http://arxiv.org/abs/https://doi.org/10.1063/1.524055}
  {https://doi.org/10.1063/1.524055} \BibitemShut {NoStop}%
\bibitem [{\citenamefont {Nica}\ and\ \citenamefont
  {Speicher}(1998)}]{nsDMJ92}%
  \BibitemOpen
  \bibfield  {author} {\bibinfo {author} {\bibfnamefont {A.}~\bibnamefont
  {Nica}}\ and\ \bibinfo {author} {\bibfnamefont {R.}~\bibnamefont
  {Speicher}},\ }\bibfield  {title} {\enquote {\bibinfo {title} {Commutators of
  free random variables},}\ }\href {\doibase 10.1215/S0012-7094-98-09216-X}
  {\bibfield  {journal} {\bibinfo  {journal} {Duke Math. J.}\ }\textbf
  {\bibinfo {volume} {92}},\ \bibinfo {pages} {553--592} (\bibinfo {year}
  {1998})}\BibitemShut {NoStop}%
\bibitem [{\citenamefont {Lehner}(2002)}]{lEJC23}%
  \BibitemOpen
  \bibfield  {author} {\bibinfo {author} {\bibfnamefont {F.}~\bibnamefont
  {Lehner}},\ }\bibfield  {title} {\enquote {\bibinfo {title} {Free cumulants
  and enumeration of connected partitions},}\ }\href {\doibase
  https://doi.org/10.1006/eujc.2002.0619} {\bibfield  {journal} {\bibinfo
  {journal} {European Journal of Combinatorics}\ }\textbf {\bibinfo {volume}
  {23}},\ \bibinfo {pages} {1025 -- 1031} (\bibinfo {year} {2002})}\BibitemShut
  {NoStop}%
\bibitem [{\citenamefont {Freed}(1968)}]{fJCP49}%
  \BibitemOpen
  \bibfield  {author} {\bibinfo {author} {\bibfnamefont {J.~H.}\ \bibnamefont
  {Freed}},\ }\bibfield  {title} {\enquote {\bibinfo {title} {Generalized
  cumulant expansions and spin‐relaxation theory},}\ }\href {\doibase
  10.1063/1.1669833} {\bibfield  {journal} {\bibinfo  {journal} {The Journal of
  Chemical Physics}\ }\textbf {\bibinfo {volume} {49}},\ \bibinfo {pages}
  {376--391} (\bibinfo {year} {1968})},\ \Eprint
  {http://arxiv.org/abs/https://doi.org/10.1063/1.1669833}
  {https://doi.org/10.1063/1.1669833} \BibitemShut {NoStop}%
\bibitem [{\citenamefont {Yoon}, \citenamefont {Deutch},\ and\ \citenamefont
  {Freed}(1975)}]{ydfJCP62}%
  \BibitemOpen
  \bibfield  {author} {\bibinfo {author} {\bibfnamefont {B.}~\bibnamefont
  {Yoon}}, \bibinfo {author} {\bibfnamefont {J.~M.}\ \bibnamefont {Deutch}}, \
  and\ \bibinfo {author} {\bibfnamefont {J.~H.}\ \bibnamefont {Freed}},\
  }\bibfield  {title} {\enquote {\bibinfo {title} {A comparison of generalized
  cumulant and projection operator methods in spin‐relaxation theory},}\
  }\href {\doibase 10.1063/1.430417} {\bibfield  {journal} {\bibinfo  {journal}
  {The Journal of Chemical Physics}\ }\textbf {\bibinfo {volume} {62}},\
  \bibinfo {pages} {4687--4696} (\bibinfo {year} {1975})},\ \Eprint
  {http://arxiv.org/abs/https://doi.org/10.1063/1.430417}
  {https://doi.org/10.1063/1.430417} \BibitemShut {NoStop}%
\bibitem [{\citenamefont {Skolnik}\ and\ \citenamefont
  {Mazziotti}(2013)}]{smPRA88}%
  \BibitemOpen
  \bibfield  {author} {\bibinfo {author} {\bibfnamefont {J.~T.}\ \bibnamefont
  {Skolnik}}\ and\ \bibinfo {author} {\bibfnamefont {D.~A.}\ \bibnamefont
  {Mazziotti}},\ }\bibfield  {title} {\enquote {\bibinfo {title} {Cumulant
  reduced density matrices as measures of statistical dependence and
  entanglement between electronic quantum domains with application to
  photosynthetic light harvesting},}\ }\href {\doibase
  10.1103/PhysRevA.88.032517} {\bibfield  {journal} {\bibinfo  {journal} {Phys.
  Rev. A}\ }\textbf {\bibinfo {volume} {88}},\ \bibinfo {pages} {032517}
  (\bibinfo {year} {2013})}\BibitemShut {NoStop}%
\bibitem [{\citenamefont {Ziesche}(2000)}]{ziesche2000}%
  \BibitemOpen
  \bibfield  {author} {\bibinfo {author} {\bibfnamefont {P.}~\bibnamefont
  {Ziesche}},\ }\enquote {\bibinfo {title} {Cumulant expansions of reduced
  densities, reduced density matrices, and green's functions},}\ in\ \href
  {\doibase 10.1007/978-1-4615-4211-7} {\emph {\bibinfo {booktitle}
  {Many-Electron Densities and Reduced Density Matrices}}},\ \bibinfo {series
  and number} {Mathematical and Computational Chemistry},\ \bibinfo {editor}
  {edited by\ \bibinfo {editor} {\bibfnamefont {J.}~\bibnamefont
  {Cioslowski}}}\ (\bibinfo  {publisher} {Springer US},\ \bibinfo {year}
  {2000})\ Chap.~\bibinfo {chapter} {3}, pp.\ \bibinfo {pages} {XIV, 301},\
  \bibinfo {edition} {1st}\ ed.,\ \Eprint
  {http://arxiv.org/abs/978-1-4615-4211-7} {978-1-4615-4211-7} \BibitemShut
  {NoStop}%
\bibitem [{\citenamefont {Mazziotti}(1998{\natexlab{a}})}]{mCPL289}%
  \BibitemOpen
  \bibfield  {author} {\bibinfo {author} {\bibfnamefont {D.~A.}\ \bibnamefont
  {Mazziotti}},\ }\bibfield  {title} {\enquote {\bibinfo {title} {Approximate
  solution for electron correlation through the use of schwinger probes},}\
  }\href {\doibase https://doi.org/10.1016/S0009-2614(98)00470-9} {\bibfield
  {journal} {\bibinfo  {journal} {Chemical Physics Letters}\ }\textbf {\bibinfo
  {volume} {289}},\ \bibinfo {pages} {419 -- 427} (\bibinfo {year}
  {1998}{\natexlab{a}})}\BibitemShut {NoStop}%
\bibitem [{\citenamefont {Mazziotti}(1998{\natexlab{b}})}]{mIJQC70}%
  \BibitemOpen
  \bibfield  {author} {\bibinfo {author} {\bibfnamefont {D.~A.}\ \bibnamefont
  {Mazziotti}},\ }\bibfield  {title} {\enquote {\bibinfo {title}
  {3,5-contracted schrödinger equation: Determining quantum energies and
  reduced density matrices without wave functions},}\ }\href {\doibase
  10.1002/(SICI)1097-461X(1998)70:4/5<557::AID-QUA2>3.0.CO;2-U} {\bibfield
  {journal} {\bibinfo  {journal} {International Journal of Quantum Chemistry}\
  }\textbf {\bibinfo {volume} {70}},\ \bibinfo {pages} {557--570} (\bibinfo
  {year} {1998}{\natexlab{b}})}\BibitemShut {NoStop}%
\bibitem [{\citenamefont {Juhász}\ and\ \citenamefont
  {Mazziotti}(2006)}]{jmJCP125}%
  \BibitemOpen
  \bibfield  {author} {\bibinfo {author} {\bibfnamefont {T.}~\bibnamefont
  {Juhász}}\ and\ \bibinfo {author} {\bibfnamefont {D.~A.}\ \bibnamefont
  {Mazziotti}},\ }\bibfield  {title} {\enquote {\bibinfo {title} {The cumulant
  two-particle reduced density matrix as a measure of electron correlation and
  entanglement},}\ }\href {\doibase 10.1063/1.2378768} {\bibfield  {journal}
  {\bibinfo  {journal} {The Journal of Chemical Physics}\ }\textbf {\bibinfo
  {volume} {125}},\ \bibinfo {pages} {174105} (\bibinfo {year} {2006})},\
  \Eprint {http://arxiv.org/abs/https://doi.org/10.1063/1.2378768}
  {https://doi.org/10.1063/1.2378768} \BibitemShut {NoStop}%
\bibitem [{\citenamefont {Pavlyukh}\ and\ \citenamefont
  {Berakdar}(2014)}]{pbPRA90}%
  \BibitemOpen
  \bibfield  {author} {\bibinfo {author} {\bibfnamefont {Y.}~\bibnamefont
  {Pavlyukh}}\ and\ \bibinfo {author} {\bibfnamefont {J.}~\bibnamefont
  {Berakdar}},\ }\bibfield  {title} {\enquote {\bibinfo {title} {Accessing
  electronic correlations by half-cycle pulses and time-resolved
  spectroscopy},}\ }\href {\doibase 10.1103/PhysRevA.90.053417} {\bibfield
  {journal} {\bibinfo  {journal} {Phys. Rev. A}\ }\textbf {\bibinfo {volume}
  {90}},\ \bibinfo {pages} {053417} (\bibinfo {year} {2014})}\BibitemShut
  {NoStop}%
\bibitem [{\citenamefont {Ramos-Cordoba}\ \emph {et~al.}(2014)\citenamefont
  {Ramos-Cordoba}, \citenamefont {Salvador}, \citenamefont {Piris},\ and\
  \citenamefont {Matito}}]{rcspmJCP141}%
  \BibitemOpen
  \bibfield  {author} {\bibinfo {author} {\bibfnamefont {E.}~\bibnamefont
  {Ramos-Cordoba}}, \bibinfo {author} {\bibfnamefont {P.}~\bibnamefont
  {Salvador}}, \bibinfo {author} {\bibfnamefont {M.}~\bibnamefont {Piris}}, \
  and\ \bibinfo {author} {\bibfnamefont {E.}~\bibnamefont {Matito}},\
  }\bibfield  {title} {\enquote {\bibinfo {title} {Two new constraints for the
  cumulant matrix},}\ }\href {\doibase 10.1063/1.4903449} {\bibfield  {journal}
  {\bibinfo  {journal} {The Journal of Chemical Physics}\ }\textbf {\bibinfo
  {volume} {141}},\ \bibinfo {pages} {234101} (\bibinfo {year} {2014})},\
  \Eprint {http://arxiv.org/abs/https://doi.org/10.1063/1.4903449}
  {https://doi.org/10.1063/1.4903449} \BibitemShut {NoStop}%
\bibitem [{\citenamefont {Raeber}\ and\ \citenamefont
  {Mazziotti}(2015)}]{rmPRA92}%
  \BibitemOpen
  \bibfield  {author} {\bibinfo {author} {\bibfnamefont {A.}~\bibnamefont
  {Raeber}}\ and\ \bibinfo {author} {\bibfnamefont {D.~A.}\ \bibnamefont
  {Mazziotti}},\ }\bibfield  {title} {\enquote {\bibinfo {title} {Large
  eigenvalue of the cumulant part of the two-electron reduced density matrix as
  a measure of off-diagonal long-range order},}\ }\href {\doibase
  10.1103/PhysRevA.92.052502} {\bibfield  {journal} {\bibinfo  {journal} {Phys.
  Rev. A}\ }\textbf {\bibinfo {volume} {92}},\ \bibinfo {pages} {052502}
  (\bibinfo {year} {2015})}\BibitemShut {NoStop}%
\bibitem [{Note1()}]{Note1}%
  \BibitemOpen
  \bibinfo {note} {For historical reason we shall use the definition of
  q-numbers as objects of a {\protect \em non commutative} algebra, as opposed
  to c-numbers that are objects of a {\protect \em commutative} algebra. The
  operators considered in the present work are generally
  q-numbers.}\BibitemShut {Stop}%
\bibitem [{\citenamefont {Kampen}(1974{\natexlab{a}})}]{vkP74a}%
  \BibitemOpen
  \bibfield  {author} {\bibinfo {author} {\bibfnamefont {N.~V.}\ \bibnamefont
  {Kampen}},\ }\bibfield  {title} {\enquote {\bibinfo {title} {A cumulant
  expansion for stochastic linear differential equations. i},}\ }\href
  {\doibase https://doi.org/10.1016/0031-8914(74)90121-9} {\bibfield  {journal}
  {\bibinfo  {journal} {Physica}\ }\textbf {\bibinfo {volume} {74}},\ \bibinfo
  {pages} {215 -- 238} (\bibinfo {year} {1974}{\natexlab{a}})}\BibitemShut
  {NoStop}%
\bibitem [{\citenamefont {Kampen}(1974{\natexlab{b}})}]{vkP74b}%
  \BibitemOpen
  \bibfield  {author} {\bibinfo {author} {\bibfnamefont {N.~V.}\ \bibnamefont
  {Kampen}},\ }\bibfield  {title} {\enquote {\bibinfo {title} {A cumulant
  expansion for stochastic linear differential equations. ii},}\ }\href
  {\doibase https://doi.org/10.1016/0031-8914(74)90122-0} {\bibfield  {journal}
  {\bibinfo  {journal} {Physica}\ }\textbf {\bibinfo {volume} {74}},\ \bibinfo
  {pages} {239 -- 247} (\bibinfo {year} {1974}{\natexlab{b}})}\BibitemShut
  {NoStop}%
\bibitem [{\citenamefont {Terwiel}(1974)}]{tP74}%
  \BibitemOpen
  \bibfield  {author} {\bibinfo {author} {\bibfnamefont {R.}~\bibnamefont
  {Terwiel}},\ }\bibfield  {title} {\enquote {\bibinfo {title} {Projection
  operator method applied to stochastic linear differential equations},}\
  }\href {\doibase https://doi.org/10.1016/0031-8914(74)90123-2} {\bibfield
  {journal} {\bibinfo  {journal} {Physica}\ }\textbf {\bibinfo {volume} {74}},\
  \bibinfo {pages} {248 -- 265} (\bibinfo {year} {1974})}\BibitemShut {NoStop}%
\bibitem [{\citenamefont {Casula}, \citenamefont {Rubtsov},\ and\ \citenamefont
  {Biermann}(2012)}]{crbPRB85}%
  \BibitemOpen
  \bibfield  {author} {\bibinfo {author} {\bibfnamefont {M.}~\bibnamefont
  {Casula}}, \bibinfo {author} {\bibfnamefont {A.}~\bibnamefont {Rubtsov}}, \
  and\ \bibinfo {author} {\bibfnamefont {S.}~\bibnamefont {Biermann}},\
  }\bibfield  {title} {\enquote {\bibinfo {title} {Dynamical screening effects
  in correlated materials: Plasmon satellites and spectral weight transfers
  from a green's function ansatz to extended dynamical mean field theory},}\
  }\href {\doibase 10.1103/PhysRevB.85.035115} {\bibfield  {journal} {\bibinfo
  {journal} {Phys. Rev. B}\ }\textbf {\bibinfo {volume} {85}},\ \bibinfo
  {pages} {035115} (\bibinfo {year} {2012})}\BibitemShut {NoStop}%
\bibitem [{\citenamefont {Hedin}(1980)}]{hPS21}%
  \BibitemOpen
  \bibfield  {author} {\bibinfo {author} {\bibfnamefont {L.}~\bibnamefont
  {Hedin}},\ }\bibfield  {title} {\enquote {\bibinfo {title} {Effects of recoil
  on shake-up spectra in metals},}\ }\href {\doibase
  10.1088/0031-8949/21/3-4/039} {\bibfield  {journal} {\bibinfo  {journal}
  {Physica Scripta}\ }\textbf {\bibinfo {volume} {21}},\ \bibinfo {pages}
  {477--480} (\bibinfo {year} {1980})}\BibitemShut {NoStop}%
\bibitem [{\citenamefont {Aryasetiawan}, \citenamefont {Hedin},\ and\
  \citenamefont {Karlsson}(1996)}]{ahkPRL77}%
  \BibitemOpen
  \bibfield  {author} {\bibinfo {author} {\bibfnamefont {F.}~\bibnamefont
  {Aryasetiawan}}, \bibinfo {author} {\bibfnamefont {L.}~\bibnamefont {Hedin}},
  \ and\ \bibinfo {author} {\bibfnamefont {K.}~\bibnamefont {Karlsson}},\
  }\bibfield  {title} {\enquote {\bibinfo {title} {Multiple plasmon satellites
  in na and al spectral functions from ab initio cumulant expansion},}\ }\href
  {\doibase 10.1103/PhysRevLett.77.2268} {\bibfield  {journal} {\bibinfo
  {journal} {Phys. Rev. Lett.}\ }\textbf {\bibinfo {volume} {77}},\ \bibinfo
  {pages} {2268--2271} (\bibinfo {year} {1996})}\BibitemShut {NoStop}%
\bibitem [{\citenamefont {Guzzo}\ \emph {et~al.}(2011)\citenamefont {Guzzo},
  \citenamefont {Lani}, \citenamefont {Sottile}, \citenamefont {Romaniello},
  \citenamefont {Gatti}, \citenamefont {Kas}, \citenamefont {Rehr},
  \citenamefont {Silly}, \citenamefont {Sirotti},\ and\ \citenamefont
  {Reining}}]{glsrgkrssrPRL107}%
  \BibitemOpen
  \bibfield  {author} {\bibinfo {author} {\bibfnamefont {M.}~\bibnamefont
  {Guzzo}}, \bibinfo {author} {\bibfnamefont {G.}~\bibnamefont {Lani}},
  \bibinfo {author} {\bibfnamefont {F.}~\bibnamefont {Sottile}}, \bibinfo
  {author} {\bibfnamefont {P.}~\bibnamefont {Romaniello}}, \bibinfo {author}
  {\bibfnamefont {M.}~\bibnamefont {Gatti}}, \bibinfo {author} {\bibfnamefont
  {J.~J.}\ \bibnamefont {Kas}}, \bibinfo {author} {\bibfnamefont {J.~J.}\
  \bibnamefont {Rehr}}, \bibinfo {author} {\bibfnamefont {M.~G.}\ \bibnamefont
  {Silly}}, \bibinfo {author} {\bibfnamefont {F.}~\bibnamefont {Sirotti}}, \
  and\ \bibinfo {author} {\bibfnamefont {L.}~\bibnamefont {Reining}},\
  }\bibfield  {title} {\enquote {\bibinfo {title} {Valence electron
  photoemission spectrum of semiconductors: Ab initio description of multiple
  satellites},}\ }\href {\doibase 10.1103/PhysRevLett.107.166401} {\bibfield
  {journal} {\bibinfo  {journal} {Phys. Rev. Lett.}\ }\textbf {\bibinfo
  {volume} {107}},\ \bibinfo {pages} {166401} (\bibinfo {year}
  {2011})}\BibitemShut {NoStop}%
\bibitem [{\citenamefont {M\"uller-Hartmann}, \citenamefont {Ramakrishnan},\
  and\ \citenamefont {Toulouse}(1971)}]{m-hrtPRB3}%
  \BibitemOpen
  \bibfield  {author} {\bibinfo {author} {\bibfnamefont {E.}~\bibnamefont
  {M\"uller-Hartmann}}, \bibinfo {author} {\bibfnamefont {T.~V.}\ \bibnamefont
  {Ramakrishnan}}, \ and\ \bibinfo {author} {\bibfnamefont {G.}~\bibnamefont
  {Toulouse}},\ }\bibfield  {title} {\enquote {\bibinfo {title} {Localized
  dynamic perturbations in metals},}\ }\href {\doibase 10.1103/PhysRevB.3.1102}
  {\bibfield  {journal} {\bibinfo  {journal} {Phys. Rev. B}\ }\textbf {\bibinfo
  {volume} {3}},\ \bibinfo {pages} {1102--1119} (\bibinfo {year}
  {1971})}\BibitemShut {NoStop}%
\bibitem [{\citenamefont {Kas}, \citenamefont {Rehr},\ and\ \citenamefont
  {Reining}(2014)}]{krrPRB90}%
  \BibitemOpen
  \bibfield  {author} {\bibinfo {author} {\bibfnamefont {J.~J.}\ \bibnamefont
  {Kas}}, \bibinfo {author} {\bibfnamefont {J.~J.}\ \bibnamefont {Rehr}}, \
  and\ \bibinfo {author} {\bibfnamefont {L.}~\bibnamefont {Reining}},\
  }\bibfield  {title} {\enquote {\bibinfo {title} {Cumulant expansion of the
  retarded one-electron green function},}\ }\href {\doibase
  10.1103/PhysRevB.90.085112} {\bibfield  {journal} {\bibinfo  {journal} {Phys.
  Rev. B}\ }\textbf {\bibinfo {volume} {90}},\ \bibinfo {pages} {085112}
  (\bibinfo {year} {2014})}\BibitemShut {NoStop}%
\bibitem [{\citenamefont {Mahan}(2008)}]{Mahan2008}%
  \BibitemOpen
  \bibfield  {author} {\bibinfo {author} {\bibfnamefont {G.~D.}\ \bibnamefont
  {Mahan}},\ }\href@noop {} {\emph {\bibinfo {title} {Many-particle physics,
  3e}}},\ edited by\ \bibinfo {editor} {\bibfnamefont {S.~I.~P.}\ \bibnamefont
  {Limited}},\ Physics of solids and liquids\ (\bibinfo  {publisher} {Kluwer
  Academic/Plenum Publishers},\ \bibinfo {year} {2008})\ p.\ \bibinfo {pages}
  {788}\BibitemShut {NoStop}%
\bibitem [{\citenamefont {Mukherjee}(1995)}]{Mukherjee1995}%
  \BibitemOpen
  \bibfield  {author} {\bibinfo {author} {\bibfnamefont {D.}~\bibnamefont
  {Mukherjee}},\ }\enquote {\bibinfo {title} {A coupled cluster approach to the
  electron correlation problem using a correlated reference state},}\ in\ \href
  {\doibase 10.1007/978-1-4615-1937-9_12} {\emph {\bibinfo {booktitle} {Recent
  Progress in Many-Body Theories: Volume 4}}},\ \bibinfo {editor} {edited by\
  \bibinfo {editor} {\bibfnamefont {E.}~\bibnamefont {Schachinger}}, \bibinfo
  {editor} {\bibfnamefont {H.}~\bibnamefont {Mitter}}, \ and\ \bibinfo {editor}
  {\bibfnamefont {H.}~\bibnamefont {Sormann}}}\ (\bibinfo  {publisher}
  {Springer US},\ \bibinfo {address} {Boston, MA},\ \bibinfo {year} {1995})\
  pp.\ \bibinfo {pages} {127--133}\BibitemShut {NoStop}%
\bibitem [{\citenamefont {Hanauer}\ and\ \citenamefont
  {Köhn}(2012)}]{hkJCP401}%
  \BibitemOpen
  \bibfield  {author} {\bibinfo {author} {\bibfnamefont {M.}~\bibnamefont
  {Hanauer}}\ and\ \bibinfo {author} {\bibfnamefont {A.}~\bibnamefont
  {Köhn}},\ }\bibfield  {title} {\enquote {\bibinfo {title} {Meaning and
  magnitude of the reduced density matrix cumulants},}\ }\href {\doibase
  https://doi.org/10.1016/j.chemphys.2011.09.024} {\bibfield  {journal}
  {\bibinfo  {journal} {Chemical Physics}\ }\textbf {\bibinfo {volume} {401}},\
  \bibinfo {pages} {50 -- 61} (\bibinfo {year} {2012})},\ \bibinfo {note}
  {recent advances in electron correlation methods and
  applications}\BibitemShut {NoStop}%
\bibitem [{\citenamefont {Honmi}\ \emph {et~al.}(2015)\citenamefont {Honmi},
  \citenamefont {Hashizume}, \citenamefont {Nakajima},\ and\ \citenamefont
  {Okamura}}]{hytsPA433}%
  \BibitemOpen
  \bibfield  {author} {\bibinfo {author} {\bibfnamefont {H.}~\bibnamefont
  {Honmi}}, \bibinfo {author} {\bibfnamefont {Y.}~\bibnamefont {Hashizume}},
  \bibinfo {author} {\bibfnamefont {T.}~\bibnamefont {Nakajima}}, \ and\
  \bibinfo {author} {\bibfnamefont {S.}~\bibnamefont {Okamura}},\ }\bibfield
  {title} {\enquote {\bibinfo {title} {Microscopic study on magnetocaloric and
  electrocaloric effects near the critical point},}\ }\href {\doibase
  https://doi.org/10.1016/j.physa.2015.03.079} {\bibfield  {journal} {\bibinfo
  {journal} {Physica A: Statistical Mechanics and its Applications}\ }\textbf
  {\bibinfo {volume} {433}},\ \bibinfo {pages} {126 -- 135} (\bibinfo {year}
  {2015})}\BibitemShut {NoStop}%
\bibitem [{\citenamefont {Suzuki}(1967)}]{suzukiJPSJ22}%
  \BibitemOpen
  \bibfield  {author} {\bibinfo {author} {\bibfnamefont {M.}~\bibnamefont
  {Suzuki}},\ }\bibfield  {title} {\enquote {\bibinfo {title} {A
  semi-phenomenological theory of the second order phase transitions in spin
  systems. i},}\ }\href {\doibase doi:10.1143/jpsj.22.757} {\bibfield
  {journal} {\bibinfo  {journal} {Journal of the Physical Society of Japan}\
  }\textbf {\bibinfo {volume} {22}},\ \bibinfo {pages} {756--761} (\bibinfo
  {year} {1967})},\ \Eprint
  {http://arxiv.org/abs/https://journals.jps.jp/doi/10.1143/JPSJ.22.757}
  {https://journals.jps.jp/doi/10.1143/JPSJ.22.757} \BibitemShut {NoStop}%
\bibitem [{\citenamefont {Suzuki}(1980)}]{sPTPS69}%
  \BibitemOpen
  \bibfield  {author} {\bibinfo {author} {\bibfnamefont {M.}~\bibnamefont
  {Suzuki}},\ }\bibfield  {title} {\enquote {\bibinfo {title} {{Fluctuation and
  Relaxation in Stochastic Systems}},}\ }\href {\doibase 10.1143/PTP.69.160}
  {\bibfield  {journal} {\bibinfo  {journal} {Progress of Theoretical Physics
  Supplement}\ }\textbf {\bibinfo {volume} {69}},\ \bibinfo {pages} {160--173}
  (\bibinfo {year} {1980})},\ \Eprint
  {http://arxiv.org/abs/http://oup.prod.sis.lan/ptps/article-pdf/doi/10.1143/PTP.69.160/5349369/69-160.pdf}
  {http://oup.prod.sis.lan/ptps/article-pdf/doi/10.1143/PTP.69.160/5349369/69-160.pdf}
  \BibitemShut {NoStop}%
\bibitem [{\citenamefont {V.~Pereverzev}, \citenamefont {Pereverzev},\ and\
  \citenamefont {Prezhdo}(2013)}]{pppJPSJ82}%
  \BibitemOpen
  \bibfield  {author} {\bibinfo {author} {\bibfnamefont {Y.}~\bibnamefont
  {V.~Pereverzev}}, \bibinfo {author} {\bibfnamefont {A.}~\bibnamefont
  {Pereverzev}}, \ and\ \bibinfo {author} {\bibfnamefont {E.}~\bibnamefont
  {Prezhdo}},\ }\bibfield  {title} {\enquote {\bibinfo {title} {Smoluchowski
  equation in cumulant approximation},}\ }\href {\doibase
  10.7566/JPSJ.82.024001} {\bibfield  {journal} {\bibinfo  {journal} {Journal
  of the Physical Society of Japan}\ }\textbf {\bibinfo {volume} {82}},\
  \bibinfo {pages} {024001} (\bibinfo {year} {2013})},\ \Eprint
  {http://arxiv.org/abs/https://doi.org/10.7566/JPSJ.82.024001}
  {https://doi.org/10.7566/JPSJ.82.024001} \BibitemShut {NoStop}%
\bibitem [{\citenamefont {Schneider}\ and\ \citenamefont
  {Freed}(2007)}]{Schneider89}%
  \BibitemOpen
  \bibfield  {author} {\bibinfo {author} {\bibfnamefont {D.~J.}\ \bibnamefont
  {Schneider}}\ and\ \bibinfo {author} {\bibfnamefont {J.~H.}\ \bibnamefont
  {Freed}},\ }\enquote {\bibinfo {title} {Spin relaxation and motional
  dynamics},}\ in\ \href {\doibase 10.1002/9780470141229.ch10} {\emph {\bibinfo
  {booktitle} {Advances in Chemical Physics}}}\ (\bibinfo  {publisher} {John
  Wiley\& Sons, Inc.},\ \bibinfo {year} {2007})\ pp.\ \bibinfo {pages}
  {387--527}\BibitemShut {NoStop}%
\bibitem [{\citenamefont {Tokuyama}(1980)}]{TokuyamaPA102}%
  \BibitemOpen
  \bibfield  {author} {\bibinfo {author} {\bibfnamefont {M.}~\bibnamefont
  {Tokuyama}},\ }\bibfield  {title} {\enquote {\bibinfo {title} {On the theory
  of fluctuations around non-equilibrium steady states: A generalized
  time-convolutionless projector formalism},}\ }\href {\doibase
  https://doi.org/10.1016/0378-4371(90)90174-Q} {\bibfield  {journal} {\bibinfo
   {journal} {Physica A: Statistical Mechanics and its Applications}\ }\textbf
  {\bibinfo {volume} {102}},\ \bibinfo {pages} {399 -- 430} (\bibinfo {year}
  {1980})}\BibitemShut {NoStop}%
\bibitem [{\citenamefont {Tokuyama}(1981)}]{TokuyamaPA109}%
  \BibitemOpen
  \bibfield  {author} {\bibinfo {author} {\bibfnamefont {M.}~\bibnamefont
  {Tokuyama}},\ }\bibfield  {title} {\enquote {\bibinfo {title}
  {Statistical-dynamical theory of nonlinear stochastic processes: Ii.
  time-convolutionless projector method in nonequilibrium open systems},}\
  }\href {\doibase https://doi.org/10.1016/0378-4371(81)90041-8} {\bibfield
  {journal} {\bibinfo  {journal} {Physica A: Statistical Mechanics and its
  Applications}\ }\textbf {\bibinfo {volume} {109}},\ \bibinfo {pages} {128 --
  160} (\bibinfo {year} {1981})}\BibitemShut {NoStop}%
\bibitem [{\citenamefont {Talkner}, \citenamefont {Lutz},\ and\ \citenamefont
  {H\"anggi}(2007)}]{tlhPRE75}%
  \BibitemOpen
  \bibfield  {author} {\bibinfo {author} {\bibfnamefont {P.}~\bibnamefont
  {Talkner}}, \bibinfo {author} {\bibfnamefont {E.}~\bibnamefont {Lutz}}, \
  and\ \bibinfo {author} {\bibfnamefont {P.}~\bibnamefont {H\"anggi}},\
  }\bibfield  {title} {\enquote {\bibinfo {title} {Fluctuation theorems: Work
  is not an observable},}\ }\href {\doibase 10.1103/PhysRevE.75.050102}
  {\bibfield  {journal} {\bibinfo  {journal} {Phys. Rev. E}\ }\textbf {\bibinfo
  {volume} {75}},\ \bibinfo {pages} {050102} (\bibinfo {year}
  {2007})}\BibitemShut {NoStop}%
\bibitem [{\citenamefont {Bachmann}, \citenamefont {Graf},\ and\ \citenamefont
  {Lesovik}(2010)}]{bJSP138}%
  \BibitemOpen
  \bibfield  {author} {\bibinfo {author} {\bibfnamefont {S.}~\bibnamefont
  {Bachmann}}, \bibinfo {author} {\bibfnamefont {G.~M.}\ \bibnamefont {Graf}},
  \ and\ \bibinfo {author} {\bibfnamefont {G.~B.}\ \bibnamefont {Lesovik}},\
  }\bibfield  {title} {\enquote {\bibinfo {title} {Time ordering and counting
  statistics},}\ }\href {\doibase 10.1007/s10955-009-9885-z} {\bibfield
  {journal} {\bibinfo  {journal} {Journal of Statistical Physics}\ }\textbf
  {\bibinfo {volume} {138}},\ \bibinfo {pages} {333--350} (\bibinfo {year}
  {2010})}\BibitemShut {NoStop}%
\bibitem [{Note2()}]{Note2}%
  \BibitemOpen
  \bibinfo {note} {As done by Kubo~\cite {kuboGenCumJPSJ17}, we could consider
  the case where the integral, in the r.h.s. of Eq-~(\ref {CumulantDefGen}) is
  substituted with a sum over the components of a $N$ dimensional vector of
  stochastic operators; however, for the sake of simplicity, we shall consider
  only the case of stochastic processes depending on a continuous parameter,
  e.g. the time. Adapting the results to the case of discrete, finite or
  infinite sets of stochastic operators is straightforward.}\BibitemShut
  {Stop}%
\bibitem [{\citenamefont {Meeron}(1957)}]{mJCP27}%
  \BibitemOpen
  \bibfield  {author} {\bibinfo {author} {\bibfnamefont {E.}~\bibnamefont
  {Meeron}},\ }\bibfield  {title} {\enquote {\bibinfo {title} {Series expansion
  of distribution functions in multicomponent fluid systems},}\ }\href
  {\doibase 10.1063/1.1743985} {\bibfield  {journal} {\bibinfo  {journal} {The
  Journal of Chemical Physics}\ }\textbf {\bibinfo {volume} {27}},\ \bibinfo
  {pages} {1238--1246} (\bibinfo {year} {1957})},\ \Eprint
  {http://arxiv.org/abs/https://doi.org/10.1063/1.1743985}
  {https://doi.org/10.1063/1.1743985} \BibitemShut {NoStop}%
\bibitem [{\citenamefont {Apresyan}(1978)}]{Apresyan1978}%
  \BibitemOpen
  \bibfield  {author} {\bibinfo {author} {\bibfnamefont {L.~A.}\ \bibnamefont
  {Apresyan}},\ }\bibfield  {title} {\enquote {\bibinfo {title} {Cumulant
  analysis of stochastic linear operators},}\ }\href {\doibase
  10.1007/BF01031669} {\bibfield  {journal} {\bibinfo  {journal} {Radiophysics
  and Quantum Electronics}\ }\textbf {\bibinfo {volume} {21}},\ \bibinfo
  {pages} {493--500} (\bibinfo {year} {1978})}\BibitemShut {NoStop}%
\bibitem [{\citenamefont {Fox}(1975)}]{fJMP16}%
  \BibitemOpen
  \bibfield  {author} {\bibinfo {author} {\bibfnamefont {R.~F.}\ \bibnamefont
  {Fox}},\ }\bibfield  {title} {\enquote {\bibinfo {title} {A generalized
  theory of multiplicative stochastic processes using cumulant techniques},}\
  }\href {\doibase 10.1063/1.522540} {\bibfield  {journal} {\bibinfo  {journal}
  {Journal of Mathematical Physics}\ }\textbf {\bibinfo {volume} {16}},\
  \bibinfo {pages} {289--297} (\bibinfo {year} {1975})},\ \Eprint
  {http://arxiv.org/abs/https://aip.scitation.org/doi/pdf/10.1063/1.522540}
  {https://aip.scitation.org/doi/pdf/10.1063/1.522540} \BibitemShut {NoStop}%
\bibitem [{\citenamefont {Arizmendi}\ \emph {et~al.}(2015)\citenamefont
  {Arizmendi}, \citenamefont {Hasebe}, \citenamefont {Lehner},\ and\
  \citenamefont {Vargas}}]{ahlvAM282}%
  \BibitemOpen
  \bibfield  {author} {\bibinfo {author} {\bibfnamefont {O.}~\bibnamefont
  {Arizmendi}}, \bibinfo {author} {\bibfnamefont {T.}~\bibnamefont {Hasebe}},
  \bibinfo {author} {\bibfnamefont {F.}~\bibnamefont {Lehner}}, \ and\ \bibinfo
  {author} {\bibfnamefont {C.}~\bibnamefont {Vargas}},\ }\bibfield  {title}
  {\enquote {\bibinfo {title} {Relations between cumulants in noncommutative
  probability},}\ }\href {\doibase https://doi.org/10.1016/j.aim.2015.03.029}
  {\bibfield  {journal} {\bibinfo  {journal} {Advances in Mathematics}\
  }\textbf {\bibinfo {volume} {282}},\ \bibinfo {pages} {56 -- 92} (\bibinfo
  {year} {2015})}\BibitemShut {NoStop}%
\bibitem [{\citenamefont {Roerdink}(1981)}]{rPA109}%
  \BibitemOpen
  \bibfield  {author} {\bibinfo {author} {\bibfnamefont {J.}~\bibnamefont
  {Roerdink}},\ }\bibfield  {title} {\enquote {\bibinfo {title} {Inhomogeneous
  linear random differential equations with mutual correlations between
  multiplicative, additive and initial-value terms},}\ }\href {\doibase
  https://doi.org/10.1016/0378-4371(81)90037-6} {\bibfield  {journal} {\bibinfo
   {journal} {Physica A: Statistical Mechanics and its Applications}\ }\textbf
  {\bibinfo {volume} {109}},\ \bibinfo {pages} {23 -- 57} (\bibinfo {year}
  {1981})}\BibitemShut {NoStop}%
\bibitem [{\citenamefont {Kutzelnigg}\ and\ \citenamefont
  {Mukherjee}(1999)}]{kmJCP6}%
  \BibitemOpen
  \bibfield  {author} {\bibinfo {author} {\bibfnamefont {W.}~\bibnamefont
  {Kutzelnigg}}\ and\ \bibinfo {author} {\bibfnamefont {D.}~\bibnamefont
  {Mukherjee}},\ }\bibfield  {title} {\enquote {\bibinfo {title} {Cumulant
  expansion of the reduced density matrices},}\ }\href {\doibase
  10.1063/1.478189} {\bibfield  {journal} {\bibinfo  {journal} {The Journal of
  Chemical Physics}\ }\textbf {\bibinfo {volume} {110}},\ \bibinfo {pages}
  {2800--2809} (\bibinfo {year} {1999})},\ \Eprint
  {http://arxiv.org/abs/https://doi.org/10.1063/1.478189}
  {https://doi.org/10.1063/1.478189} \BibitemShut {NoStop}%
\bibitem [{\citenamefont {Mazziotti}(2007)}]{MazziottiWiley2007}%
  \BibitemOpen
  \bibinfo {editor} {\bibfnamefont {D.~A.}\ \bibnamefont {Mazziotti}},\ ed.,\
  \href {\doibase 10.1002/0470106603} {\emph {\bibinfo {title}
  {Reduced-Density-Matrix Mechanics: With Application to Many-Electron Atoms
  and Molecules}}},\ \bibinfo {series} {ADVANCES IN CHEMICAL PHYSICS}\ No.\
  \bibinfo {number} {134}\ (\bibinfo  {publisher} {John Wiley {\&} Sons,
  Inc.},\ \bibinfo {year} {2007})\BibitemShut {NoStop}%
\bibitem [{\citenamefont {L\"owdin}(1955)}]{lPR97}%
  \BibitemOpen
  \bibfield  {author} {\bibinfo {author} {\bibfnamefont {P.-O.}\ \bibnamefont
  {L\"owdin}},\ }\bibfield  {title} {\enquote {\bibinfo {title} {Quantum theory
  of many-particle systems. i. physical interpretations by means of density
  matrices, natural spin-orbitals, and convergence problems in the method of
  configurational interaction},}\ }\href {\doibase 10.1103/PhysRev.97.1474}
  {\bibfield  {journal} {\bibinfo  {journal} {Phys. Rev.}\ }\textbf {\bibinfo
  {volume} {97}},\ \bibinfo {pages} {1474--1489} (\bibinfo {year}
  {1955})}\BibitemShut {NoStop}%
\bibitem [{\citenamefont {Bianucci}(2017)}]{bENTROPY19}%
  \BibitemOpen
  \bibfield  {author} {\bibinfo {author} {\bibfnamefont {M.}~\bibnamefont
  {Bianucci}},\ }\bibfield  {title} {\enquote {\bibinfo {title} {Large scale
  emerging properties from non {Hamiltonian} complex systems},}\ }\href
  {\doibase 10.3390/e19070302} {\bibfield  {journal} {\bibinfo  {journal}
  {Entropy}\ }\textbf {\bibinfo {volume} {19}} (\bibinfo {year} {2017}),\
  10.3390/e19070302}\BibitemShut {NoStop}%
\bibitem [{\citenamefont {Bianucci}(2015)}]{bJSTAT2015}%
  \BibitemOpen
  \bibfield  {author} {\bibinfo {author} {\bibfnamefont {M.}~\bibnamefont
  {Bianucci}},\ }\bibfield  {title} {\enquote {\bibinfo {title} {On the
  correspondence between a large class of dynamical systems and stochastic
  processes described by the generalized {Fokker Planck} equation with
  state-dependent diffusion and drift coefficients},}\ }\href
  {http://stacks.iop.org/1742-5468/2015/i=5/a=P05016} {\bibfield  {journal}
  {\bibinfo  {journal} {Journal of Statistical Mechanics: Theory and
  Experiment}\ }\textbf {\bibinfo {volume} {2015}},\ \bibinfo {pages} {P05016}
  (\bibinfo {year} {2015})}\BibitemShut {NoStop}%
\bibitem [{\citenamefont {Grigolini}\ and\ \citenamefont
  {Marchesoni}(1985)}]{grigo_memory}%
  \BibitemOpen
  \bibfield  {author} {\bibinfo {author} {\bibfnamefont {P.}~\bibnamefont
  {Grigolini}}\ and\ \bibinfo {author} {\bibfnamefont {F.}~\bibnamefont
  {Marchesoni}},\ }\bibfield  {title} {\enquote {\bibinfo {title} {Basic
  description of the rules leading to the adiabatic elimination of fast
  variables},}\ }in\ \href@noop {} {\emph {\bibinfo {booktitle} {Memory
  Function Approaches to Stochastich Problems in Condensed Matter}}},\ \bibinfo
  {series} {Advances in Chemical Physics}, Vol.\ \bibinfo {volume} {LXII},\
  \bibinfo {editor} {edited by\ \bibinfo {editor} {\bibfnamefont {M.~W.}\
  \bibnamefont {Evans}}, \bibinfo {editor} {\bibfnamefont {P.}~\bibnamefont
  {Grigolini}}, \ and\ \bibinfo {editor} {\bibfnamefont {G.~P.}\ \bibnamefont
  {Parravicini}}}\ (\bibinfo  {publisher} {An Interscience Publication, John
  Wiley \& Sons},\ \bibinfo {address} {New York},\ \bibinfo {year} {1985})\
  Chap.~\bibinfo {chapter} {II}, p.\ \bibinfo {pages} {556}\BibitemShut
  {NoStop}%
\bibitem [{\citenamefont {Zwanzig}(2001)}]{Zwanzig2001}%
  \BibitemOpen
  \bibinfo {editor} {\bibfnamefont {R.}~\bibnamefont {Zwanzig}},\ ed.,\ \href
  {http://ukcatalogue.oup.com/product/9780195140187.do} {\emph {\bibinfo
  {title} {Nonequilibrium Statistical Mechanics}}}\ (\bibinfo  {publisher}
  {Oxford University Press},\ \bibinfo {year} {2001})\BibitemShut {NoStop}%
\bibitem [{\citenamefont {Uchiyama}\ and\ \citenamefont
  {Shibata}(1999)}]{ucsfPRE60}%
  \BibitemOpen
  \bibfield  {author} {\bibinfo {author} {\bibfnamefont {C.}~\bibnamefont
  {Uchiyama}}\ and\ \bibinfo {author} {\bibfnamefont {F.}~\bibnamefont
  {Shibata}},\ }\bibfield  {title} {\enquote {\bibinfo {title} {Unified
  projection operator formalism in nonequilibrium statistical mechanics},}\
  }\href {\doibase 10.1103/PhysRevE.60.2636} {\bibfield  {journal} {\bibinfo
  {journal} {Phys. Rev. E}\ }\textbf {\bibinfo {volume} {60}},\ \bibinfo
  {pages} {2636--2650} (\bibinfo {year} {1999})}\BibitemShut {NoStop}%
\bibitem [{\citenamefont {Fox}(1974)}]{fJMP15}%
  \BibitemOpen
  \bibfield  {author} {\bibinfo {author} {\bibfnamefont {R.~F.}\ \bibnamefont
  {Fox}},\ }\bibfield  {title} {\enquote {\bibinfo {title} {Application of
  cumulant techniques to multiplicative stochastic processes},}\ }\href
  {\doibase 10.1063/1.1666835} {\bibfield  {journal} {\bibinfo  {journal}
  {Journal of Mathematical Physics}\ }\textbf {\bibinfo {volume} {15}},\
  \bibinfo {pages} {1479--1483} (\bibinfo {year} {1974})},\ \Eprint
  {http://arxiv.org/abs/https://doi.org/10.1063/1.1666835}
  {https://doi.org/10.1063/1.1666835} \BibitemShut {NoStop}%
\end{thebibliography}%
\end{document}